\documentstyle[lscape]{mn}

\begin{document}

\title[The Las Campanas IR Survey. II.]{The Las Campanas IR Survey. II. Photometric redshifts, comparison with models and clustering evolution.}
\author[A. E. Firth]
       {A. E. Firth$^{1}$, R. S. Somerville$^{1}$, R. G. McMahon$^{1}$, O. Lahav$^{1}$,  R. S. Ellis$^{3}$, 
\and C. N. Sabbey$^{1}$, P. J. McCarthy$^{2}$, H.-W. Chen$^{2}$, R. O. Marzke$^{2,5}$, J. Wilson$^{2}$, 
\and R. G. Abraham$^{1,4}$, M. G. Beckett$^{1,2}$, R. G. Carlberg$^{4}$, J. R. Lewis$^{1}$,
\and C. D. Mackay$^{1}$, D. C. Murphy$^{2}$, A. E. Oemler$^{2}$, S. E. Persson$^{2}$\\
       $^{1}$ Institute of Astronomy, University of Cambridge, Cambridge, CB3 0HA, UK\\
       $^{2}$ Observatories of the Carnegie Institution of Washington, Pasadena, CA 91101, USA\\
       $^{3}$ California Institute of Technology, Pasadena, CA 91125-2400, USA.\\
       $^{4}$ Department of Astronomy, University of Toronto, Toronto, M5S 3H8, Canada\\
       $^{5}$ Department of Astronomy and Physics, San Francisco State University, San Francisco, CA, USA\\
}
\date{Accepted by MNRAS, 20 December 2001}

\pagerange{\pageref{firstpage}--\pageref{lastpage}}
\pubyear{20??}

\maketitle

\label{firstpage}

\begin{abstract}
The Las Campanas IR (LCIR) Survey, using the Cambridge Infra-Red Survey Instrument\footnotemark\ (CIRSI), reaches $H \sim 21$ over nearly 1 degree$^2$.  In this paper we present results from 744 arcmin$^2$ centred on the Hubble Deep Field South for which $UBVRI$ optical data are publicly available.  Making conservative magnitude cuts to ensure spatial uniformity, we detect 3177 galaxies to $H = 20.0$ in 744 arcmin$^2$ and a further 842 to $H = 20.5$ in a deeper subregion of 407 arcmin$^2$.  We compare the observed optical-IR colour distributions with the predictions of semi-analytic hierarchical models and find reasonable agreement.  We also determine photometric redshifts, finding a median redshift of $\sim 0.55$.  We compare the redshift distributions $N(z)$ of E, Sbc, Scd and Im spectral types with models, showing that the observations are inconsistent with simple passive-evolution models while semi-analytic models provide a reasonable fit to the total $N(z)$ but underestimate the number of $z \sim 1$ red spectral types relative to bluer spectral types.  We also present $N(z)$ for samples of extremely red objects (EROs) defined by optical-IR colours.  We find that EROs with $R - H > 4$ and $H < 20.5$ have a median redshift $z_m \sim 1$ while redder colour cuts have slightly higher $z_m$. In the magnitude range $19 < H < 20$ we find that EROs with $R - H > 4$ comprise $\sim$18 per cent of the observed galaxy population, while in semi-analytic models they contribute only $\sim$4 per cent.

We also determine the angular correlation function $w(\theta)$ for magnitude, colour, spectral type and photometric redshift-selected subsamples of the data and use the photometric redshift distributions to derive the spatial clustering statistic $\xi(r)$ as a function of spectral type and redshift out to $z \sim 1.2$.  Parametrizing $\xi(r)$ by $\xi(r_c,z)=(r_c/r_{*}(z))^{-1.8}$ where $r_c$ is in comoving coordinates, we find that $r_*(z)$ increases by a factor of 1.5--2 from $z = 0$ to $z \sim 1.2$.  We interpret this as a selection effect -- the galaxies selected at $z \sim 1.2$ are intrinsically very luminous, about 1--1.5 magnitudes brighter than $L_*$.  When galaxies are selected by absolute magnitude we find no evidence for evolution in $r_*$ over this redshift range.  Extrapolated to $z = 0$, we find $r_{*}(z = 0)$ $\sim$ 6.5 $h^{-1}$Mpc for red galaxies and $r_{*}(z = 0)$ $\sim$ 2--4 $h^{-1}$Mpc for blue galaxies.  We also find that while the angular clustering amplitude of EROs with $R - H > 4$ or $I - H > 3$ is up to four times that of the whole galaxy population, the spatial clustering length $r_*(z = 1)$ is $\sim$7.5--10.5 $h^{-1}$Mpc which is only a factor of $\sim 1.7$ times $r_*(z = 1)$ for $R - H < 4$ and $I - H < 3$ galaxies lying in a similar redshift and luminosity range.  This difference is similar to that observed between red and blue galaxies at low redshifts.
\end{abstract}

\footnotetext{An instrument developed with the support of the Raymond and Beverly Sackler Foundation.}


\section{Introduction}  \label{sec.intro}
The recent availability of panoramic near-IR cameras -- such as the Cambridge Infra-Red Survey Instrument -- has opened up the window on the $1 < z < 2$ Universe.  While spectroscopic surveys such as the 2dF Galaxy Redshift Survey (Colless et al. 2001), CFRS (Le F\`evre et al. 1996), CNOC2 (Carlberg et al. 2000) and the Caltech Faint Galaxy Redshift Survey (Hogg, Cohen \& Blandford 2000) probe the $z < 1$ Universe and the Lyman-dropout technique (Steidal et al. 1996; Giavalisco et al. 1998) selects star-forming galaxies at $z \sim 3$, the redshift range $1 < z < 2$ has traditionally been `hidden' to optical astronomers due to the lack of prominent spectral features at optical wavelengths in galaxies at these redshifts.  In particular, massive evolved red galaxies become very faint at optical wavelengths above $z \sim 1$ as the \mbox{4000 \AA} break moves out of the $I$ band.  In contrast red galaxies remain prominent out to $z \sim 2$ in the near-IR.  Furthermore, rather than being sensitive to recent bursts of star formation, near-IR luminosity closely tracks total stellar mass over these redshifts and is less affected by dust obscuration, making comparisons with models more straightforward than in optically-selected surveys.

Determining the number density of massive evolved galaxies at these redshifts is an important test for galaxy evolution models.  In the traditional passive-evolution model (Eggen, Lynden-Bell \& Sandage 1962; Sandage 1986), such galaxies form in single monolithic collapses at high redshift and then evolve passively with no new star formation, thus giving rise to a constant comoving density of massive galaxies to high redshifts.  On the other hand, in hierarchical merger models massive elliptical galaxies formed relatively recently from the merging of smaller disc galaxies (White \& Rees 1978).  Thus it has been proposed that at $z \sim$ 1--2 the two models should give very different predictions for the number density of massive evolved galaxies (Kauffmann \& Charlot 1998).

Likewise, measurements of the three-dimensional clustering of galaxies in the redshift range $1 < z < 2$ fill an important gap in our knowledge of the evolution of large-scale structure between $z < 1$ and $z \sim 3$ that until recently has been filled only by deep pencil-beam surveys such as the Hubble Deep Fields (Arnouts et al. 1999; Magliocchetti \& Maddox 1999) whose small fields of view and small sample size inhibit a proper clustering analysis. 

\begin{figure}
\vspace{10cm}
\includegraphics{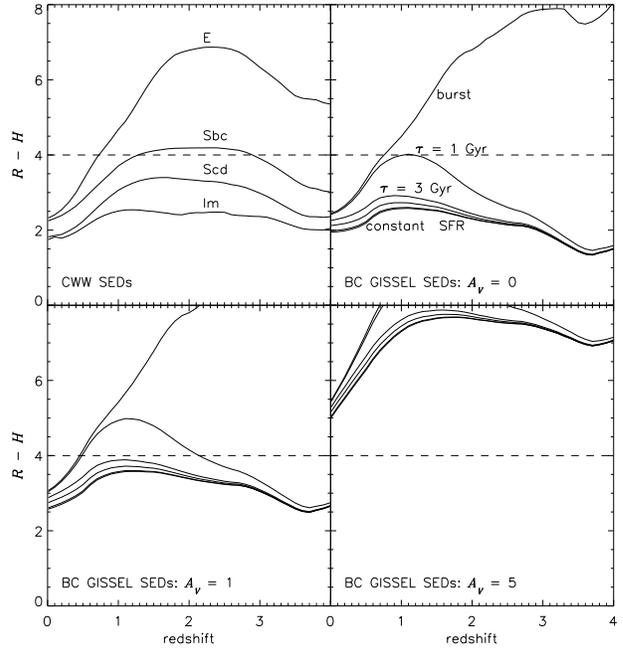}
\caption{\label{r-h} $R - H$ colour as a function of redshift for empirical SEDs (Coleman, Wu \& Weedman 1980; CWW) E, Sbc, Scd, Im and for evolving Bruzual \& Charlot (1993; BC) {\tt GISSEL'98} SEDs with single stellar population burst, exponentially decaying with time-scale $\tau$ = 1, 3, 5 and 15 Gyr and constant star formation rate, all with formation redshift $z_f = 10$, using an $\Omega_0 = 0.3$, $\Omega_{\Lambda} = 0.7$ and $H_0$ = 70 km s$^{-1}$ Mpc$^{-1}$ cosmology (see $\S$\ref{sec.models}.2 for details).  The lower panels show BC SEDs with varying amounts of reddening (Calzetti et al. 2000).  The CWW SEDs have not been reddened since they are empirical and intrinsically include some reddening.  Only the most evolved $z > 0.7$ or very dust-reddened galaxies are picked out by an $R - H > 4$ colour cut.}
\end{figure}

\begin{figure}
\vspace{10cm}
\includegraphics{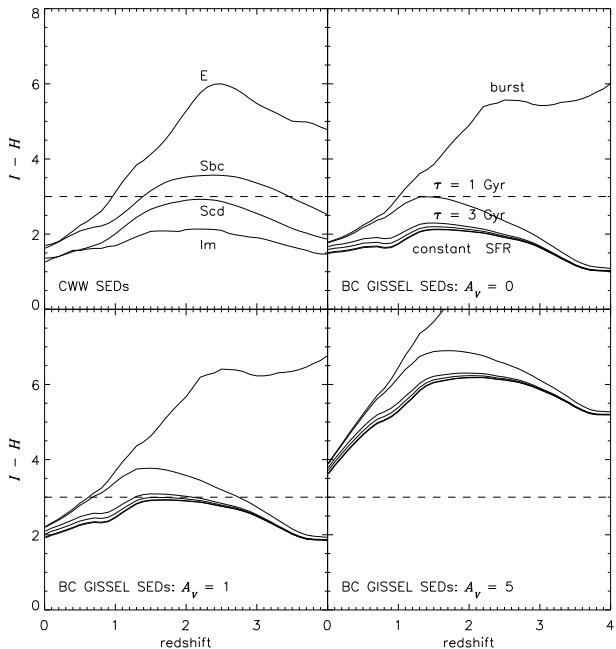}
\caption{\label{i-h} $I - H$ colour as a function of redshift for empirical SEDs (Coleman, Wu \& Weedman 1980; CWW) E, Sbc, Scd, Im and for evolving Bruzual \& Charlot (1993; BC) {\tt GISSEL'98} SEDs with single stellar population burst, exponentially decaying with time-scale $\tau$ = 1, 3, 5 and 15 Gyr and constant star formation rate, all with formation redshift $z_f = 10$, using an $\Omega_0 = 0.3$, $\Omega_{\Lambda} = 0.7$ and $H_0$ = 70 km s$^{-1}$ Mpc$^{-1}$ cosmology (see $\S$\ref{sec.models}.2 for details).  The lower panels show BC SEDs with varying amounts of reddening (Calzetti et al. 2000).  The CWW SEDs have not been reddened since they are empirical and intrinsically include some reddening.  Only the most evolved $z > 0.9$ or very dust-reddened galaxies are picked out by an $I - H > 3$ colour cut.}
\end{figure}

One result of early near-IR surveys (Elston, Rieke \& Rieke 1988; McCarthy, Persson \& West 1992; Hu \& Ridgway 1994; Cowie et al. 1994; Moustakas et al. 1997; Barger et al. 1999; Thompson et al. 1999) has been the discovery of populations of extremely red objects (EROs) defined in the literature by various extreme optical-infrared colours, typically $R - K > 5$ or 6 which roughly corresponds to $R - H > 4$ or 5 and $I - H > 3$ or 4.  These colours are characteristic of elliptical galaxies at $z \sim 1$ or greater (Fig. \ref{r-h}, \ref{i-h}) -- so that the \mbox{4000 \AA} break falls between the optical and the near-IR filters, high-redshift highly-reddened dusty star-forming galaxies, obscured AGN and low-mass stars.  Deep spectroscopy of a few such objects (Dunlop et al. 1996; Graham \& Dey 1996; Cohen et al. 1999; Cimatti et al. 1999; Liu et al. 2000) and morphological studies using the Hubble Space Telescope (Moriondo, Cimatti \& Daddi 2000; Stiavelli \& Treu 2000) suggest that 70--80 per cent are evolved massive ellipticals at $z \sim$ 1--1.5. Submillimetre observations (Cimatti et al. 1998; Smail et al. 1999; Dey et al. 1999), spectroscopy and morphology show that others, especially the reddest, are $z > 1$ dusty starbursts.  Initial estimates of the number densities of EROs varied considerably (Barger et al. 1999; Thompson et al. 1999) showing that wider field surveys would be necessary to obtain meaningful statistics.  This was later confirmed by measurements of very strong angular clustering (Daddi et al. 2000a; McCarthy et al. 2000).  The strong clustering supports the interpretation that a majority of these EROs are massive galaxies in a narrow redshift range constrained at $z \sim 1$ by the red colour cut and at $z \sim 1.5$ by the limiting magnitudes of current surveys (and/or spectral evolution). However the true significance of the measured clustering amplitude was unclear, given that any population of objects at the bright end of the luminosity function and in a restricted redshift range may be expected to appear strongly clustered.  A more informative picture comes from the three-dimensional clustering scale which may be inferred from angular clustering measurements provided the redshift distribution $N(z)$ can be estimated.

Recognizing the above we began a deep wide-field survey (1 degree$^2$) in several optical and near-IR filters to study evolved galaxies at $z > 1$ and obtain statistically significant samples of EROs.  By applying a photometric redshifts template-fitting technique (Bolzonella, Miralles \& Pell\'o 2000) we assign an approximate redshift and best-fitting spectral type to each galaxy in the sample, thus enabling us to calculate redshift distributions for different spectral types and make more detailed comparisons with models than are possible with simple colour-selected samples.  Photometric redshifts may also be used to obtain the projected clustering amplitude in redshift shells and, while they are not accurate enough to directly measure three-dimensional clustering, the estimated redshift distributions may be used along with Limber's equation to infer the three-dimensional clustering from the measured projected clustering amplitudes.  By dividing the projection axis between several redshift shells, the signal-to-noise in clustering measurements is increased relative to simple imaging surveys (see also Brunner, Szalay \& Connolly 2000; Teplitz et al. 2001; Brown, Boyle \& Webster 2001).  Furthermore, with photometric redshifts one can avoid mixing different spectral types, and intrinsic luminosities while still obtaining much larger sample sizes than are currently possible with spectroscopic surveys at these redshifts and magnitudes.

This paper presents results for one of our fields covering 744 arcmin$^2$ in which we have data in the $UBVRIH$ filters.  The field is slightly larger than that used by Daddi et al. (2000a) and has the advantage of enough optical filters to determine photometric redshifts for all objects.  A companion paper (McCarthy et al. 2001) presents red object number counts and clustering results to $H \sim 21$ for a further three fields in which we currently have deep $H$ imaging but less extensive optical coverage while a third paper (Chen et al. 2001) presents a broad overview of the survey and initial results.  The structure of this paper is as follows.  In $\S$\ref{sec.obs} we briefly describe data acquistion and reduction procedures.  In $\S$\ref{sec.cat} we describe catalogue generation and characteristics. In $\S$\ref{sec.photoz} we review the photometric redshift technique and present simulations relevant to the particular survey filters and S/N characteristics and in $\S$\ref{sec.sg} we discuss star-galaxy separation methods.  In $\S$\ref{sec.models} we introduce several galaxy evolution models: a semi-analytic hierarchical merger model and simple no-evolution and passive-evolution models, and in $\S$\ref{sec.cf} we compare number counts, colour-colour, colour-magnitude and $N(z)$ distributions between the observations and the models.  In $\S$\ref{sec.w} we present our angular clustering results and in $\S$\ref{sec.xi} we present our spatial clustering results.  Finally $\S$\ref{sec.discusion} and $\S$\ref{sec.summary} contain a discussion and summary.  We use the Vega magnitude system throughout the paper, spatial clustering lengths are expressed using comoving coordinates and we write the Hubble constant as $H_0 = 100 h$ km s$^{-1}$ Mpc$^{-1}$.

\addtocounter{footnote}{-1}

\section{The Observations}   \label{sec.obs}
\begin{figure*}
\vspace{16.2cm}
\includegraphics{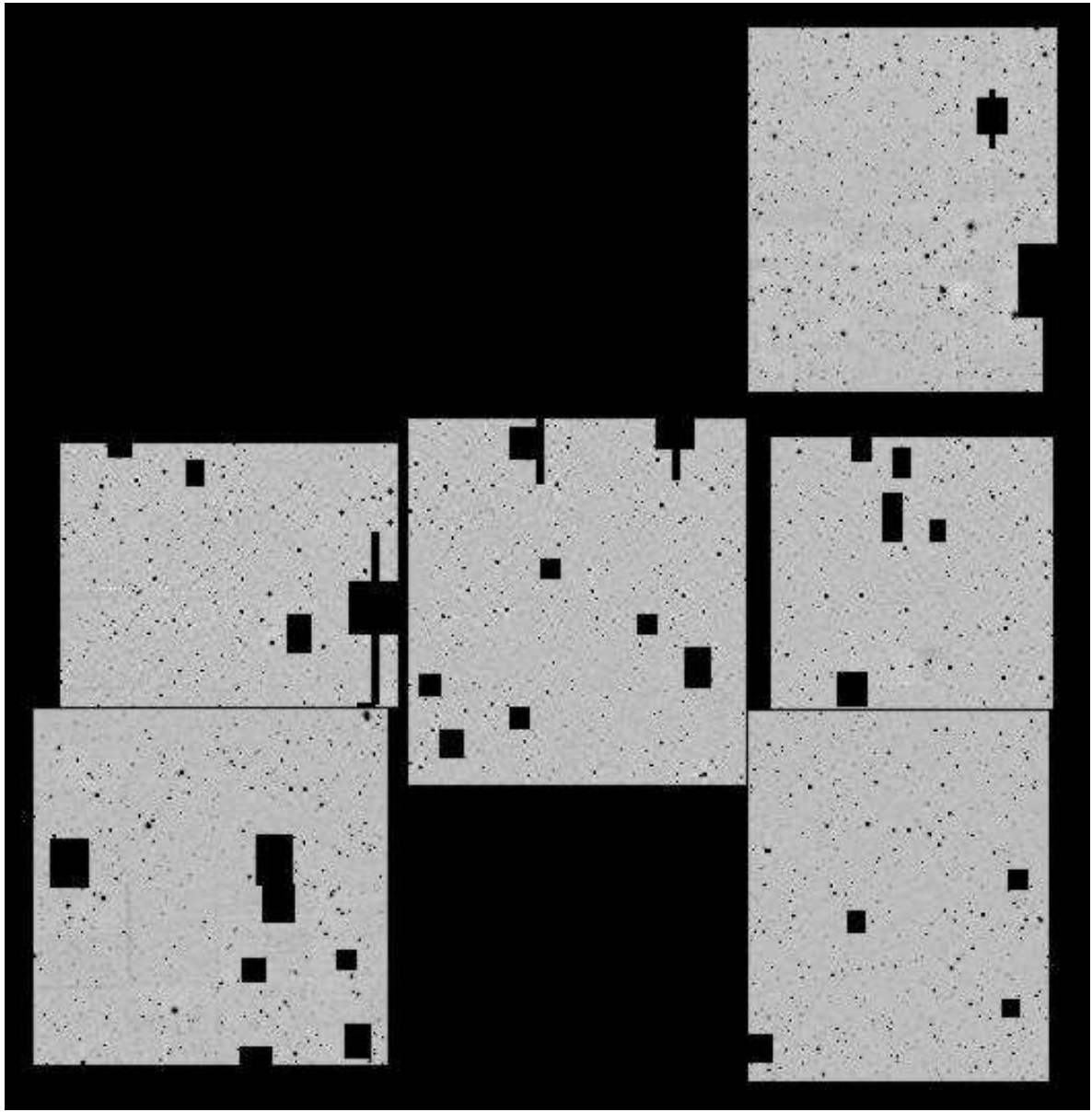}
\caption{\label{mask} The $H$-band image with the positions of the masked regions over bright stars marked.  Each of the 6 tiles measures roughly 13 $\times$ 13 arcmin$^2$.  The width of the full image corresponds to $\sim 27$ $h^{-1}$Mpc in comoving coordinates ($\sim 13.5$ $h^{-1}$Mpc in physical coordinates) at $z = 1$ for an $\Omega_{0} = 0.3$, $\Omega_{\Lambda} = 0.7$ cosmology.  Assuming this cosmology and calculating volumes to the median (photometric or spectroscopic) redshift of a survey, the volume probed by the HDFS field of the LCIR survey is approximately (40 $h^{-1}$Mpc)$^3$ ($z_m \sim 0.55$).  This compares with $\sim$ (14 $h^{-1}$Mpc)$^3$ for the Hubble Deep Field North (Williams et al. 1996) (assuming $z_m = 1.2$) or $\sim$ (200 $h^{-1}$Mpc)$^3$ for the 2dF Galaxy Redshift Survey with $z_m \sim 0.11$ (Colless et al. 2001). (Note on astro-ph version: some background structure is due to image compression.)}
\end{figure*}

The Las Campanas IR (LCIR) Survey covers 24 tiles arranged in 6 disparate groups at high galactic latitude.  Each square tile is $\sim$13 arcmin on a side, corresponding to $\sim 8.7$ $h^{-1}$Mpc in comoving coordinates ($\sim 4.4$ $h^{-1}$Mpc in physical coordinates) at $z = 1$ for an $\Omega_{0} = 0.3$, $\Omega_{\Lambda} = 0.7$ cosmology.  In this paper we present results for 6 tiles centred on the Hubble Deep Field South (HDFS) at R.A. 02:33:13, Dec. $-$60:39:27.  In these tiles we have approximately 80 minutes per pixel of $H$ imaging (Fig. \ref{mask}) obtained with the Cambridge Infra-Red Survey Instrument\footnotemark \footnotetext{An instrument developed with the support of the Raymond and Beverly Sackler Foundation.} (CIRSI) on the 2.5-m Du Pont Telescope at Las Campanas Observatory in September--October 1999 and 2000, and optical $UBVRI$ imaging obtained by the Goddard Space Flight Center with the Big Throughput Camera (BTC) at Cerro Tololo Inter-American Observatory in September 1998 and made publicly available (Palunas et al.\footnotemark \footnotetext{http://hires.gsfc.nasa.gov/$\sim$research/hdfs-btc/} 2000; see Teplitz et al. 2001 for a photometric redshift and clustering analysis of the full 0.5 degrees$^2$).  The CIRSI camera (Beckett et al. 1998) has 4 HgCdTe 1024 $\times$ 1024 detector-chips spaced at $\sim$90 per cent of a chip width so that stepping the camera four times fills in the gaps between the chips.  At the Du Pont the 4k $\times$ 4k mosaic covers 13 $\times$ 13 arcmin$^2$ with a pixel scale of $\sim$0.2 arcsec per pixel.  The BTC has a pixel scale of 0.43 arcsec per pixel.  The seeing FWHM in the reduced $H$ images is typically 0.9--1.1 arcsec.  In the optical images the seeing FWHM is typically 1.5--1.6 arcsec.

In the near-IR, high background limits the length of exposures.  We used an exposure time of 45 s, dithering by $\sim$10 arcsec after each 3--4 consecutive exposures.  Between 5 and 9 dithers were completed at one telescope pointing before moving to the next mosaic position.  Fields were observed over the course of several nights to make up the full 80 minutes per pixel exposure time.  Standards stars from Persson et al. (1998) were observed on chip 1 periodically throughout the night in photometric conditions using a 5 point dither pattern.  At least once per observing run standards were observed on all four chips.  Data taken during non-photometric conditions were either discarded or calibrated to data taken during photometric conditions.

Reduction of the infrared data involved the following steps: (1) flat-fielding using the difference of illuminated and dark dome flats, (2) subtraction of a sky frame (the running median of typically 3 adjacent frames on either side), (3) subtraction of the column and row modes (in order to remove electronic ramp effects), (4) coaddition of the resulting frames, (5) detection and masking of objects, (6) repetition of steps (2)--(3) with objects masked and (7) final coaddition and mosaicing of the resulting frames.  Further details of the observing strategy and data reduction are given in Chen et al. (2001) and Sabbey et al. (2001).

\section{Object catalogues}  \label{sec.cat}
We begin by defining our system of measuring object magnitudes.  In general one may define an object's magnitude to be the sum of the flux within a given aperture, typically defined to be some isophote (isophotal magnitudes) or a larger aperture whose dimensions are designed to enclose essentially all of the object's flux (e.g. Kron and Petrosian magnitudes).  While in principle these magnitudes measure the total flux from an object, they also suffer from decreased signal-to-noise in the faint wings of extended objects and, further, at faint magnitudes the aperture can be poorly defined.  An alternative approach is to measure the flux for every object within some circular aperture of fixed diameter.  The disadvantage of this method is that for very extended or nearby galaxies a significant fraction of the galaxy's flux falls outside the aperture.  However in this paper we are interested mainly in determining photometric redshifts, for which the important quantity is an as accurate as possible determination of each object's colours.  In particular, magnitude measures which use a different aperture in different filters are generally inappropriate.  For the observations presented here we find that a 3 arcsec diameter aperture provides a good compromise between missing flux from smaller apertures and decreased signal-to-noise in larger apertures while being sufficiently large compared with the range of seeing FWHM values (0.9--1.6 arcsec over the different filters) to be robust with respect to seeing variations across individual tiles.  

The fraction of flux scattered outside the aperture depends on the seeing.  In order to compute object colours it is necessary to correct for the different seeing FWHM in the different filters.  We determine this correction from bright stars selected in each filter and tile.  The object detection software package {\tt SExtractor version 2.1.6} (Bertin \& Arnouts\footnotemark \footnotetext{http://terapix.iap.fr/sextractor/} 1996) was used to detect objects on all images and between 30 and 60 bright non-saturated stars were selected in each filter in each of the 6 tiles using {\tt SExtractor}'s neural network star/galaxy separation parameter {\tt CLASS\_STAR}.  From these, the median seeing FWHM and magnitude offset between 3 arcsec and 10 arcsec diameter apertures was determined for each tile in each filter. Given the seeing FWHM of these images, a 10 arcsec aperture magnitude is within 0.02 magnitudes of a total magnitude for a point source.  For LCIR survey objects, the magnitudes that we quote throughout this paper are 3 arcsec aperture magnitudes corrected to 10 arcsec with the seeing correction for the relevant filter.  For compact sources these are essentially total magnitudes.  For extended objects these underestimate total magnitudes.

The degree to which total magnitudes are underestimated was investigated using the {\tt iraf}\footnotemark \footnotetext{http://iraf.noao.edu/} package {\tt noao.artdata}.  This package allows one to place simulated galaxies of various light profiles and magnitudes on to an image.  One can then measure magnitudes using the same methods that are used for the data and compare the measured magnitudes with the input magnitudes to quantify any offset. The estimated flux missed from a 3 arcsec aperture is listed, for various object profiles, in Table \ref{table.apcorr}.  As noted in the table caption, these values are insensitive to seeing provided the relevant seeing corrections are first applied to the 3 arcsec aperture magnitudes.  At $H = 20.0$ we expect typical half-light radii of $\sim$0.5 arcsec (Smail et al. 1995; Yan et al. 1998; Corbin et al. 2000) giving magnitude offsets of order 0.1--0.3 at the faint end, depending on the mix of profiles.  We emphasis that we do not correct the observations for this `aperture effect' as the half-light radius of a given galaxy is poorly determined at the faint end and we are in any case mainly interested in object colours.

\begin{table*}
\centering
\caption{\label{table.apcorr} Table of the magnitude offsets between the 3 arcsec aperture magnitudes used in this paper and actual total magnitudes for various light profiles, as measured with the {\tt iraf} task {\tt noao.artdata}.  $r_h$ is the projected radius containing half the total flux.  Object light profiles are convolved with a seeing disc of FWHM 1 arcsec before being placed on to a data image.  As with the actual observations, object magnitudes are measured in a 3 arcsec diameter aperture then corrected for seeing using the offset from 3 to 10 arcsec aperture magnitudes as determined from bright stars in the same image.  The resulting `seeing-corrected' aperture magnitude is compared with the input total magnitude.  For stars the offset is 0.  For extended objects extra flux is missed from a 3 arcsec aperture besides that scattered by the seeing disc, leading to non-zero offsets.  The same procedure repeated for different seeing FWHM values in the range covered by the data (viz. 0.9--1.6 arcsec) produces very similar offsets.}
\begin{tabular}{cccccccc}
&&\multicolumn{2}{c}{axis ratio = 0.3}&\multicolumn{2}{c}{axis ratio = 0.7}&\multicolumn{2}{c}{axis ratio = 1.0}\\
$r_h$ (arcsec)& stellar & exponential disc & de Vaucouleurs & exponential disc & de Vaucouleurs & exponential disc & de Vaucouleurs\\ \hline
-    &0&      &      & &      &     \\
0.25 & & 0.01 & 0.14 & 0.02 & 0.15 & 0.02 & 0.16\\
0.50 & & 0.06 & 0.20 & 0.09 & 0.26 & 0.12 & 0.30\\
0.75 & & 0.15 & 0.29 & 0.21 & 0.38 & 0.28 & 0.44\\
1.00 & & 0.25 & 0.38 & 0.35 & 0.50 & 0.46 & 0.58\\
1.50 & & 0.47 & 0.56 & 0.66 & 0.71 & 0.82 & 0.80\\
2.00 & & 0.68 & 0.69 & 0.95 & 0.87 & 1.15 & 0.98\\ \hline
\end{tabular}
\end{table*}

\begin{figure}
\vspace{7cm}
\includegraphics{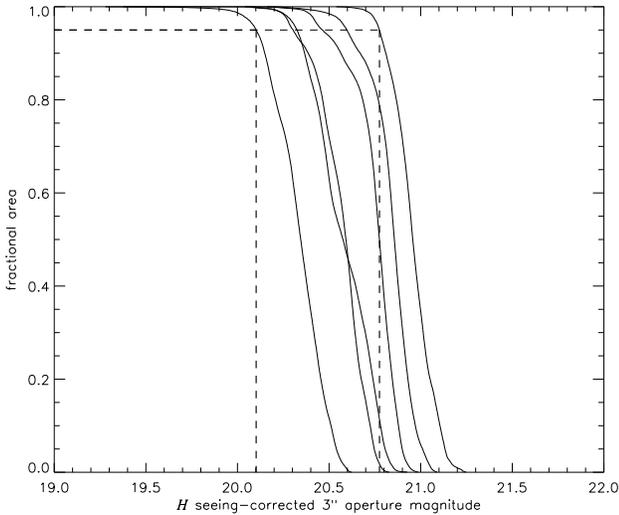}
\caption{\label{completeness} We define the limiting $H$ magnitude at a given {\it point} to be the seeing-corrected $H$ magnitude that corresponds to a 7$\sigma$ detection in a 3 arcsec aperture centred at that point.  We determine these limiting $H$ magnitudes on a grid of points across each tile and plot the corresponding cumulative fractional area versus limiting $H$ magnitude.  The six lines correspond to the six tiles.  The nominal $H$ magnitude limit in each {\it tile} is chosen so that 95 per cent of the tile's area has a limiting magnitude fainter than the tile limiting magnitude (dashed lines indicate these limits for two of the six tiles).}
\end{figure}

A first catalogue was made using {\tt SExtractor} to detect objects on the $H$ image with the threshold for object detection set at 13 contiguous pixels 1.3$\sigma$ above the local sky r.m.s.  For comparison, an object with a gaussian profile of FWHM equal to the seeing, which registers a 7$\sigma$ detection in a 3 arcsec diameter aperture, will have $\sim$26 (depending on the particular tile) pixels 1.3$\sigma$ above the sky r.m.s.  Broader profiles at the same magnitudes register fewer pixels above 1.3$\sigma$, so we also investigated the detection efficiency as a function of object profile (see below).  The coordinate transformations between the $H$ images and the optical images were calculated using the {\tt iraf} task {\tt geomap}. Between 50 and 200 stars were used in each tile to derive the transformation and the resulting r.m.s. was less than 0.2 arcsec. The {\tt iraf} task {\tt noao.digiphot.apphot.phot} was used to measure 3 arcsec aperture magnitudes at the coordinates of the object centres in each filter.  An $H$-band detection limit was estimated in each tile by measuring the sky r.m.s. in a grid of apertures over the image (with $\sigma$-clipping to remove objects).  From these values the background noise in 3 arcsec apertures and the required fluxes to register 7$\sigma$ detections were estimated at each point in the grid.  In each tile the limiting flux/magnitude was derived from the 95th percentile of these fluxes -- i.e. a source at this limiting (aperture) magnitude is sufficiently bright to register a 7$\sigma$ or greater detection in a 3 arcsec aperture over 95 per cent of the tile's area (Fig. \ref{completeness}).  These limits were applied to the catalogue giving 3 tiles (337 arcmin$^2$) at $H < 20.0$ and 3 tiles (407 arcmin$^2$) at $H < 20.5$ (where these limiting magnitudes include the seeing correction).  While some regions of the survey go deeper (see Chen et al. 2001) we make these cuts to ensure a spatially uniform magnitude limit -- making a clustering analysis more straightforward.  

\begin{figure}
\vspace{7cm}
\includegraphics{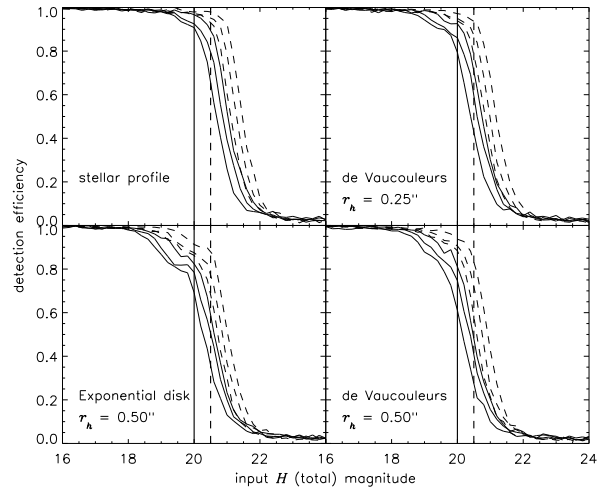}
\caption{\label{artdata} Detection efficiency as a function of input (total) magnitude for various light profiles as determined from simulations using the {\tt iraf} task {\tt noao.artdata}.  For each magnitude and light profile, 1000 simulated objects are placed onto one of the LCIR survey $H$-band tiles after convolving with the appropriate seeing disk (0.9--1.1 arcsec depending on the tile).  The fraction recovered with {\tt SExtractor}, using the detection parameters described in $\S$\ref{sec.cat}, is plotted as a function of magnitude.  The solid lines correspond to the three tiles with a nominal limiting magnitude of $H = 20.0$ and the dashed lines correspond to the three tiles with a nominal limiting magnitude of $H = 20.5$.  At $H = 20.0$ we expect typical half-flux radii of $r_h \sim 0.5$ arcsec for which the detection efficiency is 70--90 per cent, reaching 90--100 per cent for more compact galaxy profiles.}
\end{figure}

Since more extended galaxies, with the same total magnitude, yield fewer pixels above the 1.3$\sigma$ threshold than compact galaxies, there is an almost inevitable bias against extended or low surface brightness galaxies, especially at the magnitude limit.  We measured the efficiency with which {\tt SExtractor} detects objects using the {\tt iraf} package {\tt noao.artdata} (see also Gray et al. 2000; Chen et al. 2001).  For a range of magnitudes surrounding the nominal limiting magnitude in each $H$-band tile, and for a variety of light profiles, 1000 simulated objects were placed onto the tile.  The fractions recovered, using {\tt SExtractor} with the detection parameters described above, are plotted in Fig. \ref{artdata}.  At the nominal limiting magnitudes, the detection efficiency for stellar and compact sources is 90--100 per cent.  This drops to 70--90 per cent for objects with half-flux radii $r_h \sim 0.5$ arcsec.  Thus there may be some bias against such objects at the faintest magnitudes but not enough to greatly affect our results.

After applying the $H$ magnitude limits to the initial catalogue, saturated or nearly saturated objects were also deleted from the catalogue and regions around highly saturated objects were excised, for while bright objects are not very prominent in the $H$ image and are unlikely to have much effect on the estimated clustering in an $H$-selected sample, they could affect the photometry of nearby objects in the $R$ and $I$ images and hence affect photometric redshifts.  Furthermore, regions where object apertures overlap the image edges were masked.  Every object was inspected visually in $H$ and $I$ in order to remove any spurious detections.  

The final catalogue contains a total of 5401 objects to $H=20.0$ and a further 1029 to $H=20.5$.  A significant fraction are however stars (see $\S$\ref{sec.sg}).  The majority of objects have S/N $>$ 5 in $RIH$, and $>$ 2.5 in $BV$.  While the extremely red objects have low S/N in the optical filters they have prominent \mbox{4000 \AA} breaks which aid photometric redshifts.

\section{Photometric redshifts}  \label{sec.photoz}
In order to determine spectral types and approximate redshifts for the objects in our survey we use the publicly available photometric redshift code {\tt hyperz}\footnotemark \footnotetext{http://webast.ast.obs-mip.fr/hyperz/} (Bolzonella et al. 2000; BMP).  The basic procedure is to compare the measured colours of each object with a library of template spectral energy distributions (SEDs) on a grid of redshifts.  The best-fitting template SED and redshift are determined by minimising

\begin{equation}
\chi^2 = \sum_{i}\left(\frac{f_{i}-\alpha t_{i}}{\sigma_{i}}\right)^{2},
\end{equation}

\noindent where $f_{i}$ is the observed flux in filter $i$, $\sigma_{i}$ is the error in $f_{i}$, $t_{i}$ is the template flux in filter $i$ and $\alpha$ is the scaling factor normalizing the template to the observed flux.  We use the four template SEDs of Coleman, Wu \& Weedman (1980; CWW) that are distributed with {\tt hyperz}.  These have been extended into the UV and near-IR with Bruzual \& Charlot (1993; BC) spectral synthesis models and correspond roughly to the spectra of E, Sbc, Scd and Im galaxies.

\begin{figure}
\vspace{7cm}
\includegraphics{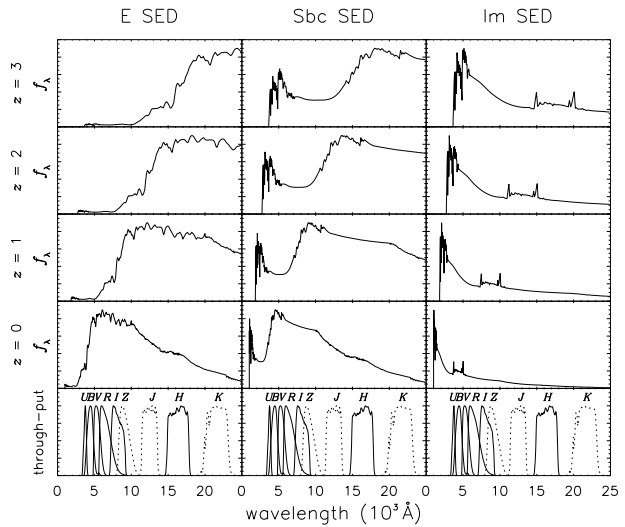}
\caption{\label{filters} Plots of template E (elliptical), Sbc (spiral) and Im (irregular) SEDs placed at various redshifts, illustrating the \mbox{912 \AA} break in blue galaxies and the \mbox{4000 \AA} break in red galaxies. The survey filters are illustrated below (solid lines represent the filters available for this paper, dotted lines represent filters available in some of our other fields).}
\end{figure}

The survey filters are displayed in Fig. \ref{filters} along with the spectral energy distributions (SEDs) of the E (elliptical), Sbc (spiral) and Im (irregular) galaxy templates redshifted to $z = 0$, 1, 2 and 3.  In our survey fields we have various combinations of the filters $UBVRIZJHK$ though in the field used in this paper we only have the filters $UBVRIH$.  This set of filters allows reasonable photometric redshift determination for all galaxies: spectral-type E galaxies have a prominent \mbox{4000 \AA} break that lies between the survey filters over the redshift range of interest; conversely, while spectral-type Im galaxies lack prominent spectral features at appropriate wavelengths, in an $H$-selected sample they have relatively high S/N in the optical filters.

\begin{figure}
\vspace{7cm}
\includegraphics{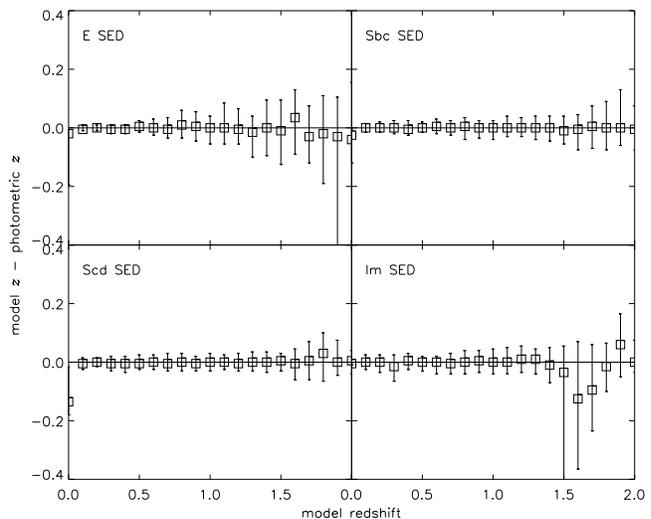}
\caption{\label{CWW1} The median offsets between model redshifts and photometric redshifts for different template SEDs with $H = 20$ and noise added to mimic the observations.  Error bars enclose the central 68 per cent of offsets.}
\end{figure}

\begin{figure}
\vspace{7cm}
\includegraphics{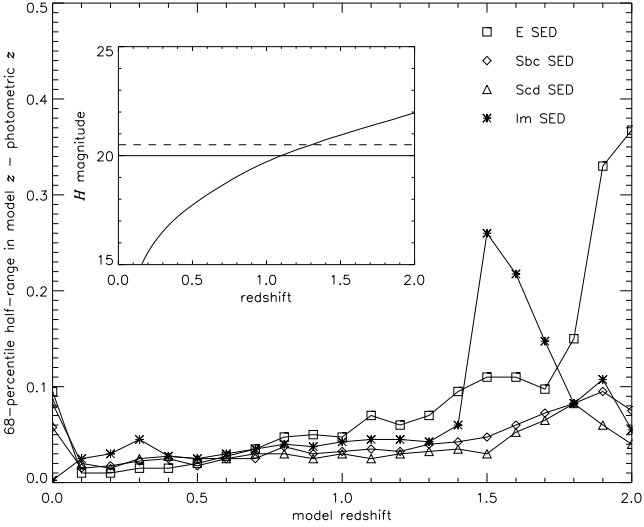}
\caption{\label{CWW2} The scatter in photometric redshifts relative to model redshifts -- measured by the half-range of the central 68 per cent of offsets -- for different template SEDs with $H = 20$ and noise added to mimic the observations.  The inset shows the expected $H$ magnitude of a non-evolving (k-correction only) 2$L_*$ E galaxy (assuming $M_* = -23.1 + 5 \log h$ in the $K$ band (Gardner et al. 1997)) as a function of redshift with the two limiting $H$ magnitudes used in this paper ($H < 20.0$ and $H < 20.5$) marked.  The relatively poor performance with respect to the E SED at $z \sim$ 2 is due to the \mbox{4000 \AA} break moving completely out of the optical filters so that there is too little information to pin-point its position.  The addition of $Z$ and $J$ filters would remove this problem.  At $z < 1.5$ -- where we expect most of the objects in this paper to lie -- the scatter in $\Delta z$ is less than 0.1.}
\end{figure}

\begin{figure}
\vspace{7cm}
\includegraphics{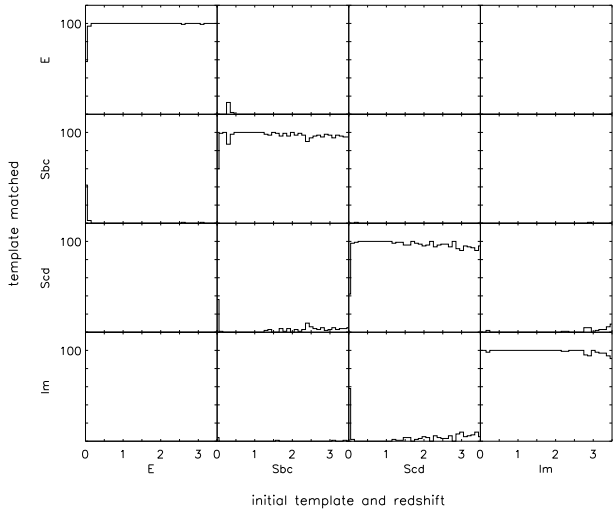}
\caption{\label{CWW3} The number of galaxies misidentified as a different spectral type as a function of redshift for different template SEDs with $H = 20$ and noise added to mimic the observations.  In the majority of cases the spectral type is identified correctly, and the few mistakes are generally within one spectral type class of the correct class.}
\end{figure}

In any application of photometric redshifts to a galaxy survey it is important to investigate the magnitude and characteristics of photometric redshift errors for the particular filter set and limiting magnitudes used in the survey.  We considered each of the four template SEDs (E, Sbc, Scd, Im) and calculated $UBVRI$ magnitudes for $H = 20$ galaxies on a grid of redshifts from 0 to 3.5.  We then added random noise in accordance with the noise present in the data at the relevant magnitude in each filter.  In addition, a minimum error term of 0.05 magnitudes r.m.s. was imposed since, given calibration errors, aperture correction errors (if the seeing FWHM varies over a tile) and other potential errors, there is necessarily some error in the photometric zeropoint of this order.  One hundred randomized replicates of each SED/redshift combination were created and photometric redshifts were estimated for the resulting catalogue.  The medians of the ranges of offsets $\Delta z = z_{\rm model} - z_{\rm photometric}$ are very small over the redshift range of interest (Fig. \ref{CWW1}).  The scatter in $\Delta z$ -- measured by the half-range of the central 68 per cent of offsets -- is plotted in Fig. \ref{CWW2}.  For $z < 1.5$, which is where we expect most of our galaxies to lie, the scatter in $\Delta z$ is less than 0.1. Also of interest is the number of galaxies of one spectral type that are misidentified as other spectral types.  This also varies with redshift.  These results for the above simulation are presented in Fig. \ref{CWW3}.

\begin{figure}
\vspace{8.1cm}
\includegraphics{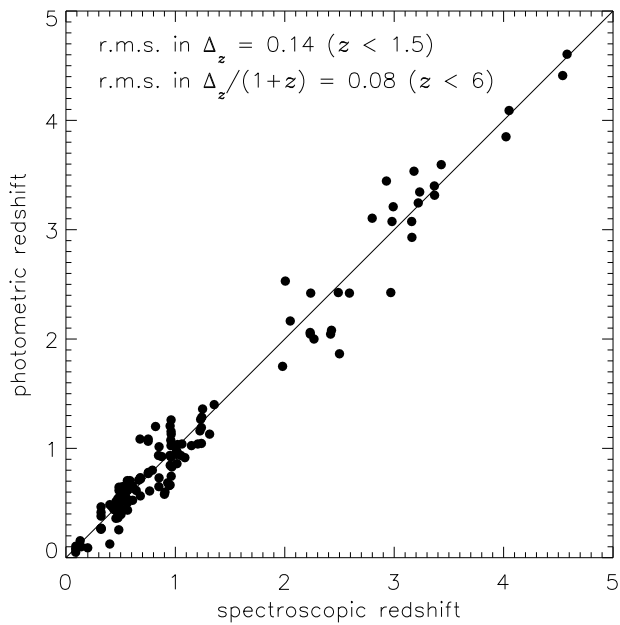}
\caption{\label{fig.hdf} A plot of photometric versus spectroscopic redshifts in the HDFN.  The spectroscopic redshifts come from the compilation of Fern\'andez-Soto et al. (2001; Table 2) omitting those galaxies with presumed incorrect spectroscopic redshifts as detailed in their Table 6.  We use the photometry of Fern\'andez-Soto et al. (1999) but resticting to only the $F300W$, $F450W$, $F606W$ and $F850W$ optical and $H$ near-infrared filters (cf. $UBVRIH$ in the LCIR survey).  The r.m.s. in $\Delta_z$, where $\Delta_z = z_{\rm spec}-z_{\rm phot}$, is 0.14 for $z < 1.5$. We note that if the $H$ filter is also omitted then the photometric redshift errors become substantially larger in this region.  The increased scatter beyond $z \sim 2$ is largely due to the combination of wavelength range covered by the filter set and our decision not to include any very blue (starburst) templates (cf. e.g. Massarotti et al. 2001).  In this paper we are not concerned with galaxies beyond $z \sim 1.5$ so we prefer to use the simpler set of 4 CWW template SEDs.}
\end{figure}

\begin{figure}
\vspace{8.1cm}
\includegraphics{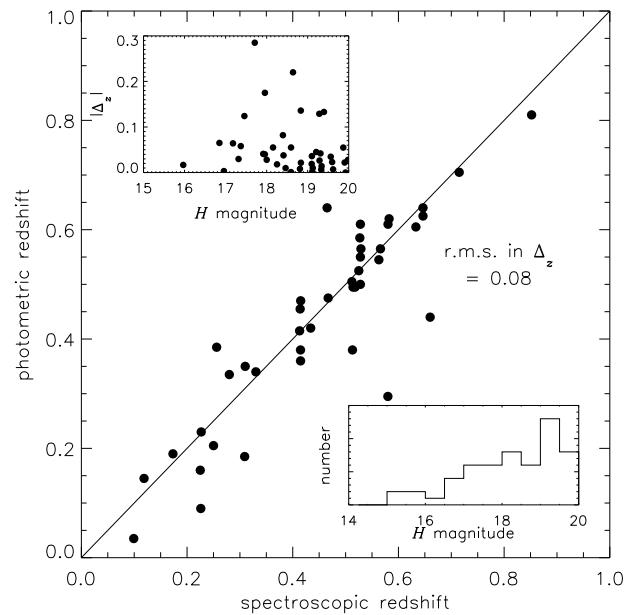}
\caption{\label{fig.hdfs} A plot of photometric versus spectroscopic redshifts in the HDFS.  Only the LCIR survey $UBVRIH$ data (i.e. the data used in the rest of this paper) has been used.  The r.m.s. in $\Delta_z$, where $\Delta_z = z_{\rm spec}-z_{\rm phot}$, is 0.08 for $z < 1.5$.  The inset at lower right shows a histogram of the $H$ magnitudes of these objects -- all of which lie in the $H < 20$ region of the LCIR survey -- showing that the spectroscopic sample includes a fair selection of faint galaxies.  The inset at upper left plots the photometric redshift errors as a function of magnitude, showing that there is no indication of an increase in photometric redshift error to fainter magnitudes.  In fact the largest photometric redshift error occurs for a galaxy with $H = 17.7$, which may indicate an incorrect spectroscopic redshift (and a consequent reduction in the true r.m.s. photometric redshift error).}
\end{figure}

These simulations represent an idealised situation -- we are assuming that real galaxies correspond exactly to the template SEDs, and we are assuming that there are no systematic or spurious errors in our data.  The component of photometric redshift errors due to these latter effects may be gauged by comparing photometric redshifts with spectroscopic redshifts.  Using 139 spectroscopic redshifts (compiled by Fern\'andez-Soto et al. 2001; see also Fern\'andez-Soto, Lanzetta \& Yahil 1999) in the Hubble Deep Field North (HDFN) and the $F300W$, $F450W$, $F606W$ and $F850W$ Hubble Space Telescope imaging (Williams et al. 1996) and $JHK$ ground-based imaging (Dickinson et al. in preparation\footnotemark \footnotetext{http://www.stsci.edu/ftp/science/hdf/clearinghouse/ irim/irim\_hdf.html}), the r.m.s. in $\frac{\Delta_z}{1+z_{\rm spec}}$, where $\Delta_z = z_{\rm spec}-z_{\rm phot}$, is $\sim 0.07$ (see also BMP; Hogg et al. 1998; Connolly, Szalay \& Brunner 1998; Arnouts et al. 1999).  Since the S/N in the HDF data is very high, this value gives an indication of the amount of scatter that arises from intrinsic differences between the template SEDs and the SEDs of observed galaxies.  In the LCIR survey there are fewer filters and lower S/N data.  Fig. \ref{fig.hdf} compares photometric and spectroscopic redshifts in the HDFN using just the optical $F300W$, $F450W$, $F606W$ and $F850W$ and near-infrared $H$ filters (i.e. one fewer filter than used in the LCIR survey HDFS field).  The r.m.s. in $\Delta_z$, for $z < 1.5$, is $\sim$0.14.  Since our HDFS field overlaps redshift surveys in the Hubble Deep Field South (Cristiani et al. 2000 and references therein; Glazebrook et al. in preparation\footnotemark)\footnotetext{See http://www.aao.gov.au/hdfs/Redshifts/RedShifts.} we may directly compare our photometric redshifts, using the LCIR survey photometry, with spectroscopic redshifts (Fig. \ref{fig.hdfs}).  The r.m.s. in $\Delta_z$, for $z < 1.5$, is $\sim$0.08.  The inset plots in Fig. \ref{fig.hdfs} show that the spectroscopic sample includes faint galaxies up to the survey limiting magnitude and there is no evidence for a significant increase in photometric redshift errors for fainter galaxies.  This error estimate assumes that the HDFS spectroscopic identifications are all correct and therefore may in fact be pessimistic.

In summary, we expect our photometric redshift errors to be small on the scale of the 0.5 bins in redshift that we adopt in $\S$\ref{sec.w}--\ref{sec.xi} of this paper, and small enough not to greatly affect the results of $\S$\ref{sec.cf}--\ref{sec.xi} (see also these sections for Monte Carlo simulations of the errors).

\section{Star/galaxy separation}  \label{sec.sg}
\begin{figure}
\vspace{8.5cm}
\includegraphics{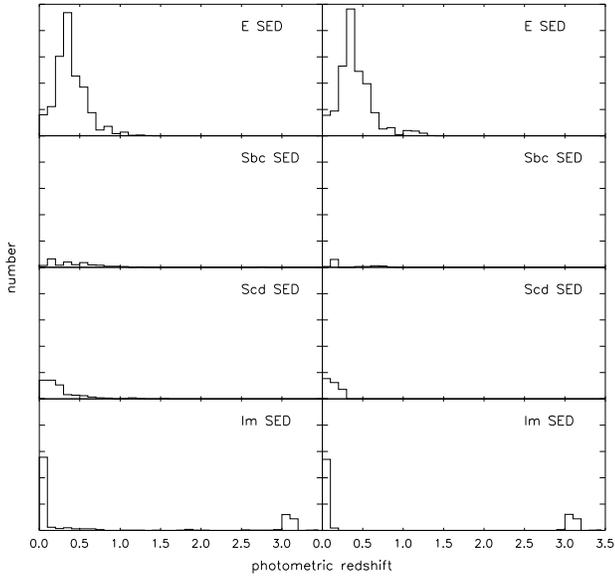}
\caption{\label{star_zsed} On the right are shown the best-fitting photometric redshifts and galaxy spectral types to objects that are {\it photometrically} identified as stars.  This illustrates that there are certain galaxy spectral types and redshifts with which stars are more likely to be confused (and {\it vice versa}). Using photometric techniques to separate stars may lead to excess stellar contamination in, for example, the sample of $z < 0.7$ spectral-type E galaxies with a corresponding misestimate of their correlation function.  Using morphological separation techniques avoids this bias.  On the left are shown the best-fitting photometric redshifts and galaxy spectral types to objects that are {\it morphologically} identified as stars.  This illustrates the actual SED/redshift distribution of potential stellar contamination, if photometric separation techniques were used, based on the actual distribution of stellar spectral types in the observations.  Since our photometric and morphological separation techniques agree in 87 per cent of cases the plots are in fact very similar.  However to avoid any remaining bias we only use {\it morphological} star/galaxy separation in this paper.}
\end{figure}

Separation of stars from galaxies is important if one hopes to compare the observations with model galaxy catalogues but it is even more important for accurately calculating the correlation function.  Since stars are uncorrelated with galaxies, any contamination from stars will reduce the correlation signal -- the reduction is a factor of $(1 - f)^2$ where $f$ is the stellar contamination fraction (Roche et al. 1997).  There are two obvious approaches open to us.  Firstly, {\tt SExtractor} uses a neural net and the measured seeing FWHM to morphologically classify stars, outputting the parameter {\tt CLASS\_STAR} ranging from 0 (galaxies) to 1 (stars).  Separation is reasonably clear-cut at bright magnitudes but less so towards the magnitude limit.  Secondly we can use photometric template fitting.  To do this we modified {\tt hyperz} to accept 173 stellar templates from the Bruzual, Persson, Gunn \& Stryker stellar atlas\footnotemark \footnotetext{See {\tt STSDAS}: http://ra.stsci.edu/About.html.}. Objects are then classified photometrically according to the best-fitting template -- either stellar or galactic.  This method also suffers towards the magnitude limit.  We much prefer the morphological approach: since particular galaxy SED/redshift combinations are photometrically confused with stars (and {\it vice versa}) more than other combinations (e.g. Fig. \ref{star_zsed}), the use of photometric star/galaxy separation would introduce excessive stellar contamination in particular SED/redshift bins and lead to possibly severe inaccuracies in the correlation function.  Admittedly compact galaxies could be lost along with stars in a morphological classification but, in this paper, we consider this the lesser of two evils.

\begin{figure}
\vspace{8.5cm}
\includegraphics{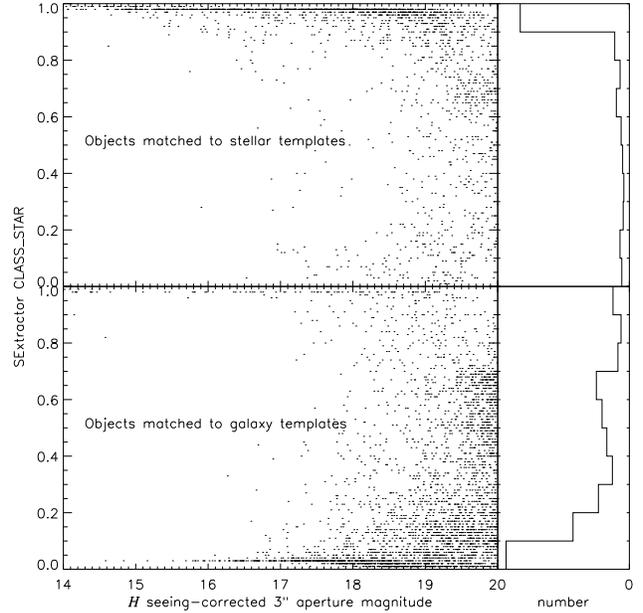}
\caption{\label{class} The $H$-band {\tt SExtractor} star/galaxy classification parameter {\tt CLASS\_STAR} as a function of $H$ magnitude for objects that are {\it photometrically} best-fitted by a stellar template (upper left) and objects that are {\it photometrically} best-fitted by a galaxy template (lower left).  The panels on the right show histograms summed over magnitude.  Even at bright $H$ magnitudes a small proportion of photometrically-identified stars have extended morphologies. Hence our photometric star/galaxy separation is too unreliable to use for clustering analyses, where colour-dependent residual stellar contamination can differentially bias the calculated correlation function in different spectral-type/redshift-selected galaxy subsamples.  We point out that this is not necessarily true of other surveys as photometric classification depends crucially on the particular filter set and S/N characteristics of the survey.  To avoid this bias we use {\it morphological} techniques only to separate stars and galaxies in this paper.}
\end{figure}

\begin{figure}
\vspace{7cm}
\includegraphics{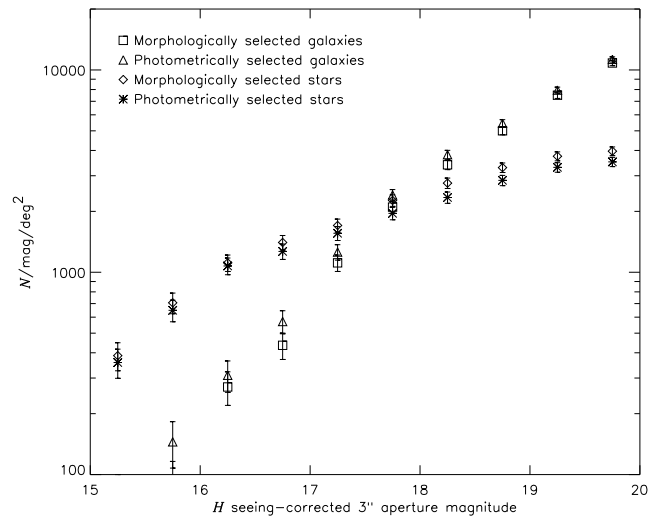}
\caption{\label{sgcounts} The total number of objects identified (a) as galaxies and (b) as stars using (i) image morphology in $VRIH$ and (ii) photometric template fitting.  Error bars are Poisson.}
\end{figure}

In fact in most cases there is good agreement between the two methods, for example Fig. \ref{class} compares {\tt SExtractor}'s morphological classification in $H$ with photometric classification.  However a small proportion of photometrically classified stars have extended $H$ images even at relatively bright magnitudes.  Since most of the objects in our $H$-selected catalogue are well-detected in $VRI$, we adopt the criteria that an object is classified as a star if {\tt CLASS\_STAR} $>$ 0.95 in any one of $VRIH$ and the mean {\tt CLASS\_STAR} in those of these filters in which the object is detected is greater than 0.5.  With these criteria the morphological and photometric classifications agree in 87 per cent of cases.  Fig. \ref{sgcounts} shows the star and galaxy number counts as a function of $H$ magnitude using the two classifiers.

\begin{figure}
\vspace{7cm}
\includegraphics{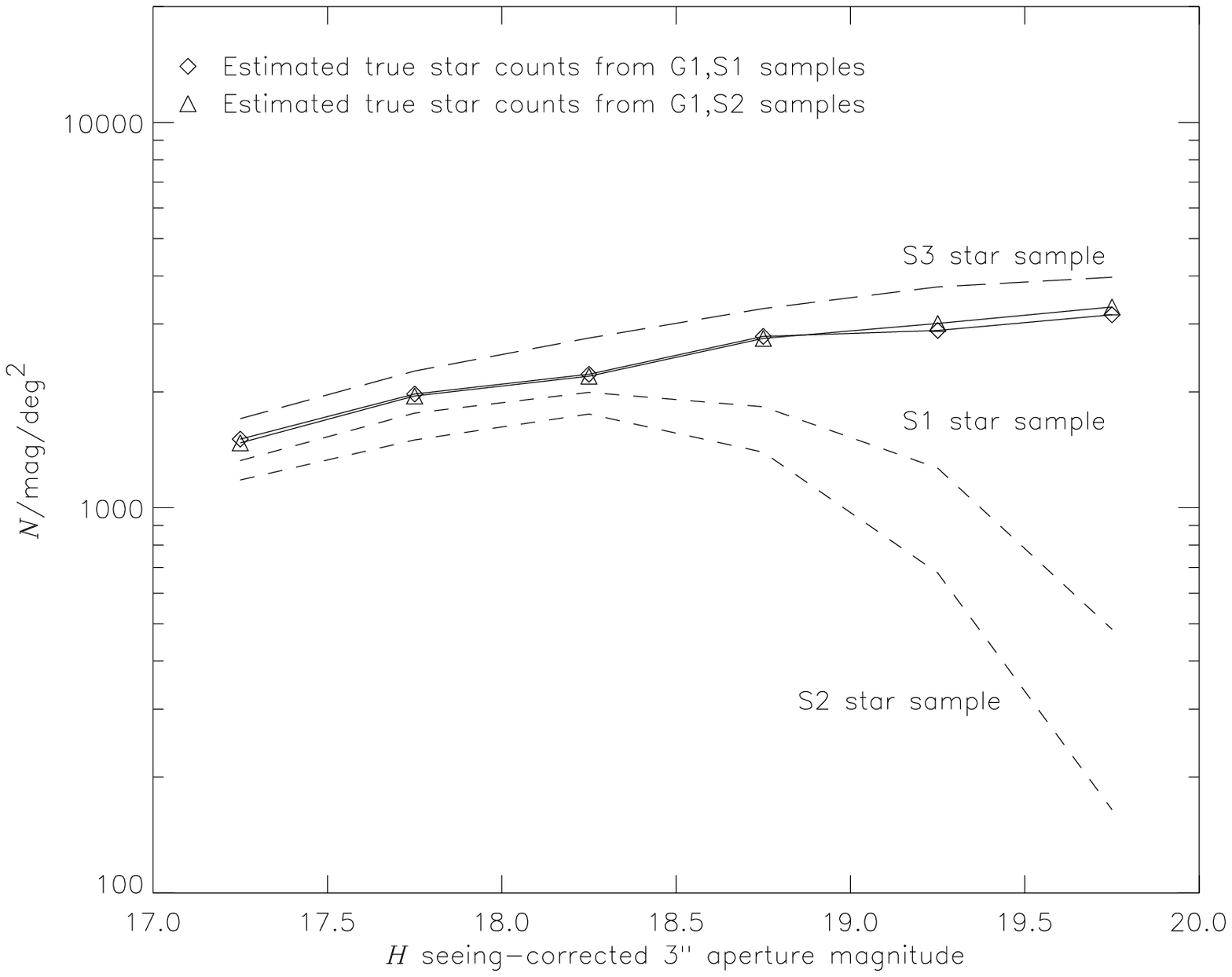}
\caption{\label{nstars} Intrinsic star number counts (solid line and symbols) as a function of $H$ magnitude in the HDFS field of the LCIR survey, as estimated from the number counts of photometrically-identified stars and galaxies and the estimated error rate in photometric star/galaxy classification.  Also plotted are the number of stars selected using the morphological criteria {\tt CLASS\_STAR} $>$ 0.90 in $H$ (S1), {\tt CLASS\_STAR} $>$ 0.95 in $H$ (S2) and {\tt CLASS\_STAR} $>$ 0.95 in any one of $VRIH$ (S3).  The latter (long dashes) consistently exceeds the estimated intrinsic star counts by about 15 per cent. We use S3 in this paper in order to err on the cautious side in removing stellar contamination.}
\end{figure}

The combination of the two techniques allows us to roughly estimate the total population of stars and hence gauge the remaining stellar contamination or the number of galaxies discarded as stars.  Using the $H$ data, which have the best seeing (1 arcsec), we select a sample G1 of galaxies with {\tt CLASS\_STAR} $<$ 0.10 in $H$, a sample S1 of stars with {\tt CLASS\_STAR} $>$ 0.90 in $H$ and a sample S2 of stars with {\tt CLASS\_STAR} $>$ 0.95 in $H$.  These criteria leave many objects unclassified but we can be fairly confident that nearly all of the selected objects are classified correctly.  Within the samples G1, S1 and S2 we calculate the fraction that are photometrically misidentified.  Now, assuming that the fraction of photometric misidentifications is reasonably independent of morphology, we use these fractions to invert the number of photometrically classified stars and galaxies to estimate the actual number of stars and galaxies.  The estimates using the samples G1, S1 and G1, S2 are plotted in Fig. \ref{nstars}.  Also plotted are the star counts in S1 and S2 and the star counts in the $VRIH$ morphological classification that we actually use (i.e. {\tt CLASS\_STAR} $>$ 0.95 in any one of $VRIH$ and the mean {\tt CLASS\_STAR} over these filters is greater than 0.5) which we label S3 in this section.   Clearly S1 and S2 miss many of the stars but S3 consistently exceeds the estimated total number of stars by about 15 per cent.  

We use S3 throughout this paper in order to err on the side of caution in removing stellar contamination but it must be remembered when comparing with models that the galaxy counts may be depleted at the 15 per cent level. Out of 5401 $H < 20.0$ objects we identify 2224 stars.  We expect the remaining stellar contamination in our galaxy sample to be of the order of a few percent or fewer.  After removing stars morphologically we then proceed to fit the remaining objects only to the grid of galaxy SEDs.

\section{Galaxy evolution models} \label{sec.models}
In order to have a basis upon which to compare and interpret the observations, we have generated several model galaxy catalogues -- a semi-analytic hierarchical merger model and several no-evolution and passive-evolution models.  In this section we briefly describe these models.

\subsection{Semi-analytic hierarchical merger models} 
The hierarchical paradigm of galaxy formation, set within the
framework provided by the Cold Dark Matter theory of structure
formation (e.g. Blumenthal et al. 1984), has proven to be very
successful at reproducing many observed properties of
galaxies. Testing and refining this paradigm is one of the major goals
of modern observational cosmology. 

Semi-analytic modelling is an attempt to use simple recipes to
parameterize the main physical processes of galaxy formation within
the hierarchical paradigm (for an introduction see Kauffmann, White
\& Guiderdoni 1993; Cole et al. 1994; Somerville 1997; 
Kauffmann et al. 1999a; Somerville \& Primack 1999 (SP); Cole et
al. 2000 and references therein). Monte Carlo techniques can be used
to efficiently produce mock galaxy catalogues representing large volumes
of space, and can be run with effectively arbitrary `mass
resolution'. In addition to model spectra, magnitudes, and colours,
this approach provides predictions of many physical properties of the
galaxies, such as the distribution of stellar ages and metallicities,
stellar and cold gas mass, bulge-to-disc ratio, etc. The free
parameters of the models are typically set by requiring that
fundamental observed properties of nearby galaxies (such as luminosity
functions, gas content, etc.) be reproduced at redshift zero. The
normalized models can then be used to predict galaxy properties at any
desired redshift.

We use the current version of the code developed by Somerville (1997),
which has been shown to produce good agreement with many properties of
local (SP) and high-redshift (Somerville, Primack \& Faber 2001; SPF)
galaxies. We now briefly summarize the main ingredients of the models,
and specify how they differ from the previously published models of SP
and SPF.

The formation and merging of dark matter haloes as a function of time
is represented by a `merger tree', which we construct using the
method of Somerville \& Kolatt (1999). The number density of haloes of
various masses at a given redshift is determined by an improved
version of the Press-Schechter model (Sheth \& Tormen 1999), which
mostly cures the usual discrepancy with N-body simulations. The
cooling of gas, formation of stars, and reheating and ejection of gas
by supernovae within these haloes is modelled by simple
recipes. Chemical evolution is traced assuming a constant yield of
metals per unit mass of new stars formed. Metals are cycled through
the cold and hot gas phase by cooling and feedback, and the stellar
metallicity of each generation of stars is determined by the metal
content of the cold gas at the time of its formation. All cold gas is
assumed to initially cool into, and form stars within, a rotationally
supported disc; major mergers between galaxies destroy the discs and
create spheroids. New discs may then be formed by subsequent cooling
and star formation, producing galaxies with a range of bulge-to-disc
ratios. Galaxy mergers also produce bursts of star formation,
according to the prescription described in SPF. Thus the star
formation history of a single galaxy is typically quite complex and is
a direct consequence of its gas accretion and merger history and its
environment.

These star formation histories are convolved with stellar population
models to produce model spectra or to calculate magnitudes and
colours. We have used the multi-metallicity stellar population synthesis 
models of Devriendt, Guiderdoni \& Sadat (1999) with a Salpeter IMF to 
calculate the stellar part of the spectra.  The effect of dust 
extinction is modelled using an approach similar to
that of Guiderdoni \& Rocca-Volmerange (1987) and Devriendt \&
Guiderdoni (2000). Here, the face-on optical depth $\tau_{\lambda}$ of
a galactic disc is assumed to be proportional to the column density of
metals in the cold ISM:
\begin{equation}
\tau_{\lambda} = \left ( \frac{A_{\lambda}}{A_V}\right )_{Z_{\odot}}
\left (\frac{Z_g}{Z_{\odot}}\right)^s \left(\frac{\langle N_{H}
\rangle}{2.1 \times 10^{21} \rm{cm}^2} \right ), \label{eq.samdust}
\end{equation} 
where the mean column density of cold gas is
\begin{equation}
\langle N_{H} \rangle = \frac{M_{\rm gas}}{1.4 \mu m_p \pi r_{t}^2}.
\end{equation}
The gas truncation radius $r_{t}$ is taken to be 3.5 times the
modelled exponential disc scale-length.  This produces an average
$B$-band face-on optical depth of $\tau_{B} = 0.8$ for $L_{*}$ spiral
galaxies, in agreement with observations (e.g. Wang \& Heckman
1996). The quantity $A_{\lambda}/A_V$ is the solar metallicity
extinction curve, which we take to be the standard Galactic extinction
curve given by Cardelli, Clayton \& Mathis (1989). The power-law scaling with 
the gas metallicity is intended to account for the metallicity dependence
of the shape of the extinction curve (see Guiderdoni \&
Rocca-Volmerange 1987).  We assign random inclinations to the galaxies
and use a `slab' model to compute the extinction as a function of
inclination (see Somerville 1997 or Somerville et al. 2001b for
details). Note that spheroids are assumed to be dust-free.

As described in SP, we set the free parameters of the models by
reference to a subset of local galaxy data; in particular, we require
a typical $L_{*}$ galaxy to obey the observed $I$-band Tully-Fisher
relation and to have a gas fraction of $0.1$ to $0.2$, consistent with
observed gas contents of local spiral galaxies. If we assume that
mergers with mass ratios greater than $\sim$ 1:3 form spheroids, we
find that the models produce the correct morphological mix of spirals,
S0s and ellipticals at the present day (we use the mapping between
bulge-to-disc ratio and morphological type from Simien \& de
Vaucouleurs 1986).  This critical value for spheroid formation is what
is predicted by N-body simulations of disc collisions (cf. Barnes \&
Hernquist 1992).

In this paper, we assume the currently favoured $\Lambda$CDM cosmology
with $\Omega_0=0.3$, $\Omega_{\Lambda}=0.7$, and $H_0 = 70$
km s$^{-1}$ Mpc$^{-1}$. It was shown in SP that with the proper choice of 
values for the parameters controlling the star formation, feedback, and 
chemical yield, this model produces reasonable agreement with the observed
$B$-band and $K$-band luminosity functions, Tully-Fisher relation,
metallicity-luminosity relation, and optical colours of local
galaxies. We use the same fiducial model here, with a few minor
modifications: we incorporate self-consistently the modelled
metallicity of the hot gas in the cooling function, and use the
multi-metallicity SEDs (instead of solar metallicity) with a Salpeter
(instead of Scalo) IMF. Another minor detail is that material ejected
by supernovae feedback is eventually returned to the haloes as
described in the updated models of SPF. We find that these minor
modifications do not significantly change our previous results for
local galaxies. An updated set of predictions for local galaxy
luminosity functions and galaxy counts will be presented in Somerville
et al. (2001b).

We note that the model is in no way tuned to fit the results of the LCIR survey.

\subsection{No-evolution and passive-evolution models}
While the semi-analytic model is a physically-motivated attempt to trace the initial conditions following the Big Bang to the galaxy population observed today, a more traditional approach (e.g. Tinsley 1977; King \& Ellis 1985; Yoshii \& Takahara 1988; Pozzetti, Bruzual \& Zamorani 1996 (PBZ); Gardner 1998 (G98)) has been to take the $z = 0$ galaxy population and extrapolate it backwards in time.  These are commonly referred to as no-evolution (where the galaxy population remains unchanged with redshift) and passive-evolution or pure luminosity evolution (PLE) (where galaxies are assigned some formation redshift and star formation history and their spectral energy distributions evolve accordingly).  Such models have been used to model galaxy number counts and have led to, for example, the identification of the so-called faint blue galaxy problem -- where it was noted that the observed $B$-band number counts displayed an excess of faint blue galaxies relative to no-evolution models but could be fitted by PLE models.  These PLE models subsequently failed to fit the observed redshift distribution of these galaxies which led PLE proponents to make various refinements such as non-number conservation (merging), fading dwarf galaxies, dust modelling and new cosmologies.  While recent PLE models have been successful at fitting optical counts and redshift distributions (PBZ; but see also Pozzetti et al. 1998) there have been discrepancies in the near-IR redshift distributions (PBZ; Kauffmann \& Charlot 1998) and a no-evolution model has often produced a better fit to near-IR number counts (PBZ; Metcalfe et al. 1995; Chen et al. 2001; but see also Totani 2001a).  Such models have also been compared with ERO number counts (Zepf 1997; Daddi, Cimatti \& Renzini 2000b).  It is useful therefore to reassess the ability of these models to fit the near-IR data using the number counts, ERO counts, {\it and} photometric redshift distributions of the large data set presented in this paper.

Following PBZ, G98 and others, we take a local luminosity function (LF) divided into spectral types.  For each spectral type, the LF and our adopted $\Omega_0 = 0.3$, $\Omega_\Lambda = 0.7$, $H_0$ = 70 km s$^{-1}$ Mpc$^{-1}$ cosmology give the galaxy distribution function $dN = N(z,M_{F})dzdM_{F}$ where $M_{F}$ is the rest-frame absolute magnitude in the filter used to define the LF.  Fitting a particular SED to each spectral type, we use {\tt hyperz} software (BMP) to calculate the observed magnitude in each survey filter for a galaxy of given redshift and absolute magnitude $M_{F}$.  Integrating $dN$ over $z$ gives the number counts as a function of magnitude in each filter.  Integrating over $M_{F}$ up to the faint limit imposed at each $z$ by the limiting magnitude in the survey filter gives the redshift distribution for galaxies brighter than the survey limit.  The normalization of the number counts comes directly from the LFs' $\phi_*$ rather than normalizing to observed galaxy number counts in some apparent magnitude range (cf. PBZ).

We used the $B_J$-band LF of Madgwick et al. (2001 (M01); see also Folkes et al. 1999) based on 75000 galaxies in the 2dF Galaxy Redshift Survey\footnotemark \footnotetext{http://www.mso.anu.edu.au/2dFGRS/}.  This LF has been divided into 4 spectral types based on a principal component analysis of the galaxy spectra.  These spectral types (1--4) correspond {\it approximately} to the morphological types E/S0/Sa, Sb, Scd and Scd (again) of Kennicutt (1992).  For our no-evolution model we used the same extended CWW SEDs described in $\S$\ref{sec.photoz}, fitting CWW E to type 1, CWW Sbc to type 2, CWW Scd to type 3 and CWW Im to type 4.  In this model the SEDs are fixed with respect to redshift.  For our passive-evolution model the SEDs evolve with redshift.  Here we used Bruzual \& Charlot (1993; BC) {\tt GISSEL'98} models with Miller \& Scalo (1979) initial mass function and solar metallicity.  We fitted the M01 types 2 and 3 with BC models with exponentially decaying star formation rates (SFR) with time-scales $\tau$ = 5 (Sb) and 15 (Sc) Gyr respectively and we fitted the M01 type 4 to a BC model with continuous star formation (Im).  As we are particularly interested in the evolution of red galaxies in this paper, we used two alternative models for the M01 type 1 -- one with $\tau$ = 1 Gyr (E) and the second being a single stellar population (SSP) burst (B).  We used a formation redshift $z_f$ = 10 for all these models, and to improve the fit of these spectra to empirical spectra at $z = 0$, dust reddening modelled by a uniform dust screen following a Calzetti et al. (2000 (CABKKS); also Calzetti, Kinney \& Storchi-Bergmann 1994) extinction law with $A_V$ = 0.3 was applied to all of the synthetic spectra.  The passive-evolution model (using the SSP burst for type 1) provides a good fit to the compilation of deep $B$-band counts in Metcalfe et al. (1995).

Since using the $B_J$-band M01 LFs to model the number counts and redshift distributions in an $H$-selected survey involves extrapolating over a factor of four in wavelength, we generated further no-evolution and passive-evolution models based on the $K$-band LF of Gardner et al. (1997 (G97); see also Gardner's own number count model {\tt ncmod} at http://hires.gsfc.nasa.gov./$\sim$gardner/ncmod/).  This LF is not divided into spectral or morphological types but following G98 we divide the normalization $\phi_*$ between five spectral types in the ratio 0.51:0.36:0.1:0.03:0.004.  In the no-evolution model we match these five types to the CWW types E, E, Sbc, Scd and Im respectively and in the passive-evolution model we match the five types to the BC types E or B, Sa, Sb, Sc and Im as described above, where Sa is an extra type with exponentially decaying star formation rate with time-scale $\tau$ = 3 Gyr.  The k- and e- corrections are in any case much less dependent on spectral type when converting from rest-frame $K$ magnitudes to observed-frame $H$ magnitudes than when converting from the $B_J$-band so the results (at least the total $H$-band number counts) are not very sensitive to the exact spectral type mix.  In the following section we find that the models based on the $K$-band LF generally provide a better fit to the observations than those based on the $B_J$-band LFs.

\section[]{Comparisons between\\ observations and models}  \label{sec.cf}
We begin ($\S$\ref{sec.cf}.1) by comparing colour-colour and colour-magnitude diagrams between the observations and the various models.  This is a useful way to visualize the data -- showing the relative positions of different populations and the effect of redshift and reddening.  In $\S$\ref{sec.cf}.2 we compare number counts, redshift distributions and ERO statistics and discuss how these are consistent with or contradict the various models. 

As noted in $\S$\ref{sec.cat} we have generally used aperture magnitudes for the observations in this paper.  This complicates direct comparison with models, which generate total magnitudes.  Ideally one would like to compare the observations with simulated images generated by adding sky and photon noise to simulated galaxies, however our semi-analytic model does not produce a scale-length for the bulge component nor does our passive-evolution model include a formalism for the evolution of scale-length so we are unable to properly take this approach at present (cf. Totani et al. 2001a).  Again as noted in $\S$\ref{sec.cat} the difference between total magnitudes and seeing-corrected 3 arcsec aperture magnitudes is expected to be of the order of 0.1--0.3 magnitudes at the faint limit of this survey.  In most of this section we account for this by comparing both $H < 20.0$ and $H < 20.5$ selected data catalogues with $H < 20.0$ selected model catalogues.  The real comparison is somewhere between the two.  We make an exception for our number counts comparison, where for each object we use the brightest of isophotal and seeing-corrected 3 arcsec aperture magnitudes since here the increase of the aperture effect towards brighter magnitudes (objects that, on average, have greater angular extent) becomes important while at faint magnitudes isophotal magnitudes will underestimate fluxes.

For further analyses of the multicolour catalogue -- including other LCIR survey fields -- see Chen et al. (2001) and Marzke et al. (in preparation).

\subsection{Colour distributions}
\begin{figure}
\vspace{7cm}
\includegraphics{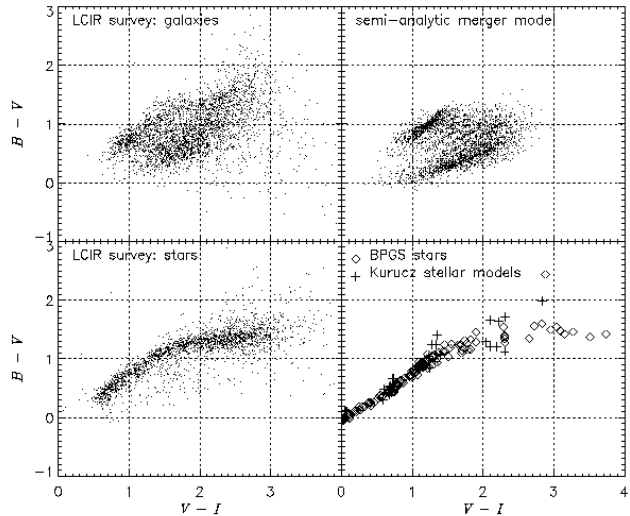}
\caption{\label{bvi} A comparison of $BVI$ colours of LCIR survey galaxies with the semi-analytic model and of LCIR survey stars with stars from the Bruzual, Persson, Gunn \& Stryker (BPGS) stellar atlas and Kurucz stellar atmosphere models.  Both LCIR survey and semi-analytic model have been cut at $H = 20.0$.  The Kurucz models cover a range of temperatures from 3500 to 50000 K, metallicities $\log (Z/Z_{\odot})$ = 1, 0 and $-$5, and surface gravities $\log g$ = 0, 2.5 and 5.  The number of points in the semi-analytic model has been scaled to match the observations.}
\end{figure}

\begin{figure}
\vspace{7cm}
\includegraphics{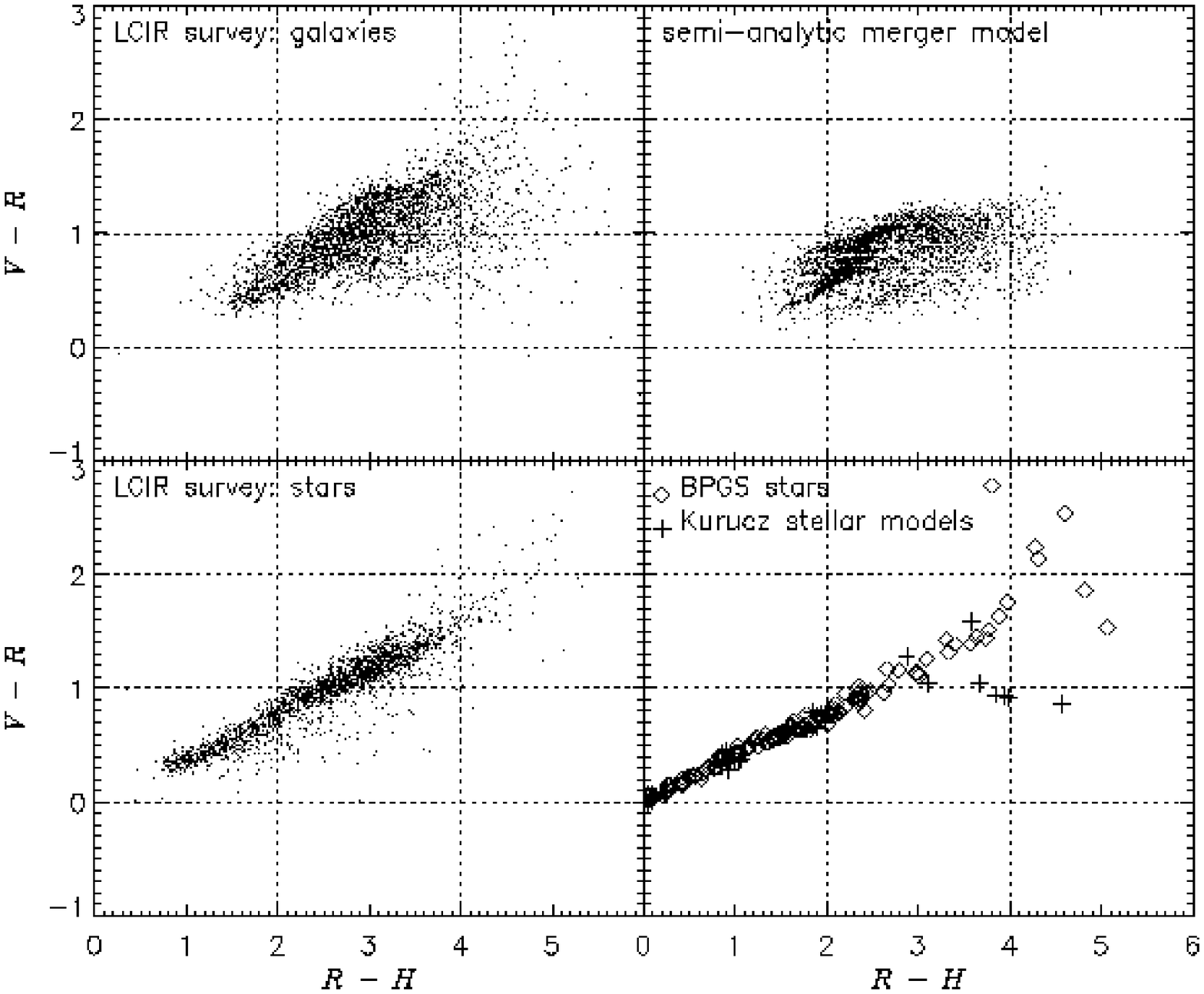}
\caption{\label{vrh} A comparison of $VRH$ colours of LCIR survey galaxies with the semi-analytic model and of LCIR survey stars with stars from the Bruzual, Persson, Gunn \& Stryker (BPGS) stellar atlas and Kurucz stellar atmosphere models.  Both LCIR survey and semi-analytic model have been cut at $H = 20.0$.  The Kurucz models cover a range of temperatures from 3500 to 50000 K, metallicities $\log (Z/Z_{\odot})$ = 1, 0 and $-$5, and surface gravities $\log g$ = 0, 2.5 and 5.  The number of points in the semi-analytic model has been scaled to match the observations.}
\end{figure}

\begin{figure}
\vspace{7cm}
\includegraphics{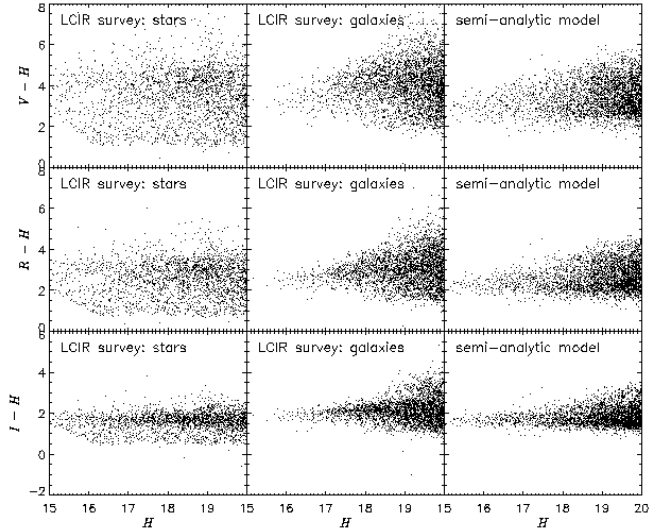}
\caption{\label{cm} Comparison of $V - H$, $R - H$ and $I - H$ colours as a function of $H$ magnitude for LCIR survey stars, LCIR survey galaxies and galaxies from the semi-analytic model.  The number of points in the semi-analytic model has been scaled to match the observations.}
\end{figure}

\begin{figure}
\vspace{7cm}
\includegraphics{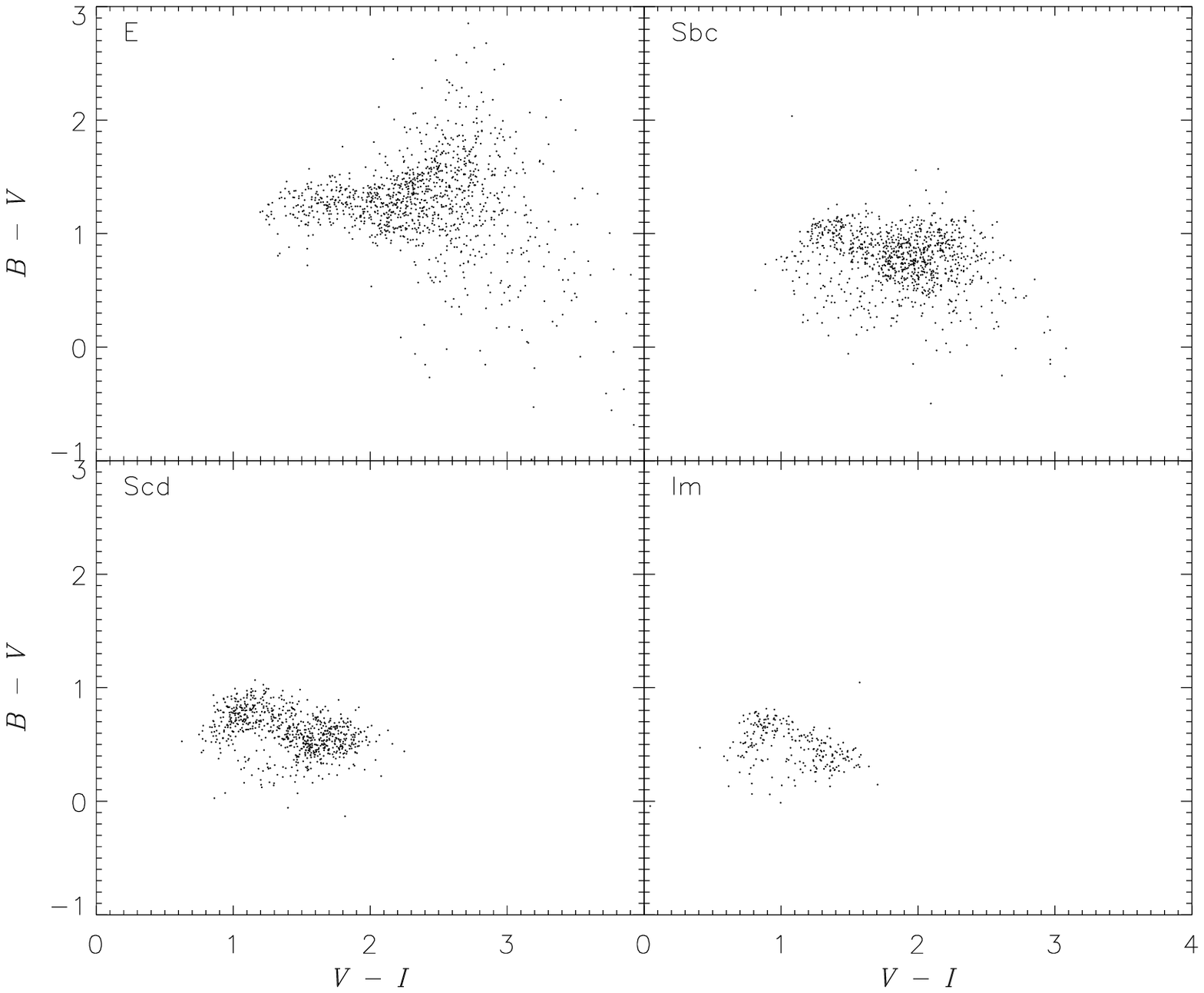}
\caption{\label{bvi2} $BVI$ colours for $H < 20.0$ LCIR survey galaxies for different photometrically-determined spectral types.}
\end{figure}

\begin{figure}
\vspace{7cm}
\includegraphics{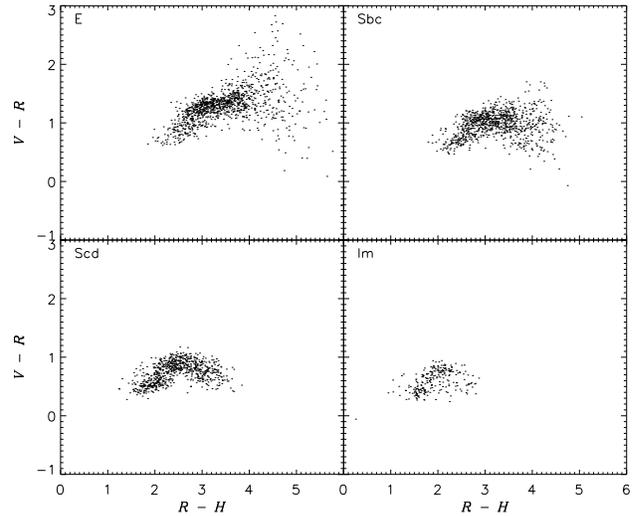}
\caption{\label{vrh2} $VRH$ colours for $H < 20.0$ LCIR survey galaxies for different photometrically-determined spectral types.}
\end{figure}

We begin by comparing colour-colour and colour-magnitude diagrams.  Fig. \ref{bvi} and Fig. \ref{vrh} compare respectively $BVI$ and $VRH$ colours for $H < 20.0$ LCIR survey galaxies with semi-analytic model galaxies and LCIR survey stars with stars from the Bruzual, Persson, Gunn \& Stryker (BPGS) stellar atlas and with Kurucz (1993) stellar atmosphere models\footnotemark sampled through the survey filters.  The Kurucz models cover a range of temperatures -- 3500, 4500, 6000, 9000, 12000, 20000 and 50000 K, metallicities -- $\log (Z/Z_{\odot})$ = 1, 0 and $-$5, and surface gravities -- $\log g$ = 0, 2.5 and 5.  Agreement between the stellar tracks provides a check on the data zeropoint calibration.  The agreement between the observations and the semi-analytic model is fair over the main body of the distribution, but it is noted firstly that there is considerable scatter in the observations (the photometric errors are mostly $<$ 0.2 magnitudes in $RIH$ but up to 0.4 magnitudes in $BV$) and secondly that the observations include a component of red and extremely red objects not present in the semi-analytic model.  In addition, there is much greater scatter in the $V - R$ colours of red $R - H$ objects in the observations than in the semi-analytic model.  In Fig. \ref{cm}, which compares $V - H$, $R - H$ and $I - H$ colours as a function of $H$ magnitude for LCIR survey stars and galaxies and galaxies in the semi-analytic model, the extremely red $R - H > 4$ and $I - H > 3$ galaxies are very evident in the observations but are largely absent from the semi-analytic model.  In Fig. \ref{bvi2} and Fig. \ref{vrh2} we compare the positions in $BVI$ and $VRH$ colour-colour space of different photometrically-determined spectral types in the catalogue showing that the red objects are mostly fitted by the E template, even those with quite blue $B - V$ and $V - R$ colours.
 
\footnotetext{See {\tt STSDAS}: http://ra.stsci.edu/About.html.}

\begin{figure}
\vspace{16.7cm}
\includegraphics{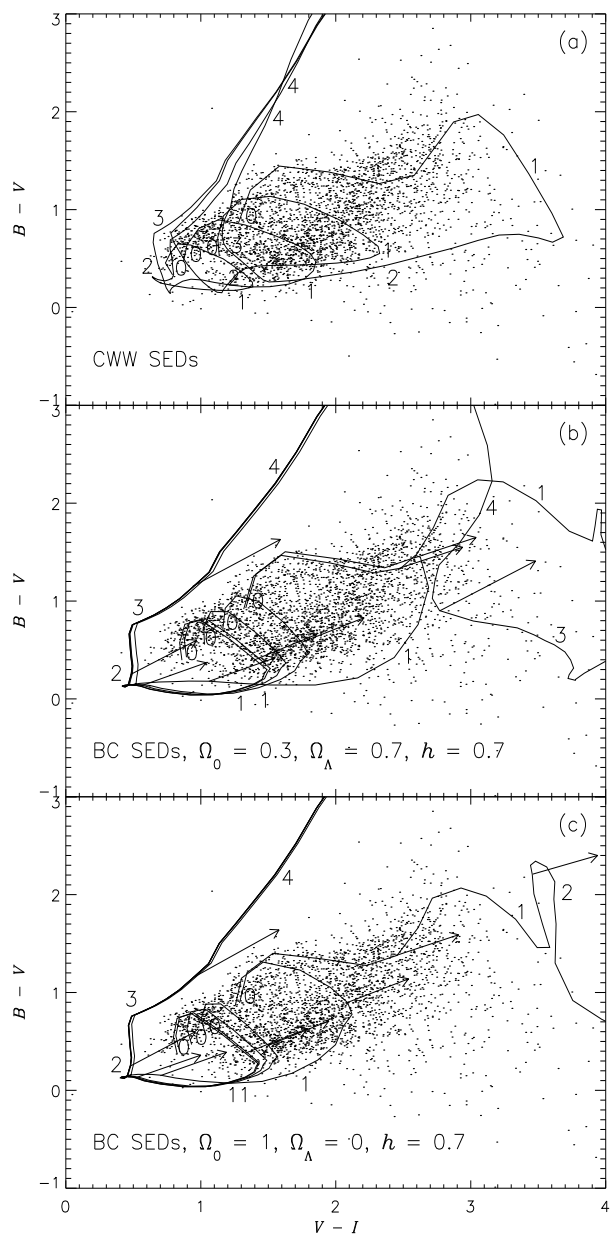}
\caption{\label{bvi.tracks} $BVI$ tracks of (a) the four empirical CWW SEDs: E, Sbc, Scd and Im, and (b, c) six BC {\tt GISSEL'98} SEDs (single stellar population (SSP) burst, exponentially decaying star formation with time-scale $\tau$ = 1, 3, 5 and 15 Gyr and constant star formation rate (SFR), all with formation redshift $z_f = 10$) in two cosmologies plotted on the LCIR survey galaxies.  Unit redshift intervals are marked and the effect of applying $A_V$ = 1.0 reddening (CABKKS) is indicated by arrows.  At $z \sim 1$ the SED towards the right corresponds to the E or SSP burst galaxy while that towards the left corresponds to the Im or constant SFR galaxy.  Note that above $z \sim 1.5$ the CWW E SED becomes increasingly unrealistic as there is limited time for such a galaxy to form.}
\end{figure}

\begin{figure}
\vspace{16.7cm}
\includegraphics{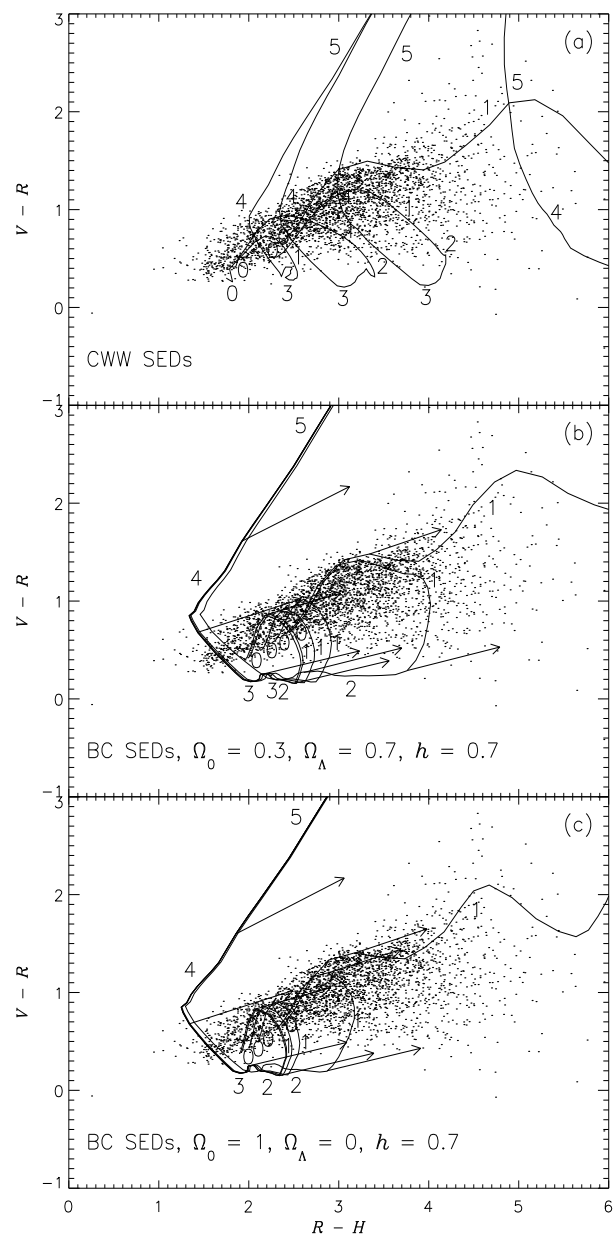}
\caption{\label{vrh.tracks} $VRH$ tracks of (a) the four empirical CWW SEDs: E, Sbc, Scd and Im, and (b, c) six BC {\tt GISSEL'98} SEDs (single stellar population (SSP) burst, exponentially decaying star formation with time-scale $\tau$ = 1, 3, 5 and 15 Gyr and constant star formation rate (SFR), all with formation redshift $z_f = 10$) in two cosmologies plotted on the LCIR survey galaxies.  Unit redshift intervals are marked and the effect of applying $A_V$ = 1.0 reddening (CABKKS) is indicated by arrows.  At $z \sim 1$ the SED towards the right corresponds to the E or SSP burst galaxy while that towards the left corresponds to the Im or constant SFR galaxy.  Note that above $z \sim 1.5$ the CWW E SED becomes increasingly unrealistic as there is limited time for such a galaxy to form.}
\end{figure}

In order to interpret the colour-colour diagrams it is instructive to plot as a function of redshift the tracks of various template SEDs from the no-evolution and passive-evolution models described in $\S$\ref{sec.models}.2.  Fig. \ref{bvi.tracks} and Fig. \ref{vrh.tracks} show $BVI$ and $VRH$ diagrams of observed galaxies with (a) the tracks of four CWW SEDs (E, Sbc, Scd and Im) over-plotted, (b) the tracks of six BC SEDs over-plotted, specifically one SED with constant SFR, one SED with an SSP burst and four SEDs with exponentially decaying SFRs with time constants $\tau$ = 1, 3, 5 and 15 Gyr, assuming a fixed redshift of formation $z_f = 10$ and the cosmology $\Omega_0 = 0.3$, $\Omega_\Lambda = 0.7$, $H_0$ = 70 km s$^{-1}$ Mpc$^{-1}$ and (c) the same BC SEDs but with an $\Omega_0 = 1$, $\Omega_\Lambda = 0$ cosmology.  The latter tracks evolve more rapidly to higher redshift as there is less time in an $\Omega_0 = 1$ universe.  Unit redshift intervals are labelled.  These diagrams illustrate that a possible interpretation of the extremely red objects are $z \sim 1$ CWW E or BC burst models while the red objects with bluer $B - V$ and $V - R$ colours could be the passively evolving progenitors of red galaxies seen at $z \sim$ 1--2 with some residual star formation (somewhere between BC SSP burst and $\tau$ = 1 Gyr models).  The addition of some dust reddening to the BC models (as indicated by the arrows) is probably necessary to match observed galaxies.  Adding large amounts of dust reddening ($A_V \gg \; \sim$2, depending on the reddening law adopted) could move later spectral types into the ERO region of colour-colour space.  The blue tail seen in the semi-analytic model in the $BVI$ diagram corresponds to $z \sim$ 1.5--2 blue spectral types -- galaxies which are missing, or redder, in the observations.

The only consequences of the aperture effect in these plots is the inclusion of extra objects in the semi-analytic model around the $H = 20.0$ limit, and a small shift in the colour-magnitude plots.

\clearpage

\clearpage

\subsection{Number counts and redshift distributions}
While the previous subsection was a general overview of the data, in this subsection we compare specific predictions of the models with a view to determining which provide the closest match to the observations, and in order to identify possible explanations for any discrepancies.  A generic prediction of hierarchical structure formation is that massive elliptical galaxies form from the merging of smaller disc galaxies and hence formed relatively recently (White \& Rees 1978).  In contrast, the traditional passive-evolution model has massive elliptical galaxies forming in single monolithic collapses at high redshift and then evolving passively to the present day with no further star formation (Eggen, Lynden-Bell \& Sandage 1962).  Thus, as Kauffmann \& Charlot (1998) point out, the number density of massive galaxies at $z \sim 1$ ($z > 1.5$ for $\Lambda$CDM) can provide a sensitive test to differentiate between the two models.  In particular a near-IR survey such as this is crucial since infrared light is a much more robust tracer of total stellar mass than optical light out to $z \sim$ 1--2.  Kauffmann \& Charlot (1998) compared the redshift distributions of the $K < 18$ sample of Songaila et al. (1994) and the $K < 19$ sample of Cowie et al. (1996) with predictions of hierarchical merger models and pure luminosity evolution (PLE) models, finding that the PLE model greatly overestimated the $z \ge 1$ cumulative counts while the hierarchical merger model provided a much better fit to the observations.  However the sample sizes were small ($\sim$ 170 galaxies in all), may suffer from spectroscopic incompleteness at $z \sim 1$, and in any case it is useful to repeat the test with a currently more favoured cosmology.  

Since a large fraction of EROs are thought to be $z \sim 1$ massive evolved galaxies, the relative efficiency with which they may be isolated using just one optical and one near-IR colour, led several groups to use the number densities of EROs as a test for various models, sometimes with the conclusion that PLE models greatly overpredict the number densities of EROs (Zepf 1997; Barger et al. 1999) and sometimes with the conclusion that PLE models are in good agreement with the observations (Daddi et al. 2000b; DCR).  One outcome of these surveys was the result that the ERO population is very strongly clustered so that field-to-field variations lead to the requirement of wide-field surveys in order to get good estimates of the number density of these objects (Daddi et al. 2000a (D00); McCarthy et al. 2000).  In particular, the observations presented here cover a comparable area at a comparable depth (using $H - K \sim 1$) to those presented by D00, while the addition of our photometric redshifts allows us to estimate $N(z)$ for the EROs and in particular to isolate $z \sim$ 1--1.5 galaxies of any spectral type, thus allowing more detailed comparisons with models.  From the observed clustering amplitude, we expect that our estimated ERO number counts are representative to within 15 per cent for $R - H > 4$ and $I - H > 3$ samples (see $\S$\ref{sec.w}).

\begin{figure}
\vspace{9.5cm}
\includegraphics{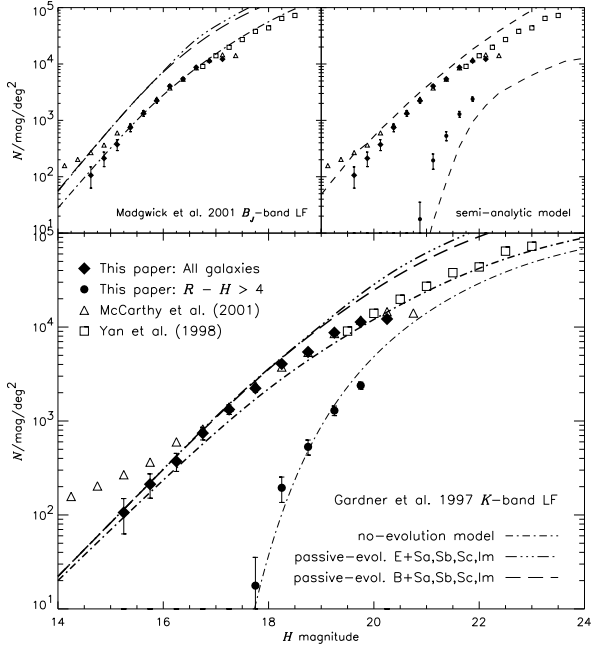}
\caption{\label{hcounts} Galaxy number counts as a function of $H$ magnitude in the HDFS field of the LCIR survey (see McCarthy et al. 2001 and Chen et al. 2001 for deeper completeness-corrected counts from the full LCIR survey).  
For each galaxy we use the brightest of its corrected isophotal and corrected aperture magnitudes (as explained in the text).  This results in the slight artificial turnover in the last magnitude bin.  Also note that the galaxy counts are expected to be depleted at the 15 per cent level due to the conservative star/galaxy separation criteria employed.  Also plotted are the number counts of McCarthy et al. (2001) from 0.62 degrees$^2$ and the deep NICMOS counts of Yan et al. (1998).  Several models are also plotted (see $\S$\ref{sec.models} for details).  The models based on the $K$-band LF are a closer match to the $H$-band observations than those based on the $B_J$-band LF, and the observed counts are most consistent with the no-evolution model.  The semi-analytic model follows the observed slope except at the faintest magnitudes but is normalized $\sim$50 per cent too high.  Also plotted are EROs satisfying $R - H > 4$.  These are well-fitted by the E/S0 galaxies with $R - H > 4$ from the no-evolution model (plotted) but the semi-analytic model has far too few $R - H > 4$ galaxies compared with the observations.}
\end{figure}

\begin{figure}
\vspace{7cm}
\includegraphics{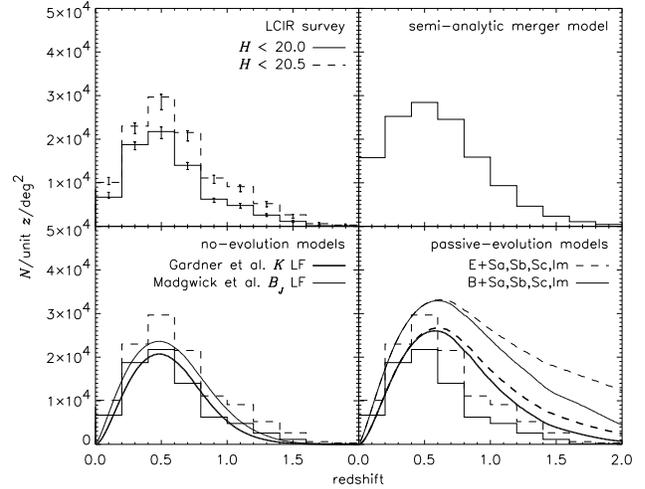}
\caption{\label{zhist} The photometric redshift distribution of LCIR survey galaxies (upper left) and the corresponding redshift distribution of galaxies in several model catalogues: semi-analytic merger model (upper right), no-evolution model (lower left) and passive-evolution with two models for evolving $z = 0$ E/S0 spectral types (lower right).  In the latter, the bold curves are calculated using the $K$-band LF while the thin curves use the $B_J$-band LF; the observations are replotted as histograms to facilitate comparison.  Error bars are 90 per cent percentiles from bootstrap resampling followed by Monte Carlo simulations in which random photometric errors were added to the observed photometry and photometric redshifts were recalculated.  As noted in the text the observations are expected to underestimate the counts slightly due to some galaxies being mistakenly removed as stars.  The semi-analytic and no-evolution models are plausible fits to the observations but the passive-evolution models predict too many high-redshift galaxies.}
\end{figure}

We begin by comparing $H$ number counts $N(M)$ in Fig. \ref{hcounts} (also Table \ref{table.hcts}).  Here, in contrast to the rest of the paper, we use for each object the brightest of its isophotal magnitude corrected to total magnitudes by assuming a gaussian profile ({\tt SExtractor}'s {\tt MAG\_ISOCOR} parameter) which we found to be a good measure of total magnitude at the bright end, and its seeing-corrected 3 arcsec aperture magnitude.  We do this because the aperture effect varies with magnitude, being larger for close bright extended galaxies.  This is unimportant for a simple $H < 20.0$ sample but is important for number counts as a function of magnitude.  Conversely, just using isophotal magnitudes would severely underestimate the flux of the faintest objects.  Our galaxy number counts agree fairly well with those of McCarthy et al. (2001) from 0.62 degrees$^2$ and the deep NICMOS counts of Yan et al. (1998).  The faint end ($18 < H < 20$) slope, $d \log N(H)/dH$, is $0.38$.  This compares with 0.31 for $H > 20.0$ (Yan et al. 1998) and 0.47 for $H < 19$ (Martini 2001a).  The no-evolution models fit the observed number counts very well.  The passive-evolution models based on the $B_J$-band luminosity function (LF) do not fit the observations -- possibly due to the uncertainties involved in extrapolating from rest-frame $B_J$-band magnitudes to observed-frame $H$-band with model SEDs that are only approximate matches to the spectral types identified by M01.  Extrapolating from the $K$-band is more robust and less dependent on the details of the model SEDs.  However, despite better agreement at bright magnitudes, the slope at $H \sim 20$ in the $K$-band LF passive-evolution models is too steep compared with the observations (but cf. Totani et al. 2001a).  The semi-analytic model matches the slope fairly well, except in the faintest magnitude bins, but the normalization is about 50 per cent too high.  As noted in $\S$\ref{sec.sg}, $\sim$15 per cent of this difference could be due to galaxies discarded from the LCIR survey as stars.  The remainder may indicate a problem with the semi-analytic model normalization since it is present even in the lowest photometric redshift bin (Fig. \ref{zhist}, next paragraph).  Correct normalization has been a common problem with number counts models as $\phi_*$ for local LFs is often uncertain to within a factor of $\sim$2 -- therefore a common practice elsewhere has been to renormalize the models to observed counts at relatively bright magnitudes.

\begin{table}
\centering
\caption{\label{table.hcts} Galaxy number counts as a function of $H$ magnitude in the HDFS field of the LCIR survey (see also McCarthy et al. 2001 and Chen et al. 2001). The quoted errors are Poisson.  The slight turnover in the last magnitude bin is artificial (see caption to Fig. \protect\ref{hcounts} and text for details).}
\begin{tabular}{cc}
{\it $H$ magnitude} & $N$/mag/deg$^2$\\ \hline
15.25&  110$\pm$40\\
15.75&  210$\pm$60\\
16.25&  370$\pm$80\\
16.75&  740$\pm$110\\
17.25& 1330$\pm$150\\
17.75& 2230$\pm$200\\
18.25& 4100$\pm$300\\
18.75& 5400$\pm$300\\
19.25& 8700$\pm$400\\
19.75&11300$\pm$500\\
20.25&12200$\pm$500\\
\end{tabular}
\end{table}

\begin{figure}
\vspace{7cm}
\includegraphics{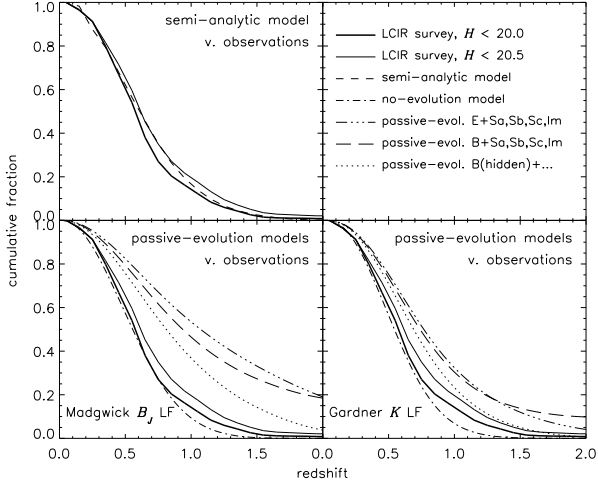}
\caption{\label{zcum} The cumulative redshift distributions of galaxies in the LCIR survey and several models.  The SSP burst SED used in one of the passive-evolution models is very bright at its formation redshift $z_f$, so with $z_f = 10$ as adopted here, it is bright enough to introduce a population of $z \sim 10$ galaxies into the sample.  This can be prevented by choosing a higher $z_f$ (e.g. $z_f \sim 30$), lengthening the duration of the star-forming burst or by invoking dust obscuration.  Thus we also plot the cumulative distribution of this model with the $z \sim 10$ population removed (B(hidden)+...).  The semi-analytic model provides the best fit to the observations.}
\end{figure}

A more sensitive discriminant between models is the redshift distribution -- as inferred from photometric redshifts here -- and in particular the redshift distribution of different spectral types.  In Fig. \ref{zhist} we compare the redshift distribution $N(z)$ of the observations with the various models.  As noted in the introduction to this section we plot both $H < 20.0$ and $H < 20.5$ observed samples so that if we were able to correct for the aperture effect then the resulting $N(z)$ would lie somewhere between the two.  We estimate errors by bootstrap-resampling the whole galaxy catalogue.  Then for each galaxy we use the S/N characteristics of the survey to add random photometric errors in each filter.  Photometric redshifts and spectral types were then recalculated for these galaxies and the new redshift distributions were calculated.  This was repeated 60 times and error bars were estimated using the central 90 per cent percentile.  Fig. \ref{zcum} illustrates the same data as cumulative fractions in order to factor out the normalization.  Apart from the normalization of the semi-analytic model, it and the no-evolution model are in broad agreement with the observations, the no-evolution model perhaps lacking galaxies in the highest ($z \sim$ 1--1.5) redshift bins and the semi-analytic model predicting an excess in the lowest redshift bin.  The latter may be due to an excess of bright galaxies in the $z = 0$ $K$-band luminosity function in the semi-analytic model, as demonstrated in SP.  The similarity between the no-evolution and semi-analytic model predictions over this redshift range results from a `conspiracy' effect for bright galaxies in the semi-analytic model: although the galaxy stellar mass function evolves significantly from $z = 1.5$ to $z = 0$, this is offset by evolution in the mass-to-light ratio which increases, even in the near-IR, as stellar populations age and become more metal rich, resulting in little evolution in the luminous component of bright galaxies.  The passive-evolution models based on the $B_J$-band LF predict far too many $z > 0.75$ galaxies, while that based on the $K$-band LF with the E/S0 population modelled by high-redshift SSP bursts provides a closer fit to the observations though the median redshift is still too high.  In the remainder of this section we consider only the no-evolution and passive-evolution models based on the  $K$-band LF as those based on the $B_J$-band LF are a poorer fit to the infrared observations.

\begin{figure}
\vspace{7cm}
\includegraphics{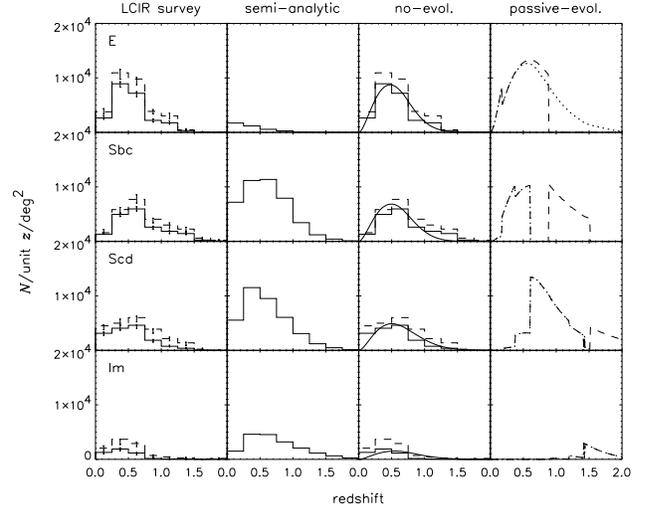}
\caption{\label{zsedhist} The redshift distributions of different CWW spectral types in the observations ($H < 20.0$ -- solid line; $H < 20.5$ -- dashed line), the semi-analytic merger model, the no-evolution model (using the $K$-band LF and with the observations plotted as histograms to facilitate comparison) and two passive-evolution models (using a single stellar population burst (dotted line) and a model with exponentially decaying star formation rate with time-scale 1 Gyr (dashed line) respectively, both with formation redshift $z_f$ = 10, to model the E/S0 population).  Error bars are 90 per cent percentiles from bootstrap resampling followed by Monte Carlo simulations in which random photometric errors were added to the observed photometry and photometric redshifts and spectral-types were recalculated.  The sharp boundaries in the passive-evolution models as BC SEDs evolve from one CWW spectral type to another are a consequence of modelling galaxy populations with discrete spectra instead of a continuous range. Compared with the observations, the semi-analytic model predicts an excess of bluer spectral types relative to redder spectral types at $z > 0.5$.}
\end{figure}

In Fig. \ref{zsedhist} we plot the redshift distributions broken down into spectral types.  For the no-evolution model we take the redshift distributions estimated from the G97 $K$-band LF matched respectively with the CWW E, Sbc, Scd and Im SEDs, with the resulting $N(z)$ distributions renormalized to the total observed counts for each corresponding spectral type.  We do this because the divisions between the galaxy types in G98 do not correspond exactly to those imposed on the observations by fitting the four CWW template SEDs, and in any case G98 identifies six types while we divide the LCIR survey galaxies into only four types.  For the semi-analytic and passive-evolution models, where the SEDs evolve with redshift so that a spectral type may look like the CWW E SED at $z=0$ but look like the CWW Scd SED at $z \sim 2$ say, we instead fit the best CWW spectral type to each galaxy using the observed-frame colours -- i.e. in a manner analogous to calculating photometric redshifts for the observations except without allowing $z$ to vary for any given galaxy.

In the LCIR survey, spectral-type E galaxies are the most common spectral type at $z < 0.75$ but at higher redshifts they begin to become less dominant relative to later spectral types and drop out by $z \sim 1.25$.  The no-evolution model agrees fairly well with observations but underpredicts the number of spectral-type Sbc galaxies at $z > 1$ -- consistent with some luminosity evolution or merging in the observations. In contrast the semi-analytic model greatly underpredicts the fraction of spectral-type E galaxies, except perhaps in the lowest ($z < 0.25$) redshift bin, while overpredicting the numbers of spectral-type Scd and Im galaxies especially towards higher redshifts.  This discrepancy could be explained by excess star formation in the semi-analytic model, for example from too frequent merging with the resulting bursts of star formation being over-prolonged; or if a large fraction of the observed $z \sim 1$ spectral-type E galaxies are in clusters then some process such as ram pressure stripping or the hot cluster environment preventing gas from cooling could act to inhibit star formation.  Some of these processes may not be modelled properly in the semi-analytic model.

Our passive-evolution model using an SSP burst model for evolving local E/S0 galaxies predicts too many spectral-type E galaxies at $z > 0.75$.  If instead we use the $\tau = 1$ Gyr model we get better agreement for spectral-type E galaxies but evolving these galaxies back gives rise to an excess of Sbc types at $0.75 < z < 1.5$ and Scd/Im types at $z > 1.5$.  We note that there is wide scope for variation of the passive-evolution models -- choice of formation redshift $z_f$, star formation history, modelling of dust, LF, initial mass function, cosmology and so on.  In particular, agreement with the observations could be improved if high-redshift bursts of star-formation are heavily obscured by dust (Smail et al. 1999; Blain et al. 1999a).

\begin{figure*}
\vspace{14cm}
\includegraphics{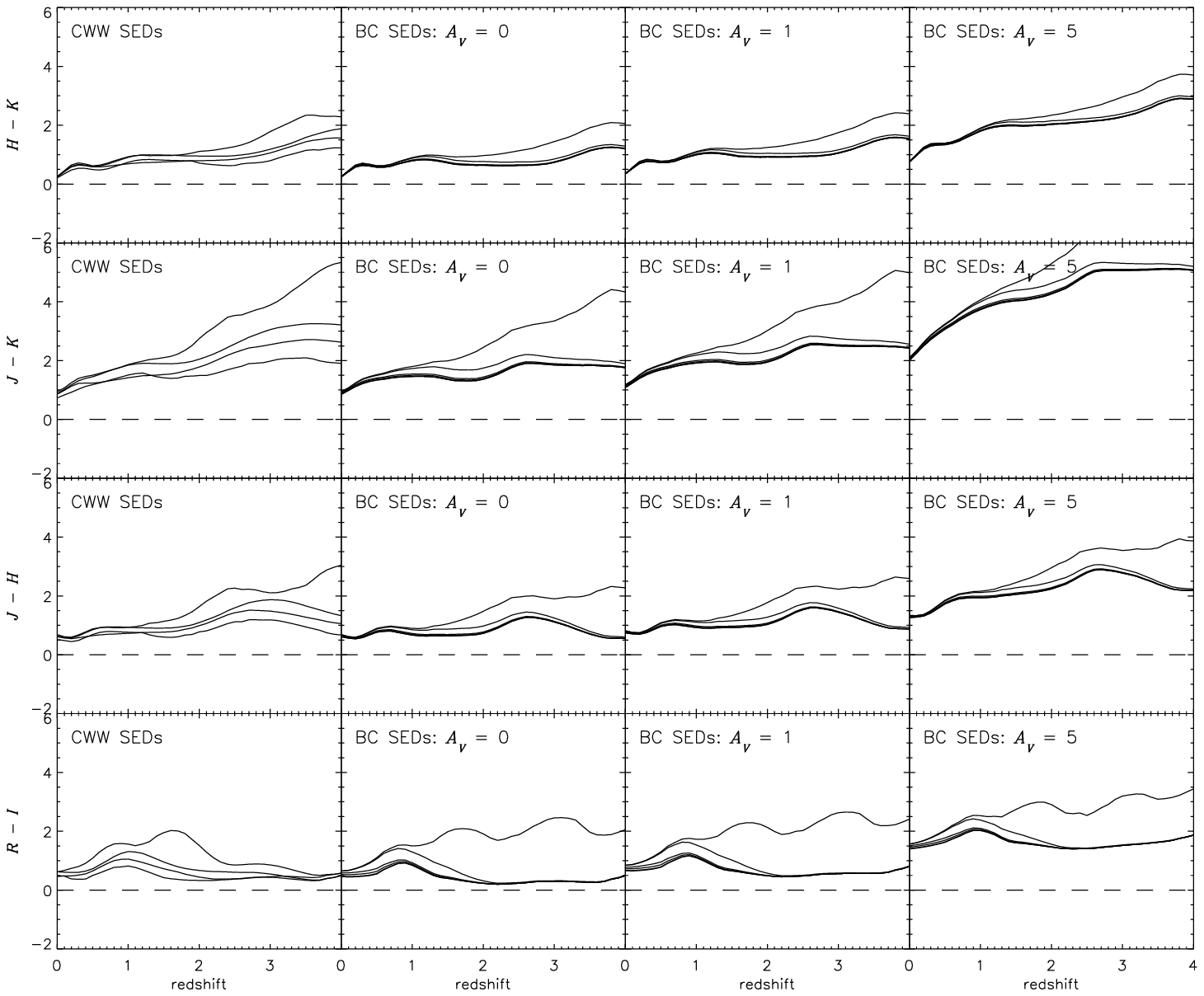}
\caption{\label{cconv} $H - K$, $J - K$, $J - H$ and $R - I$ colours for various SEDs as a function of redshift, for the purpose of conversion between the various ERO colour selection criteria used in the literature.  The SEDs include the empirical CWW templates E, Sbc, Scd, Im and evolving BC {\tt GISSEL'98} SEDs with single stellar population (SSP) burst, exponentially decaying star formation with time-scale $\tau$ = 1, 3, 5 and 15 Gyr and constant star formation rate (SFR), all with formation redshift $z_f = 10$ (see also $\S$\ref{sec.models}.2).  An $\Omega_0 = 0.3$, $\Omega_{\Lambda} = 0.7$, $H_0$ = 70 km s$^{-1}$ Mpc$^{-1}$ cosmology has been assumed.  The BC SEDs have been reddened by $A_V$ = 0.0, 1.0 and 5.0 (CABKKS).  In each panel the redder SEDs (E, SSP burst) are towards the top while the bluer SEDs (Im, constant SFR) are towards the bottom.  At $z = 4$ the galaxies are $\sim$1.1 Gyr old in this cosmology.}
\end{figure*}

The photometric redshift distributions of the EROs are useful for interpreting their angular clustering ($\S$\ref{sec.w}, $\S$\ref{sec.xi}) and surface number density.  In this paper we define EROs by several colour cuts -- $R - H > 4$, $R - H > 4.5$, $I - H > 3$ and $I - H > 3.5$.  Taking typical colours for red galaxies at $z \sim 1$ of $R - I \sim 1.5$ and $H - K \sim 1$ (Fig. \ref{cconv}), these colour cuts correspond to $R - K$ colours of roughly 5, 5.5, 5.5 and 6 respectively.  It is worth noting that these colours are less extreme than those used by many authors -- e.g. Smail et al. (1999) and Thompson et al. (1999) use $R - K > 6$ (see also Totani et al. 2001b) -- but are similar to those used by Daddi et al. (2000a).  The evidence suggests that the reddest EROs are dusty starbursts while less red ERO samples, such as ours, are dominated by `ordinary' galaxies with evolved stellar populations (Smail et al. 1999; Daddi et al. 2000b).  Furthermore, since the filter pairs $R - H$ and $I - H$ are closer together than the more traditional $R - K$, they are less sensitive to the smoother break in dusty galaxies compared with a strong \mbox{4000 \AA} break in $z \sim 1$ evolved elliptical galaxies, and the $I - H$ cut is also more sensitive to evolved galaxies that, due to small amounts of residual star formation -- as might follow late merger events -- would be missed from an $R - K$ colour cut.  

\begin{figure}
\vspace{7cm}
\includegraphics{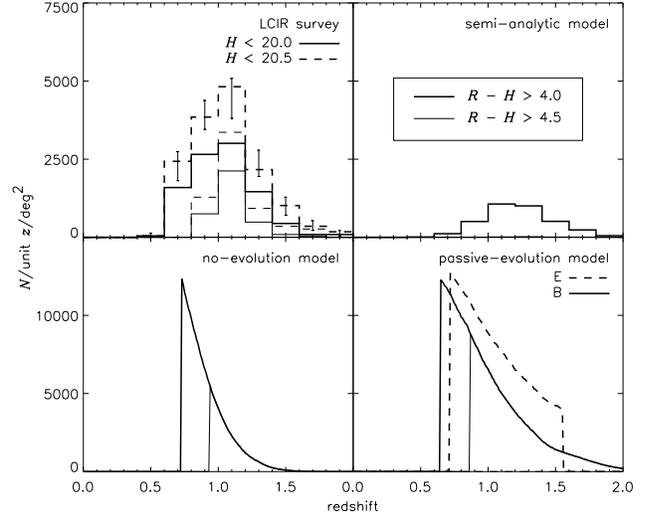}
\caption{\label{eros.rh4} The observed redshift distribution of EROs with $R - H > 4$ and with $R - H > 4.5$ in the LCIR survey for $H < 20.0$ and $H < 20.5$ (upper left), the semi-analytic model (upper right), the no-evolution model (lower left) and two passive-evolution models (lower right).  Error bars, shown on the $R - H > 4$, $H < 20.5$ sample only, are 90 per cent percentiles from bootstrap resampling of the full galaxy catalogue followed by Monte Carlo simulations in which random photometric errors were added to the observed photometry and photometric redshifts and colour cuts were recalculated.  Here we have used the G97 $K$-band LF for the no-evolution and passive-evolution models.  Note the different scale used in the lower relative to the upper panels.  The median redshift of the observed EROs is $z \sim 1$ with the redder sample being at slightly higher redshift.  The semi-analytic model greatly underpredicts the observed ERO counts while the no-evolution model is a closer fit to the observations.  The passive-evolution models generally overpredict the number of EROs, especially the model (B) where the E/S0 population is modelled by a high-redshift SSP burst (note that the $\tau = 1$ Gyr model (E) does not actually produce any $R - H > 4.5$ galaxies though).  By tuning the star formation history and formation redshift of these models it is possible to get the passive-evolution model to fit the observed ERO number counts.}
\end{figure}

\begin{figure}
\vspace{7cm}
\includegraphics{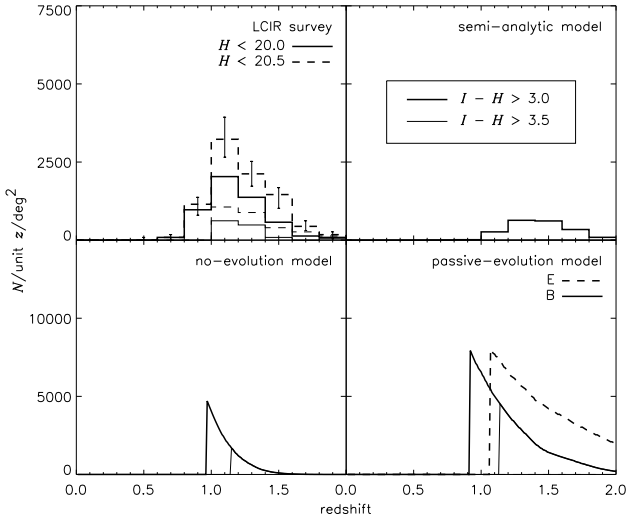}
\caption{\label{eros.ih3} The observed redshift distribution of EROs with $I - H > 3$ and with $I - H > 3.5$ in the LCIR survey for $H < 20.0$ and $H < 20.5$ (upper left), the semi-analytic model (upper right), the no-evolution model (lower left) and two passive-evolution models (lower right).  Error bars, shown on the $I - H > 3$, $H < 20.5$ sample only, are 90 per cent percentiles from bootstrap resampling of the full galaxy catalogue followed by Monte Carlo simulations in which random photometric errors were added to the observed photometry and photometric redshifts and colour cuts were recalculated.  Here we have used the G97 $K$-band LF for the no-evolution and passive-evolution models.  Note the different scale used in the lower relative to the upper panels.  The median redshift of the observed EROs is $z \sim 1.2$.  The semi-analytic model greatly underpredicts the observed ERO counts, the passive-evolution models generally predict an excess of EROs (but may be fine-tuned to fit the observations) and the no-evolution model is a better fit to the observations.}
\end{figure}

In Fig. \ref{eros.rh4} we plot $N(z)$ for EROs with $R - H > 4$ and $R - H > 4.5$ and in Fig. \ref{eros.ih3} we plot $N(z)$ for EROs with $I - H > 3$ and EROs with $I - H > 3.5$.  The median redshifts are typically 1--1.2, with the redder samples being at the higher redshifts, and the redshift range is relatively restricted, especially for the redder samples, thus providing one explanation for the strong angular clustering observed for these objects.  The observed EROs are not all best-fitted by the CWW E SED.  Instead about 70 per cent are spectral-type E while about 30 per cent are spectral-type Sbc.  The median redshift of the Sbc group is $\sim 0.3$ greater than the median redshift of the E group, consistent with a small amount of residual star formation besides a strong \mbox{4000 \AA} break in massive systems at these slightly higher redshifts (cf. Franceschini et al. 1998).  Taking a typical $H - K$ colour of 1 (Fig. \ref{cconv}), our number density of 0.5 arcmin$^{-2}$ ($H < 20.0$; 744 arcmin$^2$) to 1 arcmin$^{-2}$ ($H < 20.5$; 407 arcmin$^2$) of EROs with $R - H > 4$ agrees well with the number density of 0.5 arcmin$^{-2}$ of EROs with $R - K_s > 5$ and $K_s < 19$ of D00 (701 arcmin$^2$).  At $R - H > 5$ and $H < 20$ we find 0.06 arcmin$^{-2}$ compared with 0.04 arcmin$^{-2}$ (Thompson et al. 1999; 154 arcmin$^2$) and 0.07 arcmin$^{-2}$ (D00) for $R - K > 6$ and $K < 19$ and at $R - H > 4.3$ and $H < 19$ we find 0.03 arcmin$^{-2}$ compared with 0.05 arcmin$^{-2}$ (Martini 2001b; 185 arcmin$^2$) and 0.03 arcmin$^{-2}$ (D00) for $R - K > 5.3$ and $K < 18$.

\begin{figure}
\vspace{7cm}
\includegraphics{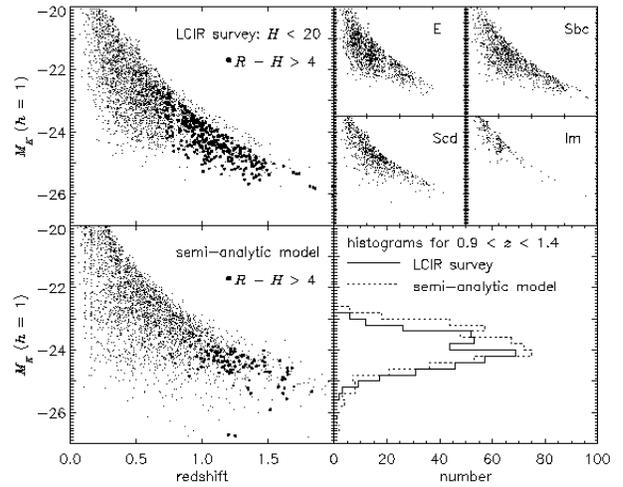}
\caption{\label{zabsk} Absolute $K$ magnitude $M_K$ as a function of redshift for galaxies in the LCIR survey (upper left) and divided by spectral type (upper right), and for galaxies in the semi-analytic model (lower left).  In the LCIR survey, $M_K$ has been estimated by extrapolating the observed $UBVRIH$ photometry to the rest-frame $K$-band using the best-fitting CWW SED for each galaxy.  The number of galaxies in the semi-analytic model has been scaled to match the observations.   EROs with $R - H > 4$ are marked by filled circles: there are many more EROs in the observations than in the model.  At lower right are histograms of $M_K$ summed over the redshift range $0.9 < z < 1.4$.  The faint cut-off is imposed by the $H < 20$ magnitude limit.  Using $M_K$ as a tracer of total stellar mass, the plot indicates that despite underpredicting the number of EROs the semi-analytic model does in fact produce enough bright galaxies (indicated by the bright-end $M_K$ magnitudes) in the redshift range occupied by the observed EROs.}
\end{figure}

The number density of EROs can provide a sensitive constraint on models (Zepf 1997; DCR) being essentially the same test as the number density of $z \sim 1$ spectral-type E galaxies.  We have listed the ERO fractions in the observations and models for various red colour cuts in Table \ref{table.erofraction}.  The no-evolution model based on the $K$-band LF predicts that 30 per cent of galaxies in the magnitude range $19 < H < 20$ have $R - H > 4$, exceeding the observed fraction of $\sim$18 per cent by two thirds.  This fraction would decrease to 17 per cent if we matched a bluer spectral type to spectral type 2 (see $\S$\ref{sec.models}.2) instead of fitting the CWW E SED to both types 1 and 2.  We plot the $R - H > 4$ counts for this model in Fig. \ref{hcounts} showing that they provide a good fit to the observed counts.  On the other hand the semi-analytic model greatly underestimates the number of EROs, predicting that only 4 per cent of galaxies in this magnitude range have $R - H > 4$ (see also DCR; Martini 2001b).  However the semi-analytic model does appear to produce enough bright galaxies at these redshifts (Fig. \ref{zabsk}) indicating that the problem lies in the modelled star formation history of massive galaxies rather than the assembly of mass through merging.  The deficit of EROs in the semi-analytic model could be due, as mentioned above, to some process inhibiting late star formation in massive galaxies that may not be modelled properly in the semi-analytic model, or it could be due to a significant fraction of the observed EROs being very dusty ULIRG-type galaxies and these are not modelled correctly by the simple dust formalism used (equation \ref{eq.samdust}) (see also Guiderdoni et al. 1998; Blain et al. 1999b).  

\begin{table*}
\centering
\caption{\label{table.erofraction} Table showing the fraction of EROs, in the magnitude range $19 < H < 20$, defined by various optical-near-IR colour cuts in the observations and in several models.  For the passive-evolution model we fitted two alternative BC SEDs to the E/S0 LF -- (B) a single stellar population burst and (E) exponentially-decaying star formation with time-scale $\tau$ = 1 Gyr, both with formation redshift $z_f = 10$.  Here we have used the G97 $K$-band LF for the no-evolution and passive-evolution models.  The (E) model produces no $R - H > 4.5$ or $I - H > 3.5$ galaxies, but could still be consistent with the observations if the reddest EROs are a separate population of dusty ULIRG-type galaxies (e.g. Smail et al. 1999 and references therein).}
\begin{tabular}{lcccc}
{\it sample} & $R - H > 4.0$ & $R - H > 4.5$ & $I - H > 3.0$ & $I - H > 3.5$ \\ \hline
LCIR survey           &  18\% &   7\% &  11\% & 2.7\%\\
semi-analytic model   & 4.3\% & 0.3\% & 2.5\% & 0.1\%\\
no-evolution model    &  30\% &  11\% &   9\% & 2.7\%\\
passive-evolution (B) &  38\% &  28\% &  26\% &  19\%\\
passive-evolution (E) &  29\% &   0\% &  21\% &   0\%\\
\end{tabular}
\end{table*}

The passive-evolution models -- in particular those where present-day ellipticals are modelled by SSP high-redshift bursts -- generally overpredict the number of EROs.  However it is possible to tune the star formation history, formation redshift and fraction of the LF's normalization  $\phi_*$ taken to correspond to the ERO population, in order to fit the observed numbers of EROs (e.g. DCR; McCarthy et al. 2001).  Dust obscuration may be invoked in these models to hide intense star-formation activity at high redshifts in order to fit the overall counts and redshift distributions (Mazzei, de Zotti \& Xu 1994; Franceschini et al. 1994, 1998;  Blain et al. 1999a) and this may be consistent with populations of high-redshift ULIRG-type galaxies being the progenitors of the $z \sim 1$ ERO population and ultimately present-day elliptical galaxies (de Zotti et al. 2001, and references therein).

While we hoped to use the $z \sim 1$ density of evolved galaxies to discriminate between passive-evolution and hierarchical merging models, the test is compromised by the considerable freedom allowed within the passive-evolution frame-work, so that the passive-evolution models can be adjusted to fit a wide range of potential observations.  A proper treatment must include constraints from a range of wavebands -- something which is beyond the scope of this paper.  Within the LCIR survey it is expected that more stringent constraints will be produced with the inclusion of further fields, some of which (a) have deeper near-IR imaging -- allowing one to study galaxies in greater numbers at $z \sim$ 1.5--2, where the number density of evolved galaxies more strongly discriminates between hierarchical merging and passive-evolution models and puts more stringent constraints on the latter, and (b) have $ZJK$ imaging -- which, through increased spectral coverage, will allow better constraints to be placed on the role of dust in these objects.  On the other hand the observations have usefully been used to identify discrepancies in the semi-analytic model that are worthy of further investigation.

\section{Angular clustering}  \label{sec.w}
The angular correlation function $w(\theta)$ gives the excess probability over a random Poisson distribution that two sources will be found in the solid angle elements $\delta \Omega_{\mbox{1}}$ and $\delta \Omega_{\mbox{2}}$ separated by an angle $\theta$ and is defined by
\begin{equation}
\delta P = n^2 \delta \Omega_1 \delta \Omega_2 [1 + w(\theta)],   \label{eq.wdef}
\end{equation}
where $n$ is the mean number density of objects.  A standard procedure for estimating $w(\theta)$ from a catalogue of objects is to generate a random catalogue of a large number of points within the same area covered by the observations and then for each $\theta$ count the number of distinct data-data ($DD$) pairs, data-random ($DR$) pairs and random-random ($RR$) pairs with separation $\theta$ to $\theta + d \theta$.  It is important that the random points are subject to the same selection effects as the data points.  Due to the relatively small size of infrared detectors and various instrumental artifacts, the infrared data are relatively patchy in terms of depth so in this paper we have adopted conservative magnitude limits such that the remaining catalogue is essentially spatially uniform (see $\S$\ref{sec.cat}).  The masked regions around bright stars are likewise removed from the random catalogue.  With $DD$, $DR$ and $RR$ as defined above, $w(\theta)$ is estimated by:
\begin{equation}
w = 4 \frac{DD \times RR}{DR^2} -1   \label{eq.H}
\end{equation}
(Hamilton 1993).  Following Hewett et al. (1982) we estimate errors analytically as $\sqrt{(1+w(\theta))/DD}$.  These are similar to 1$\sigma$ errors (Sharp 1979; but see also Mo, Jing \& B\"orner 1992).  Proceeding in this manner we calculate $w(\theta)$ for various subsamples of the observations selected by $H$ magnitude, colour, photometric redshift and spectral type.  At low redshifts, $w(\theta)$ is well-approximated by $w(\theta) = A (\theta ^ {1-\gamma} - C)$ (Groth \& Peebles 1977; Maddox et al. 1990), where $A$ is the clustering amplitude and $C$ is the integral constraint
\begin{equation}
C = \frac{1}{\Omega^2} \int \int \theta^{1-\gamma} \delta \Omega_1 \delta \Omega_2.
\end{equation}
Typically $\gamma =$ 1.6--1.8 depending to some extent on Hubble type.  We fix $\gamma = 1.8$ since this value has been frequently chosen in the literature when the quantity of data is too small to allow an accurate determination of both $A$ and $\gamma$.  We calculate $C \sim 0.141$ for the area comprising all six tiles and $C \sim 0.151$ for the area comprising the three deeper tiles ($\theta$ in arcminutes).  We then obtained the amplitudes at 1 arcmin ( $\equiv$ 0.7 $h^{-1}$Mpc in comoving coordinates at $z = 1$ and 1.1 $h^{-1}$Mpc at $z = 2$ for an $\Omega_{0} = 0.3$, $\Omega_{\Lambda} = 0.7$ cosmology) of  $w(\theta)$ by least-squares fitting over the angular range 4--600 arcsecs.

\begin{figure}
\vspace{7cm}
\includegraphics{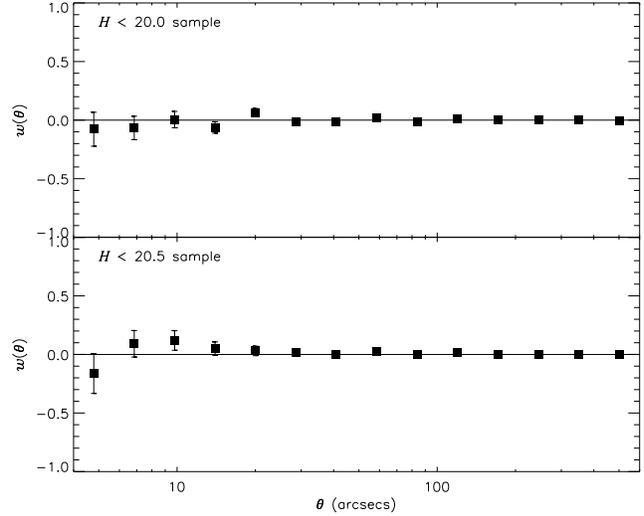}
\caption{\label{wth.stars} The angular correlation function $w(\theta)$ of stars in the $H < 20.0$ catalogue (top panel) and in the $H < 20.5$ catalogue (lower panel).  The observations are consistent with zero clustering implying that there is little loss of galaxies during star/galaxy separation and that the spatial uniformity of the data is good.}
\end{figure}

\begin{figure}
\vspace{7cm}
\includegraphics{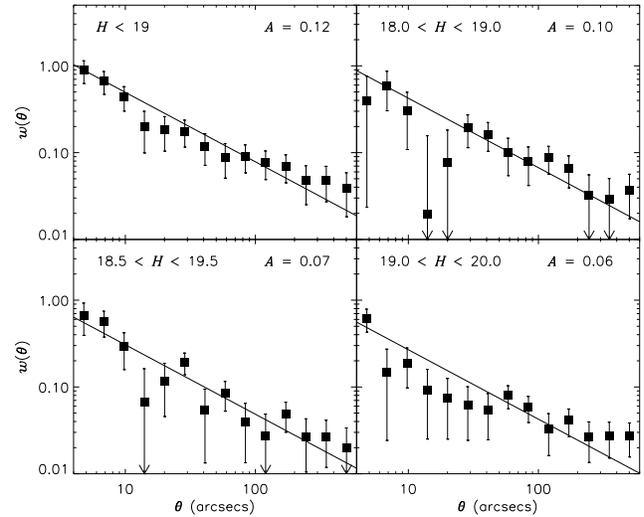}
\caption{\label{wth.hmags} The angular correlation function $w(\theta)$ of galaxies in $H$ magnitude bins.  The amplitude decreases as fainter magnitude limits encompass broader $N(z)$ distributions.}
\end{figure}

Fig. \ref{wth.stars} shows the correlation function of the stars in the $H < 20.0$ and $H < 20.5$ samples.  It is a useful check on the quality and spatial uniformity of the catalogues that the correlation function of stars is consistent with zero clustering.  Forcing a power-law fit returns an amplitude $A$ of 0.002 ($H < 20.0$) compared with 0.08 for the galaxies.  The value 0.002 is in fact consistent with contamination by the estimated $\sim$15 per cent fraction of galaxies that may have been conservatively classed as stars (see $\S$\ref{sec.sg}).  In the remaining $w(\theta)$ figures the integral constraint term has been added to the data so that the power-law fit is a straight line on a log plot.  Fig. \ref{wth.hmags} compares $w(\theta)$ for different $H$ magnitude bins.  The trend as one moves to fainter magnitudes is for the amplitude to decrease -- a fainter sample covers a greater redshift range and hence, in projection, there is greater dilution of the intrinsic clustering signal.  These results are summarized in Table \ref{table.hmags} in which we also note the increasing median redshift (from 0.42 to 0.69) with increasing $H$ magnitude.

\begin{table*}
\centering
\caption{\label{table.hmags}  The number of galaxies $N$, median photometric redshift $z_m$, median absolute $B$ and $I$ magnitudes $M_B$ and $M_I$ ($h = 1$) and amplitude $A$ of $w(\theta)$ at 1 arcmin for $H$ magnitude bins.  The approximately 1$\sigma$ errors on $A$ are derived from $\sqrt{(1+w(\theta))/DD}$ errors on $w(\theta)$.  Extra statistics are included in Table \ref{table.all} in the appendix.}
\begin{tabular}{cccccc}
\it sample & $N$ & $z_{m}$ & $M_B$ & $M_I$ & $A$\\ \hline
$H < 19$          & 1284& 0.42& $-$19.2& $-$21.1& $0.12 \pm 0.03$\\
$18.0 < H < 19.0$ &  868& 0.47& $-$19.1& $-$21.0& $0.10 \pm 0.04$\\ 
$18.5 < H < 19.5$ & 1292& 0.55& $-$19.1& $-$20.9& $0.07 \pm 0.03$\\ 
$19.0 < H < 20.0$ & 1893& 0.61& $-$19.1& $-$20.8& $0.06 \pm 0.02$\\ 
$19.5 < H < 20.5$ & 1472& 0.69& $-$18.9& $-$20.5& $0.07 \pm 0.02$\\ 
\hline
\end{tabular}
\end{table*}

\begin{figure}
\vspace{7cm}
\includegraphics{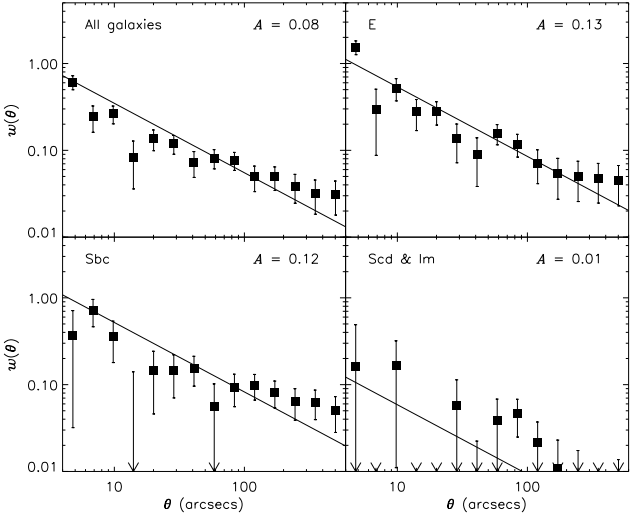}
\caption{\label{wth.seds} The angular correlation function $w(\theta)$ of the full galaxy sample in the $H < 20.0$ catalogue and of different photometrically-determined spectral type classes.  The amplitude is greatest for red spectral types.}
\end{figure}

\begin{figure}
\vspace{7cm}
\includegraphics{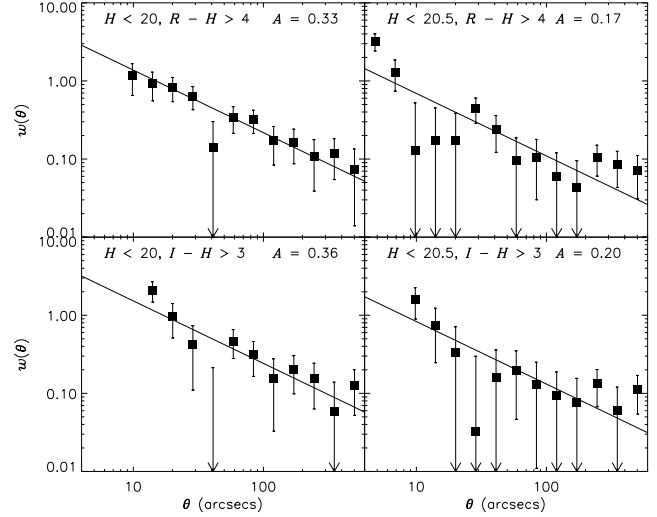}
\caption{\label{wth.eros} The angular correlation function of EROs with $R - H > 4$ and $I - H > 3$ for two $H$ limiting magnitudes.  Points have been omitted in the smallest angular separation bins where there are too few data-data pairs to obtain a decent measurement.  For the $H < 20.0$ sample, the clustering amplitude is four times that of the galaxy population as a whole.  This decreases to only about two-and-a-half times in the fainter sample, explained both by the more restricted redshift range of the brighter sample (Fig. \ref{eros.rh4} and Fig. \ref{eros.ih3}) and the greater intrinsic luminosity of these galaxies.  The clustering amplitude of the $R - H > 4$, $H < 20$ sample is in good agreement (within 10 per cent) of that measured by Daddi et al. (2000a) for $R - K_s > 5$, $K_s < 19$ objects.}
\end{figure}

\begin{figure}
\vspace{7cm}
\includegraphics{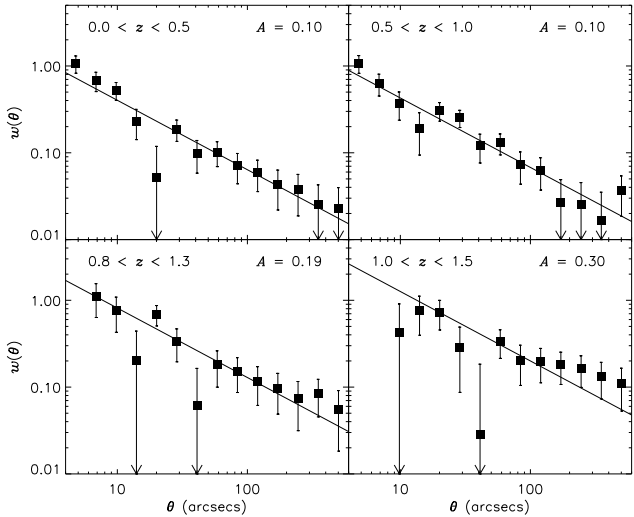}
\caption{\label{wth.zbins} The angular correlation function $w(\theta)$ of $H < 20.0$ galaxies selected by photometric redshift.  The amplitude increases dramatically to higher redshifts for this apparent magnitude limited sample since at high redshifts only the most luminous galaxies (presumably corresponding to peak fluctuations in the underlying matter density distribution) make it into the sample.  For a fainter $H < 20.5$ sample the increase to higher redshift is less dramatic.}
\end{figure}

\begin{figure}
\vspace{8.5cm}
\includegraphics{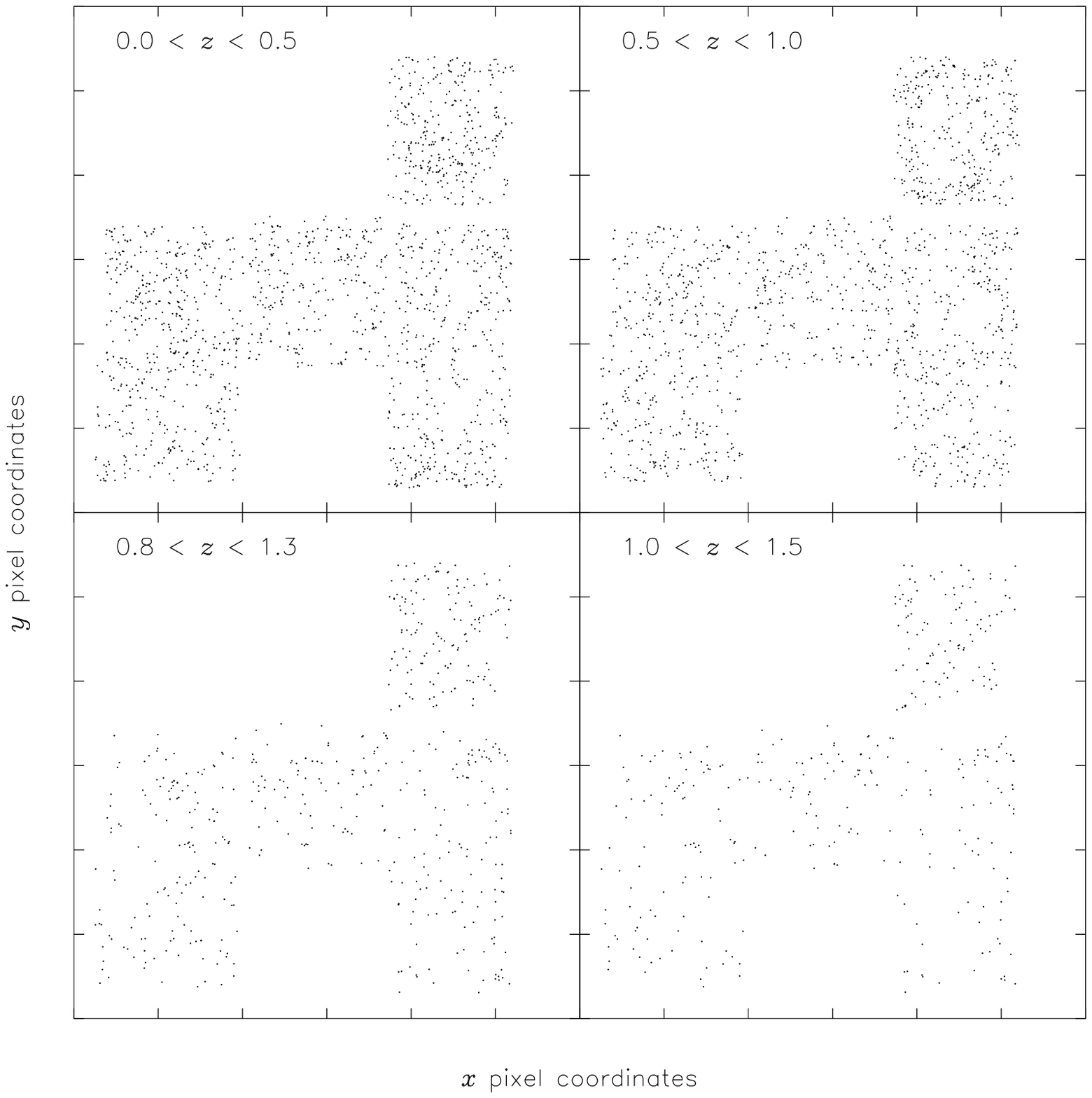}
\caption{\label{xy} The positions within the survey area of the $H < 20.0$ galaxies in four photometric redshift bins.}
\end{figure}

In Fig. \ref{wth.seds} we compare $w(\theta)$ for different photo-\linebreak metrically-determined spectral type classes -- all galaxies, E, Sbc and Scd + Im -- in the $H < 20.0$ catalogue.  While the E and Sbc classes are equally strongly clustered, the Scd + Im class is much less strongly clustered.  These results, and those for the $H < 20.5$ catalogue, are summarized in Table \ref{table.seds}.  This table also contains spatial clustering results from $\S$\ref{sec.xi} and we leave further discussion until that section.  In Fig. \ref{wth.eros} we show angular clustering results for several red colour cuts.  The $R - H > 4$ and $I - H > 3$ galaxies are four times more strongly clustered than the sample of all galaxies in the $H < 20.0$ catalogue but this drops to only two-and-a-half times in the $H < 20.5$ catalogue.  This is probably due to both the broader redshift distribution and the fainter average luminosity of the fainter sample.  Indeed if we compare the clustering of $R - H > 4$ or $I - H > 3$ galaxies with that of $R - H < 4$ or $I - H < 3$ galaxies in the same photometric redshift range (Table \ref{table.eros}) then the difference in clustering amplitude is only a factor of two in both samples.  We discuss these results further in $\S$\ref{sec.xi}.  In Fig. \ref{wth.zbins} we compare $w(\theta)$ in different photometric redshift bins from $z = 0$ to $z = 1.5$.  The projected positions of the galaxies in the different redshift bins are plotted in Fig. \ref{xy}.  The angular clustering in 0.5 redshift bins increases dramatically with redshift.  From Table \ref{table.zbins} we see that this correlates with an increase in median absolute magnitude from $M_I - 5 \log h = -19.9$ in the lowest redshift bin to $M_I - 5 \log h = -22.5$ at $z_m \sim 1.2$.  The increase in clustering amplitude with redshift is less pronounced in the fainter $H < 20.5$ sample.

\begin{table*}
\centering
\caption{\label{table.seds}  The number of galaxies $N$, median photometric redshift $z_m$, median absolute $B$ and $I$ magnitudes $M_B$ and $M_I$ ($h = 1$), amplitude $A$ of $w(\theta)$ at 1 arcmin and spatial correlation lengths $r$ in $h^{-1}$Mpc (comoving coordinates) for various spectral types.  We fix $\gamma = 1.8$.  The correlation length is calculated for two cosmologies $\Omega_0 = 1$, $\Omega_{\Lambda} = 0$ and $\Omega_0 = 0.3$, $\Omega_{\Lambda} = 0.7$.  $N(z)$ is derived from photometric redshifts. The errors on $r_{0}$ ($\epsilon = -1.2$) are approximately 1$\sigma$ and are carried through from the $\sqrt{(1+w(\theta))/DD}$ errors on $w(\theta)$.  The errors on the other correlation lengths are of similar proportions.  $r_{*}(z = 0.5)$ is listed for $\epsilon = 0$.  For $\epsilon = -1.2$, $r_{*}(z)$ equals $r_0$ for any $z$.  With median redshifts $z_m \sim 0.5$, $r_0^{\epsilon = -1.2} \equiv r_{*}^{\epsilon = -1.2}(z = 0.5) \approx r_{*}^{\epsilon = 0}(z = 0.5)$.  Extra statistics are included in Table \ref{table.all} in the appendix.}
\begin{tabular}{lccccccccc}
\it sample & $N$ & $z_{m}$ & $M_B$ & $M_I$ & $A$ & $r_0^{\epsilon = -1.2}$ & $r_0^{\epsilon = 0}$ & $r_{*}^{\epsilon = 0}(0.5)$ & $r_0^{\epsilon = -1.2}$ \\ \hline
&&&&&&\multicolumn{3}{c}{$\Omega_0 = 0.3$, $\Omega_{\Lambda} = 0.7$}&\multicolumn{1}{c}{$\Omega_0 = 1$, $\Omega_{\Lambda} = 0$}\\ \hline
$H < 20.0$\\
All galaxies & 3177& 0.52& $-$19.1& $-$20.9& 0.08$\pm$0.01& 5.5$\pm$0.4&  6.9& 5.3& 4.3$\pm$0.3\\
E            & 1194& 0.50& $-$18.7& $-$20.9& 0.13$\pm$0.03& 6.4$\pm$0.8&  8.1& 6.2& 5.1$\pm$0.7\\
Sbc          &  963& 0.61& $-$19.6& $-$21.3& 0.12$\pm$0.04& 7.7$\pm$1.3& 10.2& 7.8& 5.8$\pm$1.0\\
Scd $+$ Im   & 1020& 0.48& $-$19.1& $-$20.5& 0.01$\pm$0.04& 1.9$\pm$1.9&  2.2& 1.7& 1.5$\pm$1.5\\
\hline
$H < 20.5$\\
All galaxies & 2616& 0.57& $-$19.0& $-$20.8& 0.08$\pm$0.01& 6.1$\pm$0.4&  7.7& 5.9& 4.7$\pm$0.3\\
E            &  910& 0.53& $-$18.5& $-$20.7& 0.12$\pm$0.03& 6.6$\pm$0.9&  8.3& 6.4& 5.2$\pm$0.7\\
Sbc          &  735& 0.65& $-$19.5& $-$21.3& 0.12$\pm$0.04& 8.2$\pm$1.4& 11.0& 8.4& 6.1$\pm$1.1\\
Scd $+$ Im   &  971& 0.52& $-$19.0& $-$20.4& 0.05$\pm$0.03& 4.2$\pm$1.4&  5.1& 3.9& 3.4$\pm$1.1\\
\hline
\end{tabular}
\end{table*}

\begin{table*}
\centering
\caption{\label{table.eros}  The number of galaxies $N$, median photometric redshift $z_m$, median absolute $B$ and $I$ magnitudes $M_B$ and $M_I$ ($h = 1$), amplitude $A$ of $w(\theta)$ at 1 arcmin and spatial correlation lengths $r$ in $h^{-1}$Mpc (comoving coordinates) for various colour and photometric redshift-selected groups of red galaxies.  We fix $\gamma = 1.8$.  The correlation length is calculated for two cosmologies $\Omega_0 = 1$, $\Omega_{\Lambda} = 0$ and $\Omega_0 = 0.3$, $\Omega_{\Lambda} = 0.7$.  $N(z)$ is derived from photometric redshifts.  The errors on $r_{0}$ ($\epsilon = -1.2$) are approximately 1$\sigma$ and are carried through from the $\sqrt{(1+w(\theta))/DD}$ errors on $w(\theta)$.  The errors on the other correlation lengths are of similar proportions.  $r_{*}(z = 1.0)$ is listed for $\epsilon = 0$.  For $\epsilon = -1.2$, $r_{*}(z)$ equals $r_0$ for any $z$.  Extra statistics are included in Table \ref{table.all} in the appendix.}
\begin{tabular}{llccccccccc}
\multicolumn{2}{l}{\it sample} & $N$ & $z_{m}$ & $M_B$ & $M_I$ & $A$ & $r_0^{\epsilon = -1.2}$ & $r_0^{\epsilon = 0}$ & $r_{*}^{\epsilon = 0}(1.0)$ & $r_0^{\epsilon = -1.2}$ \\ \hline
&&&&&&&\multicolumn{3}{c}{$\Omega_0 = 0.3$, $\Omega_{\Lambda} = 0.7$}&\multicolumn{1}{c}{$\Omega_0 = 1$, $\Omega_{\Lambda} = 0$}\\ \hline
$H < 20.0$\\
\multicolumn{2}{l}{All galaxies} & 3177& 0.52& $-$19.1& $-$20.9& 0.08$\pm$0.01&  5.5$\pm$0.4&  6.9&  4.4& 4.3$\pm$0.3\\
$R - H > 4$& 0.7 $< z <$ 1.5     &  337& 1.01& $-$20.2& $-$22.3& 0.33$\pm$0.11& 11.1$\pm$2.0& 17.6& 11.1& 7.6$\pm$1.4\\
$R - H < 4$& 0.7 $< z <$ 1.5     &  480& 0.87& $-$20.2& $-$21.8& 0.15$\pm$0.08&  5.9$\pm$1.7&  8.8&  5.5& 4.1$\pm$1.2\\
$I - H > 3$& 0.9 $< z <$ 1.5     &  201& 1.13& $-$20.4& $-$22.5& 0.36$\pm$0.18& 10.5$\pm$2.9& 17.3& 10.9& 7.0$\pm$1.9\\
$I - H < 3$& 0.9 $< z <$ 1.5     &  268& 1.05& $-$20.6& $-$22.2& 0.17$\pm$0.13&  6.0$\pm$2.8&  9.6&  6.0& 4.0$\pm$1.9\\
\hline
$H < 20.5$\\
\multicolumn{2}{l}{All galaxies} & 2616& 0.57& $-$19.0& $-$20.8& 0.08$\pm$0.01&  6.1$\pm$0.4&  7.7&  4.8& 4.7$\pm$0.3\\
$R - H > 4$& 0.7 $< z <$ 1.5     &  312& 1.02& $-$19.8& $-$21.9& 0.17$\pm$0.09&  7.7$\pm$2.4& 12.1&  7.6& 5.2$\pm$1.6\\
$R - H < 4$& 0.7 $< z <$ 1.5     &  516& 0.89& $-$20.0& $-$21.6& 0.09$\pm$0.05&  4.7$\pm$1.7&  7.0&  4.4& 3.3$\pm$1.2\\
$I - H > 3$& 0.9 $< z <$ 1.5     &  170& 1.16& $-$20.3& $-$22.4& 0.20$\pm$0.16&  7.5$\pm$3.7& 12.5&  7.9& 5.0$\pm$2.5\\
$I - H < 3$& 0.9 $< z <$ 1.5     &  306& 1.08& $-$20.4& $-$22.0& 0.09$\pm$0.09&  4.4$\pm$3.3&  7.1&  4.5& 3.0$\pm$2.2\\
\hline
\end{tabular}
\end{table*}

\begin{table*}
\centering
\caption{\label{table.zbins}  The number of galaxies $N$, median photometric redshift $z_m$, median absolute $B$ and $I$ magnitudes $M_B$ and $M_I$ ($h = 1$), amplitude $A$ of $w(\theta)$ at 1 arcmin and spatial correlation lengths $r$ in $h^{-1}$Mpc (comoving coordinates) for various photometric redshift-selected samples.  We fix $\gamma = 1.8$.  The correlation length is calculated for two cosmologies $\Omega_0 = 1$, $\Omega_{\Lambda} = 0$ and $\Omega_0 = 0.3$, $\Omega_{\Lambda} = 0.7$.  $N(z)$ is derived from photometric redshifts. The errors on $r_{0}$ ($\epsilon = -1.2$) are approximately 1$\sigma$ and are carried through from the $\sqrt{(1+w(\theta))/DD}$ errors on $w(\theta)$.  The errors on the other correlation lengths are of similar proportions.  $r_{*}(z_m)$ is listed for $\epsilon = 0$.  For $\epsilon = -1.2$, $r_{*}(z)$ equals $r_0$ for any $z$.  Since galaxies are localized into redshift bins, $r_*(z_m)$ is nearly independent of an assumed $\epsilon$.  Extra statistics are included in Table \ref{table.all} in the appendix.}
\begin{tabular}{lccccccccc}
\it sample & $N$ & $z_{m}$ & $M_B$ & $M_I$ & $A$ & $r_{0}^{\epsilon = -1.2}$ & $r_{*}^{\epsilon = 0}(z_{m})$ & $r_{0}^{\epsilon = -1.2}$ & $r_{*}^{\epsilon = 0}(z_{m})$ \\ \hline
&&&&&&\multicolumn{2}{c}{$\Omega_0 = 0.3$, $\Omega_{\Lambda} = 0.7$}&\multicolumn{2}{c}{$\Omega_0 = 1$, $\Omega_{\Lambda} = 0$}\\ \hline
$H < 20.0$\\
0.0 $< z <$ 0.5 & 1458& 0.32& $-$18.1& $-$19.9& 0.10$\pm$0.03& 3.3$\pm$0.5& 3.3& 2.8$\pm$0.4& 2.7\\
0.5 $< z <$ 1.0 & 1330& 0.64& $-$19.5& $-$21.3& 0.10$\pm$0.03& 4.0$\pm$0.6& 4.0& 3.0$\pm$0.4& 2.9\\
0.8 $< z <$ 1.3 &  510& 1.00& $-$20.2& $-$22.1& 0.19$\pm$0.07& 6.6$\pm$1.3& 6.5& 4.5$\pm$0.9& 4.5\\
1.0 $< z <$ 1.5 &  343& 1.16& $-$20.7& $-$22.5& 0.30$\pm$0.10& 8.0$\pm$1.5& 8.0& 5.4$\pm$1.0& 5.3\\
\hline
$H < 20.5$\\
0.0 $< z <$ 0.5 & 1056& 0.32& $-$17.8& $-$19.5& 0.13$\pm$0.03& 3.9$\pm$0.4& 3.8& 3.2$\pm$0.4& 3.2\\
0.5 $< z <$ 1.0 & 1101& 0.66& $-$19.3& $-$21.0& 0.10$\pm$0.03& 4.3$\pm$0.6& 4.2& 3.1$\pm$0.4& 3.1\\
0.8 $< z <$ 1.3 &  524& 1.00& $-$20.0& $-$21.8& 0.13$\pm$0.05& 5.4$\pm$1.2& 5.3& 3.7$\pm$0.8& 3.7\\
1.0 $< z <$ 1.5 &  369& 1.18& $-$20.5& $-$22.3& 0.17$\pm$0.08& 6.0$\pm$1.5& 6.0& 4.0$\pm$1.0& 4.0\\
\hline
\end{tabular}
\end{table*}

Though we have fitted a $1 - \gamma = -0.8$ power-law to $w(\theta)$ (in view of the small sample sizes and for consistency with other studies) there is indication that a flatter slope is more appropriate for the fainter magnitudes (Fig. \ref{wth.hmags}), later spectral types (Fig. \ref{wth.seds}) and higher redshift bins (Fig. \ref{wth.zbins}) in agreement with e.g. Loveday et al. (1995), Postman et al. (1998), Brown, Webster \& Boyle (2000) and Kauffmann et al. (1999b).

Given the observed clustering amplitude $A$ for a sample we may calculate the expected fluctuations in the number counts in a region of given angular size.  Following Peebles (1980) and Lahav \& Saslaw (1992), the variance in the number counts in square cells of size $s = a \times a$ arcmin$^2$ is 
\begin{equation}
\begin{array}{lll}
\langle (N-\bar{N})^2 \rangle & = & \bar{N} + (\frac{\bar{N}}{s})^2 \int \int_{s} w(\theta) ds_{1} ds_{2}\\
& = & \bar{N} + \bar{N}^2 C_{\gamma} a^{1-\gamma} A \label{eq.LO}\\
\end{array}
\end{equation}
where $\bar{N}$ is the mean number per cell, and $C_{\gamma} \approx 2.25$ for $\gamma = 1.8$.  Table \ref{table.stats} lists the observed $\sigma$ of the number counts over the six tiles, the expected $\sigma$ if the objects were distributed according to a Poisson distribution and the calculated $\sigma$ using equation (\ref{eq.LO}) for various samples.  For stars the observed, Poisson and calculated $\sigma$ are comparable (less so in the fainter magnitude bins where there is some galaxy contamination).  For galaxies, the observed $\sigma$ is comparable to the calculated $\sigma$ but much larger than the Poisson value -- as expected, since galaxies are clustered.  In equation (\ref{eq.wdef}) we required the mean number density $n$ of a sample of objects in order to calculate $w(\theta)$.  It is important therefore to check that the survey is large enough to provide a good estimate of $n$.  To do this we consider the calculated $\sigma$ and scale by $1/\sqrt{M}$ where $M$ is the number of tiles.  For most samples the fluctuations in number counts between areas the size of six combined tiles are less than 10 per cent.  For the ERO samples the fluctuations are of order 15 per cent (20 per cent in the deeper area comprising only three tiles) so ideally we would like a larger area, such as the complete survey will provide, to get more robust measurements for these galaxies.

\begin{table}
\centering
\caption{\label{table.stats} Table showing for various samples the mean number counts per tile, the observed standard deviation in the number counts between the six tiles, the expected standard deviation if the objects in the sample were distributed according to a Poisson distribution, and the expected standard deviation given the measured angular clustering amplitude (as described in the text).  When calculating the observed standard deviation, the observed number counts in each tile were first normalized to the mean tile area.  For stars, $\sigma_{\rm observed} \sim \sigma_{\rm Poisson} \sim \sigma_{w(\theta)}$.  For galaxies, $\sigma_{\rm observed} \sim \sigma_{w(\theta)} \gg \sigma_{\rm Poisson}$.}
\begin{tabular}{lrrrr}
{\it sample} & $\bar{N}$ & $\sigma_{\rm observed}$ & $\sigma_{\rm Poisson}$ & $\sigma_{w(\theta)}$\\ \hline
All stars           & 372 & 28.9  &  19.3  &  21.5\\
$H < 19.0$          & 237 & 22.6  &  15.4  &  16.0\\
$19.0 < H < 19.5$   & 64  &  7.9  &   8.0  &  10.0\\
$19.5 < H < 20.0$   & 70  & 15.4  &   8.4  &  11.2\\
\hline
All galaxies        & 533 & 95.8  &  23.1  &  89.9\\
E                   & 201 & 38.9  &  14.2  &  43.1\\
Sbc                 & 162 & 38.5  &  12.7  &  34.6\\
Scd + Im            & 171 & 23.4  &  13.1  &  17.3\\
\hline
$H < 19.0$          & 215 & 42.3  &  14.7  &  44.3\\
$18.0 < H < 19.0$   & 145 & 28.8  &  12.0  &  28.7\\
$18.5 < H < 19.5$   & 217 & 38.6  &  14.7  &  36.3\\
$19.0 < H < 20.0$   & 318 & 55.9  &  17.8  &  48.9\\
\hline
$0.0 < z < 0.5$     & 245 & 32.6  &  15.7  &  45.7\\
$0.5 < z < 1.0$     & 223 & 43.0  &  14.9  &  43.0\\
$0.8 < z < 1.3$     & 85  & 24.1  &   9.2  &  23.1\\
$1.0 < z < 1.5$     & 57  & 20.7  &   7.6  &  19.3\\
\hline
$R - H > 4$         & 56  & 18.5  &   7.5  &  19.7\\
$I - H > 3$         & 34  & 11.7  &   5.8  &  12.9\\
\hline
\end{tabular}
\end{table}

\section{Spatial clustering}  \label{sec.xi}
In order to interpret measurements of the angular correlation amplitude one must first deconvolve the projection effect: in a sample of galaxies with a wide redshift distribution $N(z)$, the greater distance over which the sample is projected will result in greater dilution of the intrinsic clustering as compared with a sample of galaxies with a relatively narrow $N(z)$.  The statistic of interest is spatial clustering.

Following Magliocchetti \& Maddox (1999), the angular correlation function $w(\theta)$ is related to the two-point spatial correlation function $\xi(r)$ via the relativistic Limber's equation (Peebles 1980):
\begin{equation}
w(\theta) = 2 \frac{\int_0^{\infty} \int_0^{\infty} F^{-2}(x) x^4 \Phi^2(x) \xi(r,z) dx du}{[\int_0^{\infty} F^{-1}(x) x^2 \Phi(x) dx ]^2},
\end{equation}
where $x$ is the comoving coordinate, $F(x)$ gives the correction for curvature, and the selection function $\Phi(x)$ satisfies
\begin{equation}
{\cal N} = \int_0^{\infty} \Phi(x) F^{-1}(x) x^2 dx = \frac{1}{\Omega_s} \int_0^{\infty} N(z) dz,
\end{equation}
where ${\cal N}$ is the mean surface density on a surface of solid angle $\Omega_s$ and $N(z) dz$ is the number of objects in the survey within the redshift shell $(z,z+dz)$.  To proceed we assume a power-law form for $\xi(r_c,z)$:
\begin{equation}
\xi(r_c,z) = \left( \frac{r_c}{r_* (z)} \right) ^{-\gamma},
\end{equation}
where we assume $\gamma$ = 1.8, $r_c = (1+z)r$ is the comoving separation between two sources whose physical separation is given (in the small angle approximation) by
\begin{equation}
r_c \simeq \left( \frac{u^2}{F^2} + x^2 \theta ^2 \right) ^{1/2},
\end{equation}
and $r_* (z)$ is the comoving redshift-dependent correlation length. $r_* (z)$ is related to the zero-redshift correlation length $r_0$ via
\begin{equation}
r_* (z) = r_0 (1+z)^{[\gamma - \epsilon -3]/\gamma}
\end{equation}
where $\epsilon$ paramatrizes the evolution of clustering ($\epsilon = 0$ corresponds to constant clustering in proper coordinates and $\epsilon = \gamma - 3 = -1.2$ corresponds to constant clustering in comoving coordinates).  For galaxies in a relatively narrow redshift region with median redshift $z_m$, $r_*(z_m)$ is the clustering length of the galaxies at $z_m$ and is nearly independent of the assumed value of $\epsilon$.

Combining these equations gives
\begin{equation}
w(\theta) = \frac{H_{\gamma} \int_0^{\infty} N(z)^2 P(\Omega_0, z) (x(z) \theta)^{(1-\gamma)} r_* (z)^{\gamma} F(z) dz}{\frac{c}{H_0} [\int_0^{\infty} N(z) dz]^2},
\end{equation}
where $H_{\gamma} = \Gamma[1/2]\Gamma[(\gamma -1)/2]/\Gamma[\gamma /2] = 3.68$ in the case where $\gamma = 1.8$, $H_0$ is the Hubble constant, $P=(c/H_0)(dz/dx)$, and we derive $N(z)$ from photometric redshifts.  For flat cosmologies ($\Omega_0 + \Omega_{\Lambda} = 1$), $F(z) = 1$,
\begin{equation}
x(z) = \frac{c}{H_0} \Omega_0^{-1/2} \int_0^z \frac{dz}{[(1+z)^3 + \Omega_0^{-1} -1]^{1/2}},
\end{equation}
and
\begin{equation}
P(\Omega_0, z) = \Omega_0^{1/2} [(1+z)^3 + \Omega_0^{-1} -1]^{1/2}.
\end{equation}

The results for the various samples are listed in Tables \ref{table.seds}, \ref{table.eros} and \ref{table.zbins} for the cosmologies $\Omega_0 = 1$, $\Omega_{\Lambda} = 0$ and $\Omega_0 = 0.3$, $\Omega_{\Lambda} = 0.7$.  A more complete listing of results is given in Table \ref{table.all} in the appendix.  In the following discussion we use the $\Omega_0 = 0.3$, $\Omega_{\Lambda} = 0.7$ cosmology.  

From Table \ref{table.seds}, the zero-redshift correlation length, $r_0$, of the full galaxy sample, taking $\epsilon = -1.2$, is about 5.5--6 $h^{-1}$Mpc, similar but slightly larger than typical local $r_0$ values of $\sim$5--5.5 $h^{-1}$Mpc as measured in {\it optically}-selected surveys (e.g. 5.1 $h^{-1}$Mpc Loveday et al. 1995; 5.2 $h^{-1}$Mpc Postman et al. 1998; 5.4 $h^{-1}$Mpc Willmer, da Costa \& Pellegrini 1998; 4.9 $h^{-1}$Mpc Carlberg et al. 2000; 5.9 $h^{-1}$Mpc Cabanac, de Lapparent \& Hickson 2000; 4.9 $h^{-1}$Mpc Norberg et al. 2001a)

The median redshifts of the E and Scd + Im samples are about 0.5 while the Sbc sample has a median redshift of about 0.6 and the median absolute rest-frame $B$ magnitudes (as determined from photometric redshifts) are approximately $-18.6 + 5 \log h$ (E), $-19.6 + 5 \log h$ (Sbc) and $-19.1 + 5 \log h$ (Scd + Im).  For comparison, Folkes et al. (1999) provide the following values for $L_*$ galaxies in the 2dF redshift survey: $-19.6$ (E/S0), $-19.4$ (Sb) and $-19.0$ (Scd/Irr).  Thus our spectral-type E sample reaches about one magnitude fainter than $L_*$ while the bluer samples reach about $L_*$. The correlation length $r_0$ for $\epsilon = -1.2$, is about 6.5 $h^{-1}$Mpc for the E sample, 8 $h^{-1}$Mpc for the Sbc sample and 2--4 $h^{-1}$Mpc for the Scd + Im sample.  The low correlation length of blue or late-type galaxies relative to red or elliptical galaxies is well known at low redshifts (Loveday et al. 1995 find a factor of 1.3 difference; Carlberg et al. 1997 -- factor of 2.7; Guzzo et al. 1997 -- factor of 1.5; Willmer et al. 1998 -- factor of 1.3; Brown et al. 2000 -- factor of 2; Cabanac et al. 2000 -- factor of 2.8) with some variation, depending in part on the particular selection criteria.  For optically-selected surveys, this may be understood in the context that for the same optical magnitudes, red or early-type galaxies are intrinsically more massive than blue or late-type galaxies.  For a near-IR-selected catalogue this is not so much of an issue within a given absolute magnitude range, however when we compare {\it all} red galaxies with {\it all} blue galaxies in this paper, the red galaxies are on average more luminous (in, for example, the rest-frame $I$ band) and so are again expected to be intrinsically more massive than the blue galaxies.  It is interesting that the Sbc sample is more strongly clustered than the E sample.  This may be due to the different redshift and intrinsic luminosity distributions of the two samples -- $M_{I,{\rm median}}$ for the Sbc sample is $\sim 0.5$ magnitudes brighter than $M_{I,{\rm median}}$ for the E sample -- and/or the fact that we are probing to higher redshifts than local surveys so many of the brightest $z > 0.75$ spectral-type Sbc galaxies in this survey could be progenitors of present-day ellipticals.  A clearer picture is expected to emerge with the full square degree of the LCIR survey, in which increased sample size will permit clustering measurements for different spectral types as a function of redshift and within restricted absolute magnitude ranges.

The spatial correlation lengths for the ERO samples are listed in Table \ref{table.eros}.  The median redshifts of the $R - H > 4$ and $I - H > 3$ samples are in the range $1 < z < 1.2$.  Taking $\epsilon = -1.2$, or equivalently $r_*(z \sim 1)$ for any $\epsilon$, the correlation length for the $R - H > 4$ and $I - H > 3$ samples with $H < 20.0$ is $\sim$10.5 $h^{-1}$Mpc.  This decreases to $\sim$7.5 $h^{-1}$Mpc for the $H < 20.5$ sample.  These values are about 1.5--2 times the local value of $r_0$ for all galaxies.  We note that assuming an $\Omega_0 = 1$ cosmology makes the comparison much less dramatic.  Comparing with the correlation lengths of $R - H < 4$ and $I - H < 3$ galaxies in the same redshift range as the EROs ($0.7 < z < 1.5$ for $R - H > 4$ and $0.9 < z < 1.5$ for $I - H > 3$) produces a difference in the red and blue object correlation lengths of order 1.7--1.8.  This comparison is similar to the comparison of the correlation lengths of early and late or red and blue galaxies in local surveys and the factor of 1.8 difference is typical (Hermit et al. 1996; Brown et al. 2000; Cabanac et al. 2000).  We surmise therefore that the strong clustering of EROs is not extraordinary.  By comparison Willmer et al. (1998) measure $r_0 \sim$ 8.0 $h^{-1}$Mpc for local red galaxies with $M_B - 5\log h < -19.5$ and Guzzo et al. (1997) measure $r_0 \sim$ 8.4 $h^{-1}$Mpc for local early-type galaxies with $M_{Zw} - 5\log h < -19.5$ while Norberg et al. (2001b) measure $r_0 \sim 6.1$ and 7.6 $h^{-1}$Mpc for samples of type 1 (M01) galaxies in the 2dFGRS with median magnitudes $M_{B_J} - 5\log h = -20.3$ and $-$20.8 respectively. The strong clustering is evidence that the majority of the EROs are indeed massive galaxies in a restricted redshift range.  For example, contamination at a level of $\sim$20 per cent by uncorrelated galaxies over a wider redshift range or low-mass stars would increase the inferred $r_0$ of the remaining EROs by $\sim$30 per cent.

\begin{figure}
\vspace{5cm}
\includegraphics{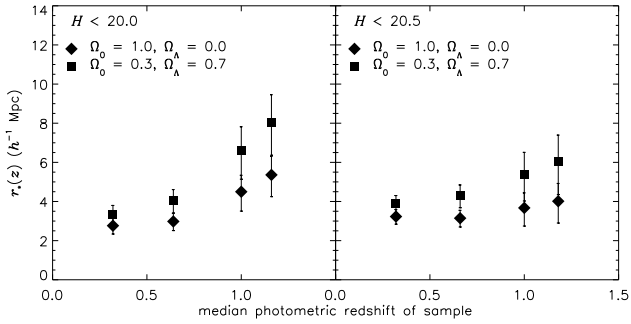}
\caption{\label{xi.zbins} The comoving correlation length, $r_*(z_m)$ ($\gamma = 1.8$), where $z_m$ is the median photometric redshift of the sample, for photometric redshift bins in the $H < 20.0$ and $H < 20.5$ catalogues.  $r_*(z_m)$ is derived via Limber's equation from the measured amplitudes of $w(\theta)$ using the photometric redshift $N(z)$ distributions and two cosmologies -- $\Omega_0 = 1$, $\Omega_{\Lambda} = 0$ and $\Omega_0 = 0.3$, $\Omega_{\Lambda} = 0.7$.  The error bars are approximately 1$\sigma$.  The increase in $r_*(z)$ to higher redshift is probably a result of the increasing average luminosity with redshift for an apparent magnitude limited survey (cf. Fig. \ref{xi.zabs}).}
\end{figure}

The photometric redshift bins $0.0 < z < 0.5$, $0.5 < z < 1.0$, $0.8 < z < 1.3$ and $1.0 < z < 1.5$ have median redshifts $z_m$ of about 0.32, 0.65, 1.0 and 1.2 respectively (Table \ref{table.zbins}).  In the $H < 20.0$ catalogue, $r_*(z_m)$ ($r_0$ for $\epsilon = -1.2$) increases dramatically over these redshift bins, being respectively about 3, 4, 6.5 and 8 $h^{-1}$Mpc (Fig. \ref{xi.zbins}).  Again we note that the increase is much weaker for an $\Omega_0 = 1$ cosmology.  The probable explanation for the increase in clustering to high redshift is simply that the galaxies visible at high redshift are on average intrinsically brighter and more massive than the galaxies at lower redshifts -- the greatest peaks of matter fluctuations are expected to be most strongly clustered (Kaiser 1984; also Kauffmann et al. 1999b; Benson et al. 2001).  Indeed the galaxies in the highest redshift bin ($1.0 < z < 1.5$) have a median absolute rest-frame $I$ magnitude of $-22.5 + 5 \log h$ compared with $-19.9 + 5 \log h$ in the lowest redshift bin and about 1--1.5 magnitudes brighter than $L_*$.  A similar dependence of $r_0$ on absolute magnitude has been measured for galaxies at low redshifts in the 2dF Galaxy Redshift Survey (Norberg et al. 2001a).  The $H < 20.5$ sample is on average fainter and the increase in $r_*$ to higher redshift is proportionally less pronounced.  If these $r_*$ values are compared with the $r_0$ values quoted in Norberg et al. (2001a) for samples with similar median absolute magnitudes to factor out this effect, then the ratio $r_*/r_0 \approx 1$ in all redshift bins.

\begin{figure}
\vspace{7cm}
\includegraphics{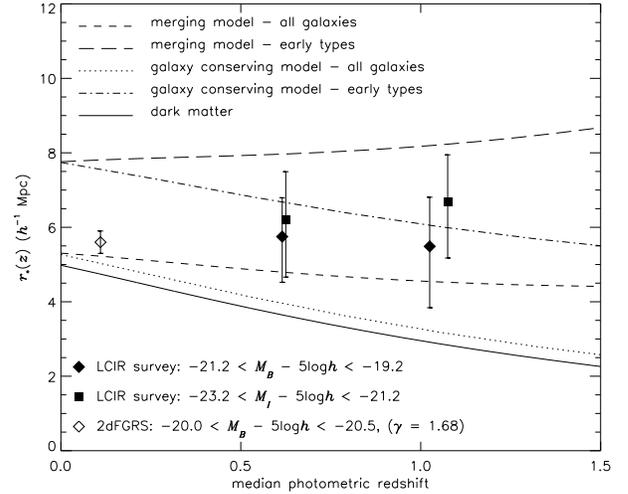}
\caption{\label{xi.zabs} The comoving correlation length, $r_*(z_m)$, where $z_m$ is the median redshift of the sample, for samples selected by absolute magnitude in two photometric redshift bins.  The correlation length $r_*(z_m)$ is derived via Limber's equation from the measured amplitudes of $w(\theta)$ assuming $\gamma = 1.8$ and using the photometric redshift $N(z)$ distributions and an $\Omega_0 = 0.3$, $\Omega_{\Lambda} = 0.7$ cosmology.  The error bars are approximately 1$\sigma$.  Results are expressed in comoving coordinates with $h = 1$.  Also plotted is a measurement from the 2dF Galaxy Redshift Survey for galaxies at $z \sim 0.1$ in a similar absolute $B$ magnitude range (using $\gamma = 1.68$) (Norberg et al. 2001a).  The lines correspond to various models of clustering evolution (assuming $\Omega_0 = 0.3$, $\Omega_{\Lambda} = 0.7$): a merging model normalized to $r_0$ of early-type galaxies in the APM (Loveday et al. 1995) (long dashes) and normalized to all galaxies in the APM (short dashes), and a galaxy conserving (Fry 1996) model similarly normalized (dot-dashed line and dotted line respectively).  The reader is referred to Moustakas \& Somerville (2001) for details of these models.  The solid line models the evolution of clustering in the underlying dark matter.  The observations are most consistent with the merging model, though the error bars still allow considerable variation.}
\end{figure}

Of more interest is the evolution of $r_*(z)$ for galaxies in a fixed absolute magnitude range and in particular for galaxies selected in the near-IR where absolute magnitude traces total stellar mass. Unfortunately the sample size available for this paper limits what we can do at this stage, however Fig. \ref{xi.zabs} shows $r_*(z)$ for two photometric redshift bins with median redshifts $z_m \sim$ 0.6 and 1.0 and $-$23.2 $<$ $M_I - 5 \log h$ $<$ $-$21.2.  These results are also summarized in Table \ref{table.absmag}.  Though there is still some increase in average absolute magnitude with redshift within the selected range, the spatial correlation length does not change significantly with redshift.  Isolating only the spectral-type E and Sbc galaxies (to reduce the effect of the variation with redshift of the spectral type composition of the sample) further reduces the correlation lengths in the higher redshift bin relative to the lower redshift bin.  This nearly flat evolution of $r_*$ for luminous red-selected galaxies to $z \sim 1$ is consistent with model predictions for massive red or early-type galaxies (e.g. Kauffmann et al. 1999b) and the merging model shown in Fig. \ref{xi.zabs}.  In contrast, previous {\it optically}-selected surveys have noted a drop in $r_*$ over the same redshifts, ranging from a slight decrease (Carlberg et al. 2000; Hogg et al. 2000; Teplitz et al. 2001) to a decrease of up to $\sim$50 per cent (Le F\`evre et al. 1996; Adelberger 2000; Cabanac et al. 2000).  Depending on the selection band, optical surveys include a mixture of red and blue galaxies at low redshifts but become increasingly dominated by bluer spectral types at higher redshifts as red galaxies drop out of the optical filters.  This selection effect could lead to a decrease in the observed clustering.  Conversely, near-IR selection avoids this effect.  Likewise, models (Kauffmann et al. 1999b) predict that the typical halo mass of galaxies selected in a fixed rest-frame absolute magnitude range decreases more rapidly with increasing redshift for bluer selection bands, leading to a corresponding decrease in $r_*$ to $z \sim 1.5$ for these galaxies, which is less pronounced for redder rest-frame selection bands.

Also in Fig. \ref{xi.zabs} we plot $r_*(z)$ for galaxies selected to have absolute $B$ magnitude in the range $-21.2 < M_B - 5 \log h < -19.2$ along with a measurement at $z \sim 0.1$ for galaxies in a similar absolute $B$ magnitude range in the 2dF Galaxy Redshift Survey (Norberg et al. 2001a).  Since we use $\gamma = 1.8$ while Norberg et al. use $\gamma = 1.68$ for this particular sample, it is useful also to calculate $\sigma_8$ -- the r.m.s. fluctuations in the galaxy distribution at the scale of 8 $h^{-1}$Mpc -- which is independent of $\gamma$.  Following Peebles (1980) and Magliocchetti et al. (2000), we calculate $\sigma_8$ from $r_*$ and $\gamma$ using
\begin{equation}
\sigma_8(z) = \left[ \left( \frac{r_*(z)}{8} \right)^{\gamma} c_{\gamma} \right]^{1/2},\label{eq.sigma8}\\
\end{equation}
where 
\begin{equation}
c_{\gamma} = \frac{72}{(3-\gamma)(4-\gamma)(6-\gamma)2^{\gamma}}.\label{eq.sigma8_c}\\
\end{equation} 
The resulting values are $\sigma_8$ = 0.95$\pm$0.07, 1.01$\pm$0.17 and 0.97$\pm$0.24 at $z_m$ $\approx$ 0.1, 0.6 and 1.0 respectively, showing little evolution with redshift.

A potential source of error in our spatial clustering estimates arises from the differences between the photometric redshift distribution of a sample of galaxies and the sample's actual redshift distribution.  For example, a sample chosen to have photometric redshifts in the interval $0.5 < z < 1.0$ will include some galaxies with actual redshifts outside this range which will dilute the measured clustering signal leading to an underestimate of $r_{*}$.  Other situations can result in biases in the opposite direction.  Since particular SED/redshift combinations give more robust photometric redshifts than others the effect can vary depending on the sample.  It is possible to address this problem to some extent with Monte Carlo simulations using simulated catalogues based on the data or on models with noise added to mimic the observations and, in Limber's equation, using $N(z)$ based on the original redshifts for photometrically-selected subsamples.  We found that the results presented here for spectral-type, magnitude and colour selected samples did not change by more than a few percent with respect to this effect while the values of $r_*$ in photometric redshift bins increased (as expected) by $\sim$10--20 per cent. This does not affect our conclusions significantly.

\begin{table*}
\centering
\caption{\label{table.absmag}  The number of galaxies $N$, median photometric redshift $z_m$, median absolute $B$ and $I$ magnitudes $M_B$ and $M_I$ ($h = 1$), amplitude $A$ of $w(\theta)$ at 1 arcmin and spatial correlation lengths $r$ in $h^{-1}$Mpc (comoving coordinates) for absolute $B$ and $I$ magnitude limited samples in two photometric redshift bins.  The $M_B$ samples are selected to have $-21.2 < M_B - 5 \log h < -19.2$ while the $M_I$ samples are selected to have $-23.2 < M_I - 5 \log h < -21.2$.  We fix $\gamma = 1.8$.  The correlation length is calculated for two cosmologies $\Omega_0 = 1$, $\Omega_{\Lambda} = 0$ and $\Omega_0 = 0.3$, $\Omega_{\Lambda} = 0.7$.  $N(z)$ is derived from photometric redshifts.  The errors on $r_{0}$ ($\epsilon = -1.2$) are approximately 1$\sigma$ and are carried through from the $\sqrt{(1+w(\theta))/DD}$ errors on $w(\theta)$.  The errors on the other correlation lengths are of similar proportions.  $r_{*}(z_m)$ is listed for $\epsilon = 0$.  For $\epsilon = -1.2$, $r_{*}(z)$ equals $r_0$ for any $z$.  Since galaxies are localized into redshift bins, $r_*(z_m)$ is nearly independent of an assumed $\epsilon$.  Extra statistics are included in Table \ref{table.all} in the appendix.}
\begin{tabular}{llcccccccc}
\multicolumn{2}{l}{\it sample} & $N$ & $z_{m}$ & $M_B$ & $M_I$ & $r_0^{\epsilon = -1.2}$ & $r_{*}^{\epsilon = 0}(z_{m})$ & $r_0^{\epsilon = -1.2}$ & $r_{*}^{\epsilon = 0}(z_{m})$ \\ \hline
&&&&&&\multicolumn{2}{c}{$\Omega_0 = 0.3$, $\Omega_{\Lambda} = 0.7$}&\multicolumn{2}{c}{$\Omega_0 = 1$, $\Omega_{\Lambda} = 0$}\\ \hline
$M_B$: all SEDs&  0.3 $< z <$ 0.8& 458& 0.62& $-$19.7& $-$21.4& 5.7$\pm$1.1& 5.8& 4.2$\pm$0.8& 4.2\\
$M_B$: all SEDs&  0.8 $< z <$ 1.4& 497& 1.03& $-$20.1& $-$21.9& 5.5$\pm$1.5& 5.5& 3.7$\pm$1.0& 3.7\\
$M_B$: E $+$ Sbc& 0.3 $< z <$ 0.8& 272& 0.62& $-$19.7& $-$21.7& 7.2$\pm$1.6& 7.2& 5.3$\pm$1.2& 5.3\\
$M_B$: E $+$ Sbc& 0.8 $< z <$ 1.4& 330& 1.03& $-$20.1& $-$22.0& 6.1$\pm$2.1& 6.0& 4.1$\pm$1.4& 4.1\\
\hline
$M_I$: all SEDs&  0.3 $< z <$ 0.8& 339& 0.62& $-$19.8& $-$21.6& 6.2$\pm$1.4& 6.2& 4.6$\pm$1.0& 4.6\\
$M_I$: all SEDs&  0.8 $< z <$ 1.5& 518& 1.07& $-$20.3& $-$22.0& 6.7$\pm$1.4& 6.7& 4.5$\pm$0.9& 4.5\\
$M_I$: E $+$ Sbc& 0.3 $< z <$ 0.8& 272& 0.61& $-$19.7& $-$21.7& 7.0$\pm$1.6& 7.0& 5.2$\pm$1.2& 5.2\\
$M_I$: E $+$ Sbc& 0.8 $< z <$ 1.5& 355& 1.05& $-$20.2& $-$22.1& 7.0$\pm$1.9& 7.0& 4.8$\pm$1.3& 4.8\\
\hline
\end{tabular}
\end{table*}

\section{Discussion}  \label{sec.discusion}
In this paper we have presented an analysis of 4019 $H < 20.5$ galaxies in 744 arcmin$^2$ of the LCIR optical/near-IR survey.  Photometric redshifts place the majority of these galaxies in the redshift range $0 < z < 1.5$ with a median redshift of $\sim 0.55$. Our analysis includes comparisons of colour distributions, number counts, $N(z)$ distributions for different spectral types and ERO number counts with a semi-analytic hierarchical merger model and no-evolution and passive-evolution models.  We also calculate angular and spatial clustering statistics for different magnitude, colour, spectral type and photometric redshift-selected subsamples out to $z \sim 1.2$.  The main results are summarized below.

\begin{itemize}
\item The LCIR number counts (Fig. \ref{hcounts}) agree well with previous surveys.  The faint end ($18 < H < 20$) slope in the $H$ number counts is 0.38 which compares with 0.31 for $H > 20$ (Yan et al. 1998) and 0.47 for $H < 19$ (Martini 2001a).  The number density of EROs with $R - H > 4$ and $H < 20$ is 0.5 arcmin$^{-2}$ which agrees well with 0.5 arcmin$^{-2}$ for $R - K_s > 5$ and $K_s < 19$ in Daddi et al. (2000a).
\item Both the total $H$ number counts and ERO number counts are well-fitted by the no-evolution model, however the no-evolution model underestimates the number of galaxies in the highest redshift bins, indicating some luminosity evolution or merging in the observations.
\item The passive-evolution models fail to fit the observed $N(z)$, predicting too many bright galaxies at high redshifts.  Similarly, passive-evolution models generally predict too many EROs, especially if the majority of massive elliptical galaxies are hypothesised to form at high redshifts and evolve passively thereafter, though it is certainly possible that this scenario is true for some fraction of the elliptical galaxy population.  Fine-tuning of star formation histories, formation redshifts and the inclusion of dust-obscured star-formation may be used to improve the match between these models and the observations (e.g. Daddi et al. 2000b) but need to be tested for consistency with observations at all wavelengths.
\item The semi-analytic model fits the observed total $N(z)$ but predicts too few spectral-type E galaxies and too few EROs relative to bluer spectral types.  For example in the magnitude range $19 < H < 20$, 18 per cent of observed galaxies have colours $R - H > 4$ while in the semi-analytic model the fraction is only 4.3 per cent.  However the semi-analytic model does appear to produce enough bright galaxies at $z \sim 1$.  Hence the deficit of EROs could be due, for example, to some process inhibiting late star formation in massive galaxies at $z \sim 1$ or to some of the EROs being very dusty ULIRG-like galaxies, either of which may not be modelled properly.  Existing spectroscopic, sub-mm and morphological observations suggest that both dusty-starbursts and evolved ellipticals are present in the ERO population with the larger fraction being evolved ellipticals.  While strong clustering of EROs indicates that they are mostly massive galaxies in a restricted redshift range, it does not necessarily distinguish between evolved elliptical galaxies at $z \sim 1$ and massive ULIRG-type galaxies at $z \sim 1$ or greater, both of which may be progenitors of present day massive elliptical galaxies (Blain et al. 1999b; de Zotti et al. 2001; Scott et al. 2001).  Other evidence such as the tightness of cluster elliptical colour-magnitude relations (Ellis et al. 1997) and the slow evolution of the fundamental plane zeropoint in cluster and field ellipticals (van Dokkum et al. 2001) also indicate that late star formation is inhibited in some massive ellipticals though other $z < 1$ elliptical galaxies (particularly the less massive ones) show features of recent star formation (e.g. Franceschini et al. 1998; Abraham et al. 1999; Menanteau, Abraham \& Ellis 2001; Im et al. 2001) possibly induced by merger events.
\item For the full galaxy sample we find $r_0 \sim$ 5.5--6 $h^{-1}$Mpc ($\epsilon = -1.2$), which is  slightly larger than the value of $r_0 \sim$ 5--5.5 $h^{-1}$Mpc typically measured in low-redshift {\it optically}-selected surveys (Loveday et al. 1995; Postman et al. 1998; Carlberg et al. 2000; Norberg et al. 2001a). The enhanced clustering in the LCIR survey is expected as a result of the increased fraction of red galaxies in a near-IR-selected sample as opposed to an optically-selected sample.
\item We find that redder (spectral-type E + Sbc) galaxies are more strongly clustered than bluer (spectral-type Scd + Im) galaxies with a factor of $\sim 2$ difference in $r_0$.  This is consistent with the results of low-redshift galaxy surveys (Loveday et al. 1995; Hermit et al. 1996; Cabanac et al. 2000; Brown et al. 2000).
\item We find that the angular clustering amplitude of EROs defined by colours $R - H > 4$ or $I - H > 3$ is a factor of 2.5--4 times that of the whole galaxy population at the same limiting $H$ magnitude (Fig. \ref{wth.eros}).  This is partly due to the restricted redshift range ($z \sim$ 0.8--1.4) of these galaxies (Fig. \ref{eros.rh4} \& \ref{eros.ih3}).  The corresponding spatial correlation lengths are $r_*(z=1) \sim$ 7.5--10.5 $h^{-1}$Mpc ($\sigma_8$ = 1.3--1.7).  At $z \sim 1$, only galaxies at the bright end of the luminosity function make it into the sample and for such galaxies a fairly large correlation length is to be expected (Norberg et al. 2001a).  The correlation lengths of the EROs are a factor of $\sim$1.8 times those of $R - H < 4$ and $I - H < 3$ galaxies selected in the same redshift range -- a difference comparable with that found between red and blue galaxies at low redshift.
\item The correlation length $r_*(z)$ in photometric redshift bins increases dramatically from $z = 0$ to $z \sim 1.2$ (Fig. \ref{xi.zbins}) which we interpret to be a result of the systematic increase in average luminosity of the sample to higher redshifts as a result of selecting by apparent magnitude.  When $r_*(z)$ is calculated for galaxies selected by absolute rather than apparent magnitude there is no significant increase in $r_*(z)$ with redshift over this range (Fig. \ref{xi.zabs}).  This is consistent with semi-analytic model predictions for red galaxies (Kauffmann et al. 1999b).
\end{itemize}

\begin{figure}
\vspace{7cm}
\includegraphics{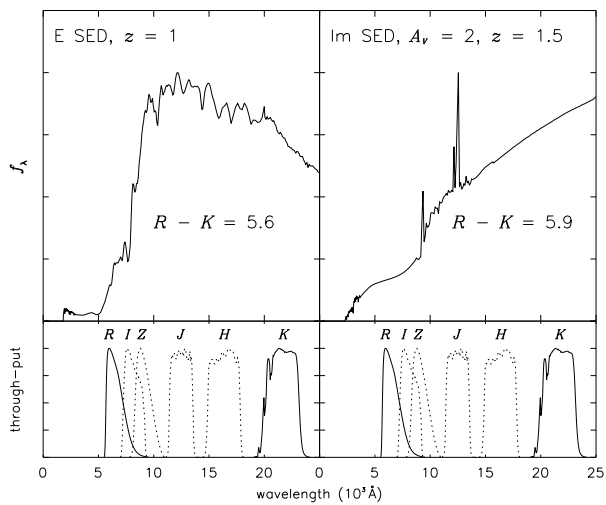}
\caption{\label{EvDSB} Comparison of a typical elliptical galaxy spectrum at $z = 1$ with an irregular galaxy spectrum at $z = 1.5$ with two magnitudes of extinction in the $V$ band (CABKKS).  Both have very red $R - K$ colours -- the colour most frequently used in the literature to define EROs.  On the otherhand, an $R - H$ or $I - H$ red colour cut is more sensitive to a prominent \mbox{4000 \AA} break than the smoother cut-off in dusty galaxies, while the intervening $Z$ and $J$ filters may be used to distinguish evolved galaxies from dusty non-evolved galaxies.  We note that other reddening laws, such as Seaton (1979; Milky Way), Prevot et al. (1984; SMC) and Bouchet et al. (1985; SMC), produce redder $R - K$ colours ($\sim$6.6) in the right-hand plot for the same $A_V$.}
\end{figure}

In the present paper, in calculating correlation statistics for various subsamples of the data, we are still restricted by some of the problems that our photometric redshift approach was designed to overcome -- specifically because of low sample sizes we are unable to calculate correlation statistics as a function of spectral type and redshift simultaneously and the results we do show have large error bars.  Similarly we are generally unable to impose absolute magnitude limits.  With a larger sample size (e.g. the square degree of the completed LCIR survey) it will be possible to compare the clustering of galaxies of the same spectral type and intrinsic luminosity at different redshifts.  In particular for a near-IR-selected sample this may be closely tied to the total stellar mass of galaxies rather than being sensitive to recent star-formation activity as are optically-selected samples.  The larger sample will also allow us to push out to $z \sim$ 1.5--2 with better statistics.  Addition of extra filters, giving more complete spectral coverage, will allow us to place stronger constraints on the effects of dust on our results (Fig. \ref{EvDSB}; see also Moustakas 1999; Pozzetti \& Mannucci 2000).

As follow-up to this survey we are engaged in a programme of follow-up spectroscopy with the Keck and Magellan telescopes in order to better characterize the ERO population and to confirm photometric redshift accuracy in the range $0.5 < z < 1.5$.\\
\\
\section{Summary}  \label{sec.summary}
\begin{itemize}
\item Simple passive-evolution models generally over-\linebreak estimate the observed number of high-redshift galaxies, including EROs.
\item The hierarchical merger model fits the overall $N(z)$ but under-estimates the number-density of EROs by a factor of four (Fig. \ref{zcum} \& \ref{eros.rh4}).
\item For all $\sim$4000 galaxies at $H <$ 20--20.5 we find comoving $r_0 \sim$5.5 $h^{-1}$Mpc.
\item We find that red galaxies have $r_0$ twice that of blue galaxies.
\item For EROs, $r_0$ $\sim$7.5--10.5 $h^{-1}$Mpc which is twice that of non-EROs selected in the same redshift range (Table \ref{table.eros}).
\item For an apparent $H$ magnitude-selected sample, $r_0$ increases dramatically with redshift (Fig. \ref{xi.zbins}) but for an absolute rest-frame $I$ magnitude-selected sample there is no significant increase in $r_0$ to $z \sim 1$ (Fig. \ref{xi.zabs}).
\end{itemize}

\section*{Acknowledgments}
We would like to express our appreciation to the Goddard Space Flight Center group for obtaining, calibrating and making public the wide-field $UBVRI$ images used in this paper, and to the creators of {\tt hyperz} for releasing their excellent photometric redshift code to the astronomy community.  Some {\tt hyperz} software was also used in the construction of the passive and no-evolution models in this paper.  We would also like to thank the very helpful staff at Las Campanas Observatory.  The construction of the infrared camera, CIRSI, was made possible by a generous grant from the Raymond and Beverly Sackler Foundation.  AF is supported by an Isaac Newton Studentship, L. B. Wood Travelling scholarship, ORS award and Trinity College, Cambridge.

\begin{landscape}
\begin{table}
\begin{tiny}
\hspace*{-2.3cm}
\centering
\caption{\label{table.all} The number of galaxies $N$, median photometric redshift $z_m$, median absolute $B$ and $I$ magnitudes $M_B$ and $M_I$ ($h = 1$), amplitude $A$ of $w(\theta)$ at 1 arcmin and spatial correlation lengths $r$ in $h^{-1}$Mpc (comoving coordinates) for various subsamples of the LCIR survey selected by magnitude, colour, spectral type and photometric redshift.  The correlation length is calculated for two cosmologies $\Omega_0 = 1$, $\Omega_{\Lambda} = 0$ and $\Omega_0 = 0.3$, $\Omega_{\Lambda} = 0.7$, $N(z)$ is derived from photometric redshifts and $\gamma$ is fixed to 1.8.  The errors on $r_{0}$ ($\epsilon = -1.2$) are approximately 1$\sigma$ and are carried through from the $\sqrt{(1+w(\theta))/DD}$ errors on $w(\theta)$.  The errors on the other correlation lengths are of similar proportions.  $r_0 = r_{*}(z = 0)$, $r_{*}(z_m)$, $r_{*}(z = 0.5)$ and $r_{*}(z = 1)$ are listed for $\epsilon = 0$.  For $\epsilon = -1.2$, $r_{*}(z)$ equals $r_0$ for any $z$.  The subsamples marked $M_B$ are selected to have absolute $B$ magnitudes in the range $-21.2 < M_B - 5 \log h < -19.2$ while those marked $M_I$ are selected to have absolute $I$ magnitudes in the range $-23.2 < M_I - 5 \log h < -21.2$.}
\begin{tabular}{lllccccccccccccccc}
\multicolumn{2}{l}{\it sample} & {\it catalogue} & $N$ & $z_{m}$ & $M_B$ & $M_I$ & $A$ & $r_0^{\epsilon = -1.2}$ & $r_0^{\epsilon = 0}$ & $r_{*}^{\epsilon = 0}(z_{m})$ & $r_{*}^{\epsilon = 0}(0.5)$ & $r_{*}^{\epsilon = 0}(1.0)$ & $r_0^{\epsilon = -1.2}$ & $r_0^{\epsilon = 0}$ & $r_{*}^{\epsilon = 0}(z_{m})$ & $r_{*}^{\epsilon = 0}(0.5)$ & $r_{*}^{\epsilon = 0}(1.0)$ \\ \hline
&&&&&&&&\multicolumn{5}{c}{$\Omega_0 = 0.3$, $\Omega_{\Lambda} = 0.7$}&\multicolumn{5}{c}{$\Omega_0 = 1$, $\Omega_{\Lambda} = 0$}\\ \hline
\multicolumn{2}{l}{All stars}           & $H < 20.0$& 2224& -   & -    & -    & 0.00&  -  &  -  &  -  &  -  &  -  &  -  &  -  &  -  &  -  & -  \\
\multicolumn{2}{l}{$H < 19.0$}          & $H < 20.0$& 1427& -   & -    & -    & 0.00&  -  &  -  &  -  &  -  &  -  &  -  &  -  &  -  &  -  & -  \\
\multicolumn{2}{l}{$19.0 < H < 19.5$}   & $H < 20.0$&  387& -   & -    & -    & 0.03&  -  &  -  &  -  &  -  &  -  &  -  &  -  &  -  &  -  & -  \\
\multicolumn{2}{l}{$19.5 < H < 20.0$}   & $H < 20.0$&  410& -   & -    & -    & 0.04&  -  &  -  &  -  &  -  &  -  &  -  &  -  &  -  &  -  & -  \\
\hline
\multicolumn{2}{l}{All galaxies}    & $H < 20.0$& 3177& 0.52& $-$19.1& $-$20.9& 0.08&  5.5$\pm$0.4&  6.9&  5.2&  5.3&  4.4& 4.3$\pm$0.3&  5.5&  4.2&  4.2& 3.5\\
\multicolumn{2}{l}{E}               & $H < 20.0$& 1194& 0.50& $-$18.7& $-$20.9& 0.13&  6.4$\pm$0.8&  8.1&  6.2&  6.2&  5.1& 5.1$\pm$0.7&  6.4&  4.9&  4.9& 4.0\\
\multicolumn{2}{l}{Sbc}             & $H < 20.0$&  963& 0.61& $-$19.6& $-$21.3& 0.12&  7.7$\pm$1.3& 10.2&  7.4&  7.8&  6.4& 5.8$\pm$1.0&  7.8&  5.7&  6.0& 4.9\\
\multicolumn{2}{l}{Scd $+$ Im}      & $H < 20.0$& 1020& 0.48& $-$19.1& $-$20.5& 0.01&  1.9$\pm$1.9&  2.2&  1.7&  1.7&  1.4& 1.5$\pm$1.5&  1.9&  1.4&  1.4& 1.2\\
\hline
\multicolumn{2}{l}{All galaxies}    & $H < 20.5$& 2616& 0.57& $-$19.0& $-$20.8& 0.08&  6.1$\pm$0.4&  7.7&  5.7&  5.9&  4.8& 4.7$\pm$0.3&  6.1&  4.5&  4.6& 3.8\\
\multicolumn{2}{l}{E}               & $H < 20.5$&  910& 0.53& $-$18.5& $-$20.7& 0.12&  6.6$\pm$0.9&  8.3&  6.3&  6.4&  5.3& 5.2$\pm$0.7&  6.6&  5.0&  5.0& 4.2\\
\multicolumn{2}{l}{Sbc}             & $H < 20.5$&  735& 0.65& $-$19.5& $-$21.3& 0.12&  8.2$\pm$1.4& 11.0&  7.8&  8.4&  6.9& 6.1$\pm$1.1&  8.3&  5.9&  6.3& 5.2\\
\multicolumn{2}{l}{Scd $+$ Im}      & $H < 20.5$&  971& 0.52& $-$19.0& $-$20.4& 0.05&  4.2$\pm$1.4&  5.1&  3.9&  3.9&  3.2& 3.4$\pm$1.1&  4.2&  3.2&  3.2& 2.7\\
\hline
\multicolumn{2}{l}{$H < 19$ galaxies}   & $H < 20.0$& 1284& 0.42& $-$19.2& $-$21.1& 0.12&  5.3$\pm$0.7&  6.5&  5.1&  5.0&  4.1& 4.3$\pm$0.6&  5.3&  4.2&  4.0& 3.3\\
\multicolumn{2}{l}{$18.0 < H < 19.0$}   & $H < 20.0$&  868& 0.47& $-$19.1& $-$21.0& 0.10&  5.1$\pm$1.2&  6.5&  5.0&  4.9&  4.1& 4.0$\pm$0.9&  5.1&  4.0&  3.9& 3.2\\
\multicolumn{2}{l}{$18.5 < H < 19.5$}   & $H < 20.0$& 1292& 0.55& $-$19.1& $-$20.9& 0.07&  5.0$\pm$1.1&  6.4&  4.8&  4.9&  4.0& 3.9$\pm$0.8&  5.0&  3.7&  3.8& 3.2\\
\multicolumn{2}{l}{$19.0 < H < 20.0$}   & $H < 20.0$& 1893& 0.61& $-$19.1& $-$20.8& 0.06&  5.4$\pm$0.9&  7.1&  5.1&  5.4&  4.4& 4.1$\pm$0.7&  5.5&  4.0&  4.2& 3.5\\
\multicolumn{2}{l}{$19.5 < H < 20.5$}   & $H < 20.5$& 1472& 0.69& $-$18.9& $-$20.5& 0.07&  6.1$\pm$1.0&  8.0&  5.6&  6.1&  5.0& 4.7$\pm$0.7&  6.3&  4.4&  4.8& 3.9\\
\hline
\multicolumn{2}{l}{0.0 $< z <$ 0.5} & $H < 20.0$& 1458& 0.32& $-$18.1& $-$19.9& 0.10&  3.3$\pm$0.5&  4.0&  3.3&  3.0&  2.5& 2.8$\pm$0.4&  3.3&  2.7&  2.5& 2.1\\
\multicolumn{2}{l}{0.5 $< z <$ 1.0} & $H < 20.0$& 1330& 0.64& $-$19.5& $-$21.3& 0.10&  4.0$\pm$0.6&  5.6&  4.0&  4.2&  3.5& 3.0$\pm$0.4&  4.1&  2.9&  3.1& 2.6\\
\multicolumn{2}{l}{0.8 $< z <$ 1.3} & $H < 20.0$&  510& 1.00& $-$20.2& $-$22.1& 0.19&  6.6$\pm$1.3& 10.4&  6.5&  7.9&  6.5& 4.5$\pm$0.9&  7.1&  4.5&  5.4& 4.5\\
\multicolumn{2}{l}{1.0 $< z <$ 1.5} & $H < 20.0$&  343& 1.16& $-$20.7& $-$22.5& 0.30&  8.0$\pm$1.5& 13.3&  8.0& 10.2&  8.4& 5.4$\pm$1.0&  8.9&  5.3&  6.8& 5.6\\
\hline
\multicolumn{2}{l}{0.0 $< z <$ 0.5} & $H < 20.5$& 1056& 0.32& $-$17.8& $-$19.5& 0.13&  3.9$\pm$0.4&  4.6&  3.8&  3.5&  2.9& 3.2$\pm$0.4&  3.8&  3.2&  2.9& 2.4\\
\multicolumn{2}{l}{0.5 $< z <$ 1.0} & $H < 20.5$& 1101& 0.66& $-$19.3& $-$21.0& 0.10&  4.3$\pm$0.6&  6.0&  4.2&  4.5&  3.8& 3.1$\pm$0.4&  4.4&  3.1&  3.3& 2.8\\
\multicolumn{2}{l}{0.8 $< z <$ 1.3} & $H < 20.5$&  524& 1.00& $-$20.0& $-$21.8& 0.13&  5.4$\pm$1.2&  8.5&  5.3&  6.5&  5.3& 3.7$\pm$0.8&  5.8&  3.7&  4.4& 3.7\\
\multicolumn{2}{l}{1.0 $< z <$ 1.5} & $H < 20.5$&  369& 1.18& $-$20.5& $-$22.3& 0.17&  6.0$\pm$1.5& 10.0&  6.0&  7.7&  6.3& 4.0$\pm$1.0&  6.7&  4.0&  5.1& 4.2\\
\hline
$M_B$: all SEDs&  0.3 $< z <$ 0.8& $H < 20.5$&  458& 0.62& $-$19.7& $-$21.4& 0.17&  5.7$\pm$1.1&  7.9&  5.8&  6.1&  5.0& 4.2$\pm$0.8&  5.8&  4.2&  4.5& 3.7\\
$M_B$: all SEDs&  0.8 $< z <$ 1.4& $H < 20.5$&  497& 1.03& $-$20.1& $-$21.9& 0.12&  5.5$\pm$1.5&  8.8&  5.5&  6.7&  5.5& 3.7$\pm$1.0&  5.9&  3.7&  4.5& 3.7\\
$M_B$: E $+$ Sbc& 0.3 $< z <$ 0.8& $H < 20.5$&  272& 0.62& $-$19.7& $-$21.7& 0.26&  7.2$\pm$1.6&  9.9&  7.2&  7.5&  6.2& 5.3$\pm$1.2&  7.3&  5.3&  5.6& 4.6\\
$M_B$: E $+$ Sbc& 0.8 $< z <$ 1.4& $H < 20.5$&  330& 1.03& $-$20.1& $-$22.0& 0.14&  6.1$\pm$2.1&  9.7&  6.0&  7.4&  6.1& 4.1$\pm$1.4&  6.6&  4.1&  5.0& 4.1\\
$M_I$: all SEDs&  0.3 $< z <$ 0.8& $H < 20.5$&  339& 0.62& $-$19.8& $-$21.6& 0.20&  6.2$\pm$1.4&  8.5&  6.2&  6.5&  5.4& 4.6$\pm$1.0&  6.3&  4.6&  4.8& 4.0\\
$M_I$: all SEDs&  0.8 $< z <$ 1.5& $H < 20.5$&  518& 1.07& $-$20.3& $-$22.0& 0.14&  6.7$\pm$1.4& 10.8&  6.7&  8.2&  6.8& 4.5$\pm$0.9&  7.3&  4.5&  5.6& 4.6\\
$M_I$: E $+$ Sbc& 0.3 $< z <$ 0.8& $H < 20.5$&  272& 0.61& $-$19.7& $-$21.7& 0.25&  7.0$\pm$1.6&  9.6&  7.0&  7.3&  6.1& 5.2$\pm$1.2&  7.1&  5.2&  5.4& 4.5\\
$M_I$: E $+$ Sbc& 0.8 $< z <$ 1.5& $H < 20.5$&  355& 1.05& $-$20.2& $-$22.1& 0.17&  7.0$\pm$1.9& 11.3&  7.0&  8.6&  7.1& 4.8$\pm$1.3&  7.7&  4.8&  5.9& 4.8\\
\hline
$R - H > 4$& 0.7 $< z < 1.5$& $H < 20.0$&  337& 1.01& $-$20.2& $-$22.3& 0.33& 11.1$\pm$2.0& 17.6& 11.1& 13.5& 11.1& 7.6$\pm$1.4& 12.0&  7.5&  9.2& 7.6\\
$R - H < 4$& 0.7 $< z < 1.5$& $H < 20.0$&  480& 0.87& $-$20.2& $-$21.8& 0.15&  5.9$\pm$1.7&  8.8&  5.8&  6.7&  5.5& 4.1$\pm$1.2&  6.2&  4.1&  4.7& 3.9\\
$I - H > 3$& 0.9 $< z < 1.5$& $H < 20.0$&  201& 1.13& $-$20.4& $-$22.5& 0.36& 10.5$\pm$2.9& 17.3& 10.4& 13.2& 10.9& 7.0$\pm$1.9& 11.6&  7.0&  8.8& 7.3\\
$I - H < 3$& 0.9 $< z < 1.5$& $H < 20.0$&  268& 1.05& $-$20.6& $-$22.2& 0.17&  6.0$\pm$2.8&  9.6&  5.9&  7.3&  6.0& 4.0$\pm$1.9&  6.5&  4.0&  5.0& 4.1\\
$R - H > 4$& 0.7 $< z < 1.5$& $H < 20.5$&  312& 1.02& $-$19.8& $-$21.9& 0.17&  7.7$\pm$2.4& 12.1&  7.6&  9.2&  7.6& 5.2$\pm$1.6&  8.3&  5.2&  6.3& 5.2\\
$R - H < 4$& 0.7 $< z < 1.5$& $H < 20.5$&  516& 0.89& $-$20.0& $-$21.6& 0.09&  4.7$\pm$1.7&  7.0&  4.6&  5.3&  4.4& 3.3$\pm$1.2&  4.9&  3.2&  3.7& 3.1\\
$I - H > 3$& 0.9 $< z < 1.5$& $H < 20.5$&  170& 1.16& $-$20.3& $-$22.4& 0.20&  7.5$\pm$3.7& 12.5&  7.5&  9.5&  7.9& 5.0$\pm$2.5&  8.3&  5.0&  6.4& 5.2\\
$I - H < 3$& 0.9 $< z < 1.5$& $H < 20.5$&  306& 1.08& $-$20.4& $-$22.0& 0.09&  4.4$\pm$3.3&  7.1&  4.4&  5.5&  4.5& 3.0$\pm$2.2&  4.8&  3.0&  3.7& 3.0\\
\hline
\end{tabular}
\end{tiny}
\end{table}
\end{landscape}

\bsp

\label{lastpage}

\end{document}